\documentclass[11pt]{article}
\usepackage[utf8]{inputenc}
\usepackage{amsfonts}
\usepackage{amsmath}
\usepackage{amssymb}
\usepackage{indentfirst}
\usepackage{graphicx}
\usepackage[dvipsnames]{xcolor}
\usepackage[colorlinks]{hyperref}
\usepackage{cite}
\usepackage{array}
\usepackage{microtype}
\usepackage{soul}
\usepackage[normalem]{ulem}
\usepackage{cancel}
\usepackage{soul}
\usepackage{xspace}
\usepackage{colortbl}
\usepackage{caption}
\newcolumntype{C}[1]{>{\centering\arraybackslash}p{#1}}
\definecolor{maroon}{cmyk}{0,0.87,0.68,0.32}

\numberwithin{equation}{section}
\allowdisplaybreaks

\makeatletter

\begin{document}

\begin{titlepage}
\vspace{3cm}

\baselineskip=24pt

\begin{center}
\textbf{\LARGE Non-Lorentzian Supergravity and Kinematical Superalgebras}
\par\end{center}{\LARGE \par}

\begin{center}
	\vspace{1cm}
	\textbf{Patrick Concha}$^{\ast, \bullet}$,
        \textbf{Lucrezia Ravera}$^{\dag, \ddag, \bullet}$
	\small
	\\[5mm]
    $^{\ast}$\textit{Departamento de Matemática y Física Aplicadas, }\\
	\textit{ Universidad Católica de la Santísima Concepción, }\\
\textit{ Alonso de Ribera 2850, Concepción, Chile.}
\\[2mm]
$^{\bullet}$\textit{Grupo de Investigación en Física Teórica, GIFT, }\\
	\textit{ Universidad Católica de la Santísima Concepción, }\\
\textit{ Alonso de Ribera 2850, Concepción, Chile.}
\\[2mm]
	$^{\dag}$\textit{DISAT, Politecnico di Torino - PoliTo, }\\
	\textit{ Corso Duca degli Abruzzi 24, 10129 Torino, Italy.}
	\\[2mm]
	$^{\ddag}$\textit{Istituto Nazionale di Fisica Nucleare, Section of Torino - INFN,}\\
	\textit{ Via P. Giuria 1, 10125 Torino, Italy.}
	 \\[5mm]
	\footnotesize
	\texttt{patrick.concha@ucsc.cl},
        \texttt{lucrezia.ravera@polito.it}
	\par\end{center}
\vskip 26pt
\begin{abstract}

In this paper, we present and classify the supersymmetric extensions of extended kinematical algebras, at the basis of non-Lorentzian physics theories. The diverse kinematical superalgebras are here derived by applying non- and ultra-relativistic expansion procedures involving different semigroups. We then build three-dimensional Chern-Simons non-Lorentzian supergravity theories based on such (extended) kinematical superalgebras, providing the supersymmetry transformation laws of the fields and the field equations of the models, which correspond to the vanishing of the curvature two-forms. In fact, the expansion procedure adopted allows to automatically end up with a non-degenerate bilinear invariant trace for the (extended) kinematical superalgebras. The latter is a crucial ingredient of the Chern-Simons field-theoretical formulation, as it allows to include a kinetic term for each gauge field of the theory, implying the vanishing of the curvature two-forms
as field equations.

\end{abstract}
\end{titlepage}\newpage {} 

{\baselineskip=12pt \tableofcontents{}}

\section{Introduction}

The gravitational interaction can be effectively described by Newtonian gravity for regimes of low velocities relative to the speed of light $c$ and weak gravitational fields. Notably, a non-relativistic (NR) formulation of General Relativity (GR) was introduced by Cartan in \cite{Cartan1,Cartan2}, commonly referred to Newton-Cartan (NC) gravity. The latter employs a NC geometry that can be seen as the NR analogue of Riemannian geometry for GR, providing a geometric interpretation of the Poisson equation. NR theories, also known as Galilean, are commonly derived as the $c\rightarrow\infty$ limit of relativistic ones. Conversely, the $c\rightarrow 0$ limit defines the ultra-relativistic (UR) regime, also referred to as the Carrollian limit \cite{LevyLeblond,SenGupta:1966qer}, in homage to Lewis Carroll, the author of \textit{Alice’s Adventures in Wonderland}. Both the Galilean and the Carrollian algebras, also referred as non-Lorentzian (NL), can be obtained through specific contraction of the Poincaré one, and they form part of the broader classification of kinematical Lie algebras introduced by Bacry and Lévy-Leblond in \cite{Bacry:1968zf}. 

NL symmetries have garnered increasing attention due to their emergence in a wide range of physical phenomena. On the one hand, NR theories and their corresponding geometries have been useful in the study of holography \cite{Son:2008ye,Balasubramanian:2008dm,Kachru:2008yh,Taylor:2008tg,Bagchi:2009my,Hartnoll:2009sz,Bagchi:2009pe,Christensen:2013lma,Christensen:2013rfa,Hartong:2014oma,Hartong:2014pma,Zaanen:2015oix}, Ho\v{r}ava-Lifshitz gravity \cite{Horava:2009uw,Hartong:2015wxa,Hartong:2015zia,Taylor:2015glc,Hartong:2016yrf,Devecioglu:2018apj}, Quantum Hall effect \cite{Hoyos:2011ez,Son:2013rqa,Abanov:2014ula,Geracie:2015dea,Gromov:2015fda}, asymptotic symmetries \cite{Barnich:2010eb,Barnich:2012aw,Bagchi:2010zz}, post-Newtonian corrections \cite{Dautcourt:1996pm,VandenBleeken:2017rij,Hansen:2020wqw}, among others. On the other hand, UR symmetries have received a renewed interest since they manifest within the context of tachyon condensation \cite{Gibbons:2002tv}, tensionless strings \cite{Bagchi:2013bga,Bagchi:2015nca,Bagchi:2016yyf,Bagchi:2017cte,Bagchi:2018wsn}, asymptotic structures \cite{Perez:2021abf,Perez:2022jpr,Fuentealba:2022gdx}, holography in asymptotically flat spacetimes \cite{Duval:2014uva,Hartong:2015xda,Hartong:2015usd,Bagchi:2016bcd,Donnay:2022aba,Saha:2022gjw,Saha:2023hsl,Saha:2023abr}, and black hole horizon \cite{Donnay:2019jiz,Ciambelli:2019lap,Grumiller:2019fmp,deBoer:2023fnj,Ecker:2023uwm}.

In particular, NL gravity theories in three spacetime dimensions can be explored using the Chern-Simons (CS) formalism. The three-dimensional CS approach serves as a laboratory model for studying the properties of higher-dimensional gravity theories, as it shares with them intriguing features, e.g. concerning black hole solutions and their associated thermodynamics \cite{Banados:1992wn}. Notably, three-dimensional GR with negative cosmological constant can be formulated as a CS action for the $\mathfrak{so}\left(2,2\right)$ algebra, commonly referred to as the AdS algebra \cite{Achucarro:1987vz,Witten:1988hc,Zanelli:2005sa}. In the limit of vanishing cosmological constant, the underlying symmetry is given by the $\mathfrak{iso}\left(2,1\right)$, namely Poincaré, algebra. While the Carrollian regime of relativistic CS gravity can be consistently derived as a UR limit \cite{Matulich:2019cdo,Bergshoeff:2016soe,Concha:2022muu}, the NR counterpart presents significant challenges. Specifically, both the Newton-Hooke \cite{Bacry:1968zf,derome1972,Gomis:2020wxp,Ravera:2022buz} and the Galilean algebras, which emerge as NR limits of the AdS and Poincaré algebras, respectively, suffer from degeneracy issues of their invariant tensor, key ingredients of field-theoretic CS formulations. In fact, a CS action with a non-degenerate invariant bilinear trace admits a kinetic term for each gauge field which, in three spacetime dimensions, implies the vanishing of the curvature two-forms as field equations in the CS field-theoretical framework. One way to avoid degeneracy in the NR realm is to add two central charges to the corresponding NR algebras, reproducing the extended Newton-Hooke \cite{Aldrovandi:1998im,Gibbons:2003rv,Brugues:2006yd,Alvarez:2007fw,Papageorgiou:2010ud,Duval:2011mi,Duval:2016tzi} and the extended Bargmann algebra \cite{Papageorgiou:2009zc,Bergshoeff:2016lwr}. Both NR algebras can be obtained through a contraction process from the AdS$\,\oplus\, \mathfrak{u}\left(1\right)^{2}$ and the Poincaré$\,\oplus\,\mathfrak{u}\left(1\right)^{2}$ algebra, respectively \cite{Matulich:2019cdo}.

More recently, an alternative approach to obtain NR symmetries, based on Lie algebra expansion procedures \cite{Hatsuda:2001pp,deAzcarraga:2002xi,Izaurieta:2006zz,deAzcarraga:2007et}, has been extensively discussed in \cite{Concha:2019lhn,Penafiel:2019czp,Gomis:2019nih,Bergshoeff:2020fiz,Concha:2022muu,Caroca:2022byi,Concha:2022you,Concha:2022jdc,Concha:2023bly}. Notably, the original ``cube" of Bacry and Lévy-Leblond, exhibiting in a compact way the links among algebraic structures, was extended in \cite{Concha:2023bly} to include non-degenerate kinematical Lie algebras through the semigroup expansion ($S$-expansion) method \cite{Izaurieta:2006zz}. This approach allows to recover centrally extended NR algebras by expanding relativistic algebras, without introducing $\mathfrak{u}\left(1\right)$ generators as required in the contraction process. The $S$-expansion method has been useful not only for deriving the commutation relations of an expanded algebra but also for obtaining the non-vanishing components of their invariant tensor in terms of the original ones \cite{Izaurieta:2006zz}. Furthermore, the CS action for a given expanded Lie algebra can be recovered directly from the CS action of the original algebra by appropriately expanding the corresponding gauge fields.

Although supersymmetric extensions of gravity theories have been largely studied, their NR regime has only recently been approached, mainly due to the lack of clarity about possible NR supermultiplets \cite{Bergshoeff:2022iyb}. Furthermore, the proper derivation of a NR superalgebra cannot be reproduced as a straightforward limit of a relativistic superalgebra as in the bosonic case, and instead requires alternative strategies. The first successful approach to NR supergravity was based on NC geometry in three spacetime dimensions \cite{Andringa:2013mma, Bergshoeff:2015uaa}. Later, a non-vanishing torsion was introduced in NC supergravity through a NR superconformal tensor calculus \cite{Bergshoeff:2015ija}. A third attempt for constructing NR supergravity was possible via the CS formalism in three spacetime dimensions. A CS supergravity action based on a supersymmetric extension of the Bargmann algebra was first presented in \cite{Bergshoeff:2016lwr}. This approach was subsequently applied to develop extended Newtonian supergravity \cite{Ozdemir:2019orp}, Maxwellian extended Bargmann supergravity \cite{Concha:2019mxx}, extended Newton-Hooke, extended Lifshitz and extended Schrödinger supergravity \cite{Ozdemir:2019tby}. The CS formalism has also been useful to construct UR supergravity models based on the Carroll superalgebra \cite{Ravera:2019ize} and its $\mathcal{N}$-extensions \cite{Ali:2019jjp}. More recently, the Lie algebra expansion procedure \cite{deAzcarraga:2002xi,Izaurieta:2006zz} has been used to derive different NR regimes of known three-dimensional relativistic CS supergravity models \cite{deAzcarraga:2019mdn,Concha:2020tqx,Concha:2020eam,Concha:2021jos,Concha:2021llq,Ravera:2022buz,Concha:2024vql}. Despite differences in the methods employed to construct NR supergravity theories, they are all defined in three spacetime dimensions and require $\mathcal{N}=2$ supersymmetry to express the time translation generators as bilinear combinations of supersymmetry generators. To our knowledge, aside from recent developments in ten and eleven spacetime dimensions,\footnote{Recent results have been obtained in the study of NL version of ten-dimensional minimal supergravity \cite{Bergshoeff:2021tfn} and 11-dimensional supergravity \cite{Bergshoeff:2023igy,Bergshoeff:2024nin}.} no consistent NL supergravity models exists in higher spacetime dimensions, despite extensive investigation into relativistic supergravity theories.

In this work, motivated by the numerous applications of NL symmetries and by the absence of a general framework for exploring NR (super)gravity in spacetime dimensions higher than three, we present a systematic approach to derive NL regimes of relativistic kinematical superalgebras. To this aim, we combine the $S$-expansion method \cite{Izaurieta:2006zz} with the CS formalism to construct three-dimensional NL supergravity actions. The derivation of consistent NL superalgebras is constrained by the non-degeneracy criterion, which allows us to identify ``good" NL supersymmetries for constructing CS actions. As a result, we extend the original cube of Bacry and Lévy-Leblond \cite{Bacry:1968zf} into a supersymmetric one, where the (extended) kinematical superalgebras are characterized by a non-degenerate bilinear invariant trace.

This paper is organized as follows: In section \ref{sec2}, we briefly review the extended kinematical Lie algebras which admit a non-degenerate bilinear invariant trace \cite{Concha:2023bly}. Sections \ref{sec3} and \ref{sec4} contain our main results. In section \ref{sec3}, we classify the supersymmetric extensions of the extended kinematical algebras by applying non- and ultra-relativistic expansions through appropriate semigroups. Section \ref{sec4} is devoted to the construction of NL supergravity theories based on the aforementioned extended kinematical superalgebras. Section \ref{sec5} concludes our work with some comments about our results and possible future developments.

\section{Extended kinematical Lie algebras}\label{sec2}

\begin{center}
 \begin{figure}[h!]
  \begin{center}
        \includegraphics[width=9.3cm, height=8cm]{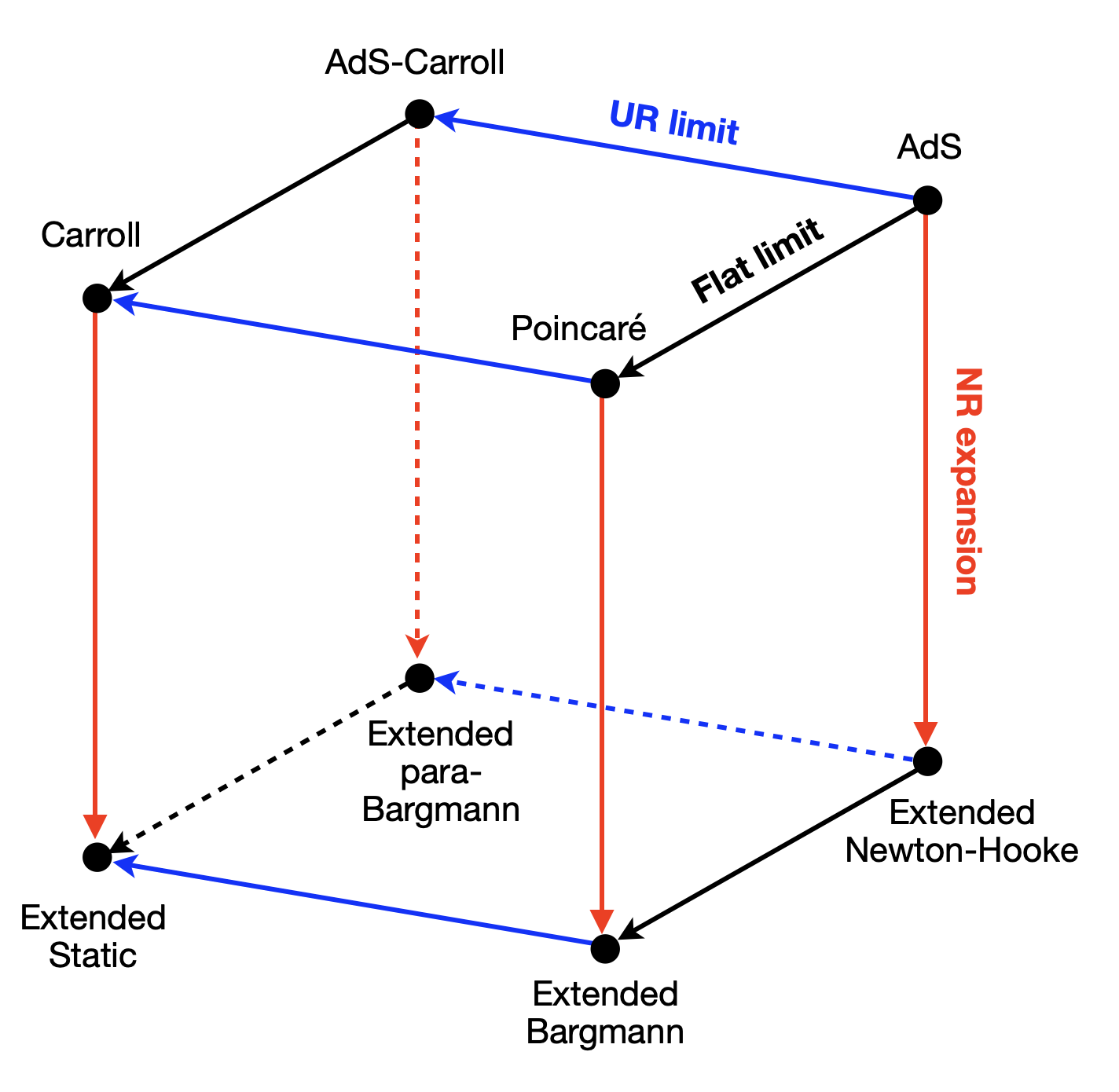}
        \captionsetup{font=footnotesize}
        \caption{This cube summarizes the different regimes starting from the AdS algebra. Both relativistic and non-Lorentzian algebras admit a non-degenerate bilinear invariant trace \cite{Concha:2023bly}.} 
        \label{fig1}
         \end{center}
        \end{figure}
    \end{center}

Diverse strategies have been implemented to address the degeneracy issue present in the NR regime of the original kinematical Lie algebra \cite{Bacry:1968zf}. Notably, the Lie algebra expansion method based on semigroups \cite{Izaurieta:2006zz} has demonstrated effectiveness not only to derive NR symmetries without degeneracy but also to construct the corresponding NR CS actions \cite{Concha:2023bly}. In this section, following \cite{Concha:2023bly}, we briefly review the different expansions applied, starting from the $\mathfrak{so}\left(2,2\right)$ algebra, which reproduce both the UR limit and the NR expansion (see Figure \ref{fig1}). 

The commutation relations of $\mathfrak{so}\left(2,2\right)$ are given by
\begin{align}
\left[ \tilde{J}_{A},\tilde{J}_{B}\right] &=\epsilon _{ABC}\tilde{J}^{C}\,, &  
\left[ \tilde{J}_{A},\tilde{P}_{B}\right] &=\epsilon _{ABC}\tilde{P}^{C}\,,  &
\left[ \tilde{P}_{A},\tilde{P}_{B}\right] &=\frac{1}{\ell^{2}}\epsilon _{ABC}\tilde{J}^{C}\,,  \label{AdS}
\end{align}%
where $\tilde{J}_{A}$ correspond to the Lorentz generators and $\tilde{P}_{A}$ are the spacetime translations. Here, the AdS radius (length parameter) $\ell$ is related to the cosmological constant through $\Lambda = - \frac{1}{\ell^2}$. In the vanishing cosmological constant limit $\ell\rightarrow\infty$, the AdS algebra reproduces the Poincaré one. Before applying the expansion, a decomposition of the relativistic rigid (Lorentz) indices $A=0,1,2$ is required, by considering a time-space splitting such that $A=\{0,a\}$ with $a=1,2$. Then, implementing this decomposition, the AdS algebra reads 
\begin{align}
\left[ J,G_{a}\right] &=\epsilon_{ab}G_{b}\,, &  
\left[ J,P_{a}\right] &=\epsilon_{ab}P_{b}\,,  &
\left[ G_{a},G_{b}\right] &=-\epsilon_{ab}J\,,  \notag\\
\left[ H,G_{a}\right] &=\epsilon_{ab}P_{b}\,, &  
\left[ G_{a},P_{b}\right] &=-\epsilon_{ab}H\,,  &
\left[ H,P_{a}\right] &=\frac{1}{\ell^2}\epsilon_{ab}G_{b}\,, \notag \\
\left[ P_{a},P_{b}\right] &=-\frac{1}{\ell^{2}}\epsilon_{ab}J\,, \label{AdS2}
\end{align}%
with $\epsilon_{ab}\equiv\epsilon_{0ab}$ and $\epsilon^{ab}\equiv\epsilon^{0ab}$. Here we have relabeled the AdS generators as follows:
\begin{align}
    J&=\tilde{J}_{0}\,, & G_{a}&=\tilde{J}_{a}\,, & H&=\tilde{P}_{0}\,, & P_{a}&=\tilde{P}_{a}\,. \label{split}
\end{align}

Let $\mathfrak{so}\left(2,2\right)=V_0+V_1$ be a subspace decomposition of the $\mathfrak{so}\left(2,2\right)$ algebra whose content depends on the desired expansion as shown in Table \ref{Table1}.
\begin{table}[h!]
\renewcommand{\arraystretch}{1.4}
\centering
    \centering
    \begin{tabular}{|c||C{2.5cm}|C{2.5cm}|}
    \hline
     \rowcolor[gray]{0.9}  Subspaces   & NR expansion & UR limit  \\ \hline
        $V_0$ &  $J, H$& $J, P_{a}$\\
       \rowcolor[gray]{0.9} $V_1$ & $G_{a}, P_{a}$& $H, G_{a}$\\
     \hline
         \end{tabular}
         \captionsetup{font=footnotesize}
    \caption{Subspace decomposition of the AdS algebra.}
    \label{Table1}
\end{table}

\noindent The subspaces $V_0$ and $V_1$ satisfy a $\mathbb{Z}_2$-graded Lie algebra,
\begin{align}
[V_0,V_0]&\subset V_0\,,  &[V_0,V_1]&\subset V_1\,,  &[V_1,V_1]&\subset V_0\,.\label{sd}
\end{align}
The UR limit can be reproduced considering the $S_{E}^{\left(1\right)}$ semigroup, whose elements fulfill the following multiplication rules:
\begin{equation}
\lambda _{i }\lambda _{j}=\left\{ 
\begin{array}{lcl}
\lambda _{i +j }\,\,\,\, & \mathrm{if}\,\,\,\,i +j \leq
2\,, &  \\ 
\lambda _{2}\,\,\, & \mathrm{if}\,\,\,\,i +j >2\,, & 
\end{array}%
\right.   \label{mlSE1}
\end{equation}%
where $\lambda_{2}=0_S$ is the zero element of the semigroup which satisfies $0_S\lambda_i=\lambda_i 0_S=0_S$. A subset decomposition of the semigroup having the same algebraic structure as the subspace decompositions provided in Table \ref{Table1} is given by
\begin{align}
    S_0&=\{\lambda_0,\lambda_2\}\,, & S_1&=\{\lambda_1,\lambda_2\}\,.\label{sd2}
\end{align}
Such subset decomposition is said to be in resonance with \eqref{sd}, since it satisfies
\begin{align}
S_0\cdot S_0&\subset S_0\,, \quad &S_0\cdot S_1&\subset S_1\,, \quad &S_1\cdot S_1&\subset S_0\,.\label{rc}
\end{align}
On the other hand, the NR expansion requires to consider $S_{E}^{\left(2\right)}$ as the relevant semigroup \cite{Concha:2023bly}, whose elements satisfy:
\begin{equation}
\lambda _{i }\lambda _{j}=\left\{ 
\begin{array}{lcl}
\lambda _{i +j }\,\,\,\, & \mathrm{if}\,\,\,\,i +j \leq
3\,, &  \\ 
\lambda _{3}\,\,\, & \mathrm{if}\,\,\,\,i +j >3\,, & 
\end{array}%
\right.   \label{mlSE2}
\end{equation}%
with $\lambda_{3}=0_S$ being the zero element of the semigroup.  A resonant subset decomposition of the semigroup reads
\begin{align}
    S_0&=\{\lambda_0,\lambda_2,\lambda_3\}\,, & S_1&=\{\lambda_1,\lambda_3\}\,.\label{sd22}
\end{align}
Each NL regime appearing in the cube of Figure \ref{fig1} can be obtained after applying an $S_{E}$-resonant expansion of the relativistic $\mathfrak{so}\left(2,2\right)$ algebra:
\begin{align}
    \mathfrak{G}=\left(S_0\times V_0\right)\oplus\left(S_1\times V_1\right)\,,
\end{align}
and performing a $0_S$-reduction, namely $0_S T_A=0$. The $S$-expanded generators are written in terms of the $\mathfrak{so}\left(2,2\right)$ ones and the semigroup elements as in Table \ref{Table2}.
\begin{table}[h]
\renewcommand{\arraystretch}{1.3}
\centering
    \begin{tabular}{l ||C{1.4cm}|C{1.4cm}||C{1.4cm}|C{1.4cm}|}
    &  \multicolumn{2}{|c||}{NR expansion}& \multicolumn{2}{|c||}{UR limit} \\ \hline
    $\lambda_3$  & \cellcolor[gray]{0.8} & \cellcolor[gray]{0.8}& \cellcolor[gray]{0.28} &\cellcolor[gray]{0.28}  \\ \hline 
    $\lambda_2$  & $\texttt{S}$, \ \, $\texttt{M}$ & \cellcolor[gray]{0.8}& \cellcolor[gray]{0.8} &\cellcolor[gray]{0.8}  \\ \hline 
    $\lambda_1$  & \cellcolor[gray]{0.8} & $\texttt{G}_a$, \, $\texttt{P}_a$ & \cellcolor[gray]{0.8} & $\texttt{H}$,  \ \, $\texttt{G}_a$ \\ \hline
    $\lambda_0$  &  $\texttt{J}$, \ \, $\texttt{H}$ & \cellcolor[gray]{0.8} & $\texttt{J}$, \, $\texttt{P}_a$ & \cellcolor[gray]{0.8}  \\ \hline
    & $J$, \ \ $H$ & $G_a$, \, $P_a$ &$J$, \, $P_a$ & $H$, \, $G_a$  \\ 
    \end{tabular}
    \captionsetup{font=footnotesize}
    \caption{Expanded generators in terms of the relativistic ones and the semigroup elements. The relativistic generators have been organized according to the subspace decomposition of Table \ref{Table1}.}
    \label{Table2}
    \end{table}

The explicit commutation relations of the expanded algebra are listed in Table \ref{Table3} and are derived by combining the multiplication rule of the semigroup $S_{E}$ and the commutators of the starting algebra. Let us note that the AdS-Carroll algebra can be alternatively recovered as a resonant $S_{E}^{(1)}$-expansion of $\mathfrak{so}\left(2,2\right)$. Despite the isomorphism existing between the AdS-Carroll and the Poincaré algebra, exchanging the generators $\texttt{G}_{a}$ and $\texttt{P}_{a}$ leads to quiet distinct physical implications \cite{Bacry:1968zf}. On the other hand, from the cube of Figure \ref{fig1}, the Carroll algebra can either be recovered as an $S_{E}^{(1)}$-expansion of the Poincaré algebra or as a flat limit $\ell\rightarrow\infty$ applied to the AdS-Carroll one. 
\renewcommand{\arraystretch}{1.3}
\begin{table}[h]
    \centering
    \begin{tabular}{|l||r|r|r|}
    \hline
      \rowcolor[gray]{0.9}  &    AdS-Carroll &   Extended Newton-Hooke  &  Extended AdS-static   \\ 
        $\left[ \texttt{J},\texttt{G}_{a} \right]$ &  $\epsilon_{ab}\texttt{G}_{b}$&  $\epsilon_{ab}\texttt{G}_{b}$ &  $\epsilon_{ab}\texttt{G}_{b}$\\
        \rowcolor[gray]{0.9}$\left[ \texttt{J},\texttt{P}_{a} \right]$ &   $\epsilon_{ab}\texttt{P}_{b}$ &  $\epsilon_{ab}\texttt{P}_{b}$&  $\epsilon_{ab}\texttt{P}_{b}$  \\
        $\left[ \texttt{G}_{a},\texttt{G}_{b} \right]$ &  $0$ & $-\epsilon_{ab}\texttt{S}$ & $0$  \\
        \rowcolor[gray]{0.9}$\left[ \texttt{H},\texttt{G}_{a} \right]$ &  $0$ & $\epsilon_{ab}\texttt{P}_{b}$ & $0$  \\
        $\left[ \texttt{G}_{a},\texttt{P}_{b} \right]$ &  $-\epsilon_{ab}\texttt{H}$ &  $-\epsilon_{ab}\texttt{M}$ & $-\epsilon_{ab}\texttt{M}$ \\
        \rowcolor[gray]{0.9}$\left[ \texttt{H},\texttt{P}_{a} \right]$ &  $\frac{1}{\ell^{2}}\epsilon_{ab}\texttt{G}_{b}$ &  $\frac{1}{\ell^{2}}\epsilon_{ab}\texttt{G}_{b}$ & $\frac{1}{\ell^{2}}\epsilon_{ab}\texttt{G}_{b}$  \\
        $\left[ \texttt{P}_{a},\texttt{P}_{b} \right]$ &  $-\frac{1}{\ell^{2}}\epsilon_{ab}\texttt{J}$ & $-\frac{1}{\ell^{2}}\epsilon_{ab}\texttt{S}$ & $-\frac{1}{\ell^{2}}\epsilon_{ab}\texttt{S}$  \\
     \hline
         \end{tabular}
         \captionsetup{font=footnotesize}
    \caption{Commutation relations of the (extended) kinematical algebras. The vanishing cosmological constant limit $\ell\rightarrow\infty$ reproduces the Carroll algebra, the extended Bargmann algebra, and the extended static algebra, respectively.}
    \label{Table3}
\end{table}

\noindent In the NR counterpart, we obtain extended kinematical algebras which are characterized by the presence of central charges. Such central extensions solve the degeneracy present in the NR regime of the original kinematical algebra \cite{Bacry:1968zf}. Here, the extended Newton-Hooke \cite{Aldrovandi:1998im,Gibbons:2003rv,Brugues:2006yd,Alvarez:2007fw,Papageorgiou:2010ud,Duval:2011mi,Duval:2016tzi} and extended Bargmann \cite{Papageorgiou:2009zc,Bergshoeff:2016lwr} algebras appear after performing the $S_{E}^{(2)}$-expansion of $\mathfrak{so}\left(2,2\right)$ and the Poincaré algebra, respectively. Extensions of the static algebra, which have been related to infinitely massive systems that cannot move \cite{Bacry:1968zf}, can also be derived as expansions. Here, the extended AdS-static algebra and its flat limit can be obtained either as a resonant $S_{E}^{(2)}$-expansion of the Carrollian algebras or as a resonant $S_{E}^{(1)}$-expansion of the extended NR ones. However, as one can see from Table \ref{Table4}, different subspace decompositions of the starting algebra have to be considered.
\renewcommand{\arraystretch}{1.3}
\begin{table}[h!]
\centering
    \begin{tabular}{l||C{2,0cm}|C{2,0cm}||C{2,0cm}|C{2,0cm}|}
& \multicolumn{2}{|c||}{NR expansion}& \multicolumn{2}{|c|}{UR limit} \\ \hline
$\lambda_3$ & \cellcolor[gray]{0.8} & \cellcolor[gray]{0.8} & \cellcolor[gray]{0.28} & \cellcolor[gray]{0.28} \\ \hline
$\lambda_2$ & $\texttt{S}$,\quad $\texttt{M}$ & \cellcolor[gray]{0.8} & \cellcolor[gray]{0.8}  & \cellcolor[gray]{0.8} \\ \hline
$\lambda_1$ & \cellcolor[gray]{0.8} & $\texttt{G}_a$,\ \ $\texttt{P}_a$ & \cellcolor[gray]{0.8} & $\texttt{H}$,\quad $\texttt{M}$,\quad $\texttt{G}_{a}$ \\ \hline
$\lambda_0$ & $ \texttt{J}$,\quad $\texttt{H}$ & \cellcolor[gray]{0.8} & $\texttt{J}$,\quad $\texttt{S}$,\quad $\texttt{P}_{a}$ & \cellcolor[gray]{0.8} \\ \hline
 & $\texttt{J}$,\quad $\texttt{H}$ & $\texttt{G}_a$, \, $\texttt{P}_{a}$ & $\texttt{J}$,\quad $\texttt{S}$,\quad $\texttt{P}_{a}$ & $\texttt{H}$,\quad $\texttt{M}$,\quad $\texttt{G}_{a}$
\end{tabular}
\captionsetup{font=footnotesize}
\caption{Extended (AdS-)static generators in terms of the non-Lorentzian ones and the semigroup elements. The UR limit column contains expanded kinematical generators in terms of the extended (AdS-)Carroll ones. On the other hand, the extended (AdS-)static generators are obtained from the extended Bargmann (or extended Newton-Hooke) ones in the NR expansion column.}
\label{Table4}%
\end{table}
\section{Supersymmetric extensions of extended kinematical algebras}\label{sec3}

In this section, we explore the supersymmetric extension of the kinematical Lie algebras \cite{Bacry:1968zf} in three-dimensional spacetime, using the semigroup expansion method \cite{Izaurieta:2006zz}, starting from an $\mathcal{N}=2$ relativistic superalgebra. The use of $\mathcal{N}=2$ ensures that the expanded non-relativistic superalgebras yield genuine supergravity actions, in which the time translational generators can be expressed as bilinear combinations of supersymmetry generators. Additionally, in the NR regime we will derive supersymmetric extensions of the extended kinematic algebras discussed in \cite{Concha:2023bly}, which are distinguished by a non-degenerate invariant bilinear form. The non-degeneracy of the three-dimensional CS action guarantees a kinetic term for each gauge field leading to the vanishing of the curvatures as equations of motion. 

In three spacetime dimensions, the $\mathcal{N}=2$ supersymmetric extension of the AdS algebra with a well-defined Poincaré limit $\ell\rightarrow\infty$ is given by an $\mathfrak{so}(2)$-extension of the $\mathfrak{osp}\left(2,2\right)\otimes \mathfrak{sp}\left(2\right)$ superalgebra whose (anti-)commutation relations are given by \cite{Howe:1995zm}:
\begin{eqnarray}
\left[ \tilde{J}_{A},\tilde{J}_{B}\right] &=&\epsilon _{ABC}\tilde{J}^{C}\,, \notag \\ \left[ \tilde{J}_{A},\tilde{P}_{B}\right] &=&\epsilon _{ABC}\tilde{P}^{C}\,, \notag \\
\left[ \tilde{P}_{A},\tilde{P}_{B}\right] &=&\frac{1}{\ell^{2}}\epsilon _{ABC}\tilde{J}^{C}\,, \notag \\
\left[\tilde{J}_{A},\tilde{Q}_{\alpha}^{i}\right] &=& -\frac{1}{2} \left( \gamma_{A} \right)_{\alpha}^{\ \beta} \tilde{Q}_{\beta}^{i}\,, \notag \\ 
\left[\tilde{P}_{A},\tilde{Q}_{\alpha}^{i}\right] &=& -\frac{1}{2\ell} \left( \gamma_{A} \right)_{\alpha}^{\ \beta} \tilde{Q}_{\beta}^{i}\,, \notag \\
\left[\tilde{\mathcal{T}},\tilde{Q}_{\alpha}^{i}\right] &=& \frac{1}{2} \epsilon^{ij} \tilde{Q}_{\beta}^{j}\,, \notag \\
\{ \tilde{Q}_{\alpha}^{i},\tilde{Q}_{\beta}^{j}\} &=&-\frac{1}{\ell}\delta_{ij}\left(\gamma^{A}C \right)_{\alpha \beta} \tilde{J}_{A}-\delta_{ij}\left(\gamma^{A}C \right)_{\alpha \beta} \tilde{P}_{A}-C_{\alpha \beta}\epsilon^{ij}\left(\tilde{\mathcal{U}}+\frac{1}{\ell}\tilde{\mathcal{T}}\right)\,, \label{n2SADS}
\end{eqnarray}%
where $\tilde{Q}_{\alpha}$ is a Majorana spinor charge. On the other hand, the presence of the generators $\{\tilde{\mathcal{T}},\tilde{\mathcal{U}}\}$ ensures the non-degeneracy of the bilinear invariant form in the Poincaré limit in which $\mathcal{\tilde{U}}$ becomes the central charge.  Here, $A,B,\dots =0,1,2$ denote the Lorentz indices which are raised and lowered with the Minkowski metric $\eta _{AB}=\left(-1,1,1\right)$ and $\epsilon _{ABC}$ is the Levi-Civita tensor, while $\alpha,\beta=1,2$, $i,j=1,2$ denote the number of supercharges, the gamma matrices in three dimensions are denoted by $\gamma_A$, and $C$ is the charge conjugation matrix, satisfying $C^{T}=-C$ and $C\gamma^{A}=(C\gamma^{A})^{T}$. The $\mathfrak{osp}\left(2,2\right)\otimes \mathfrak{sp}\left(2\right)$ superalgebra admits the following non-vanishing components of a non-degenerate invariant tensor:
\begin{align}
    \langle \tilde{J}_{A}\tilde{J}_{B} \rangle &=\tilde{\alpha}_0\eta_{AB}\,, & \langle \tilde{J}_{A}\tilde{P}_{B} \rangle &=\tilde{\alpha}_{1}\eta_{AB}\,, \notag \\
    \langle \tilde{P}_{A}\tilde{P}_{B} \rangle &=\frac{\tilde{\alpha}_0}{\ell^{2}}\eta_{AB}\,, & \langle \tilde{\mathcal{T}}\tilde{\mathcal{T}} \rangle &=\tilde{\alpha}_{0}\,, \notag \\
    \langle \tilde{\mathcal{T}}\tilde{\mathcal{U}} \rangle &=\tilde{\alpha}_{1}\,, & \langle \tilde{\mathcal{U}}\tilde{\mathcal{U}} \rangle &=-\frac{\tilde{\alpha}_{1}}{\ell}\,, \notag \\
    \langle \tilde{Q}_{\alpha}^{i}\tilde{Q}_{\beta}^{j}\rangle &=2\left(\alpha_1+\frac{\alpha_0}{\ell}\right)C_{\alpha\beta}\delta^{ij}\,. \label{IT}
\end{align}

Before applying the expansion, let us consider a time-space splitting of the relativistic Lorentz index $A=\{0,a\}$ with $a=1,2$. Then, the $\mathfrak{so}(2)$-extension of the $\mathfrak{osp}\left(2,2\right)\otimes \mathfrak{sp}\left(2\right)$ superalgebra is spanned by $\{J,G_a,H,P_a,T,U,Q_{\alpha}^{\pm}\}$, where we have considered the redefinitions \eqref{split} together with the following ones: 
\begin{align}
    T&=\mathcal{\tilde{T}}\,, & U&=\mathcal{\tilde{U}}\,, &
    Q_{\alpha}^{\pm}&=\frac{1}{\sqrt{2}}\left(\tilde{Q}_{\alpha}^{1}\pm(\gamma^{0})_{\alpha\beta}\tilde{Q}_{\beta}^{2}\right)\,. \label{split2}
\end{align}
Then, the $\mathfrak{so}\left(2\right)$-extension of the  $\mathfrak{osp}\left(2,2\right)\otimes \mathfrak{sp}\left(2\right)$ superalgebra can be written as
\begin{eqnarray}
\left[ J,G_{a}\right] &=&\epsilon_{ab}G_{b}\,, \qquad  \qquad \qquad
\left[ J,P_{a}\right] =\epsilon_{ab}P_{b}\,,  \qquad \qquad \qquad
\left[ G_{a},G_{b}\right] =-\epsilon_{ab}J\,,  \notag\\
\left[ H,G_{a}\right] &=&\epsilon_{ab}P_{b}\,, \qquad \qquad \quad \ \,
\left[ G_{a},P_{b}\right] =-\epsilon_{ab}H\,,  \qquad \qquad \quad \ \ \,
\left[ H,P_{a}\right] =\frac{1}{\ell^2}\epsilon_{ab}G_{b}\,, \notag \\
\left[ P_{a},P_{b}\right] &=&-\frac{1}{\ell^{2}}\epsilon_{ab}J\,,  \qquad \qquad
\left[ J,Q_{\alpha}^{\pm}\right] =-\frac{1}{2} \left( \gamma_{0} \right)_{\alpha}^{\ \beta} \tilde{Q}_{\beta}^{\pm}\,,  \qquad \
\left[ H,Q_{\alpha}^{\pm}\right] =-\frac{1}{2\ell} \left( \gamma_{0} \right)_{\alpha}^{\ \beta} \tilde{Q}_{\beta}^{\pm}\,, \notag \\
\left[ G_{a},Q_{\alpha}^{\pm}\right] &=&-\frac{1}{2} \left( \gamma_{a} \right)_{\alpha}^{\ \beta} \tilde{Q}_{\beta}^{\mp}\,, \quad \
\left[ P_{a},Q_{\alpha}^{\pm}\right] =-\frac{1}{2\ell} \left( \gamma_{a} \right)_{\alpha}^{\ \beta} \tilde{Q}_{\beta}^{\mp}\,, \qquad 
\left[ T,Q_{\alpha}^{\pm}\right] =\pm\frac{1}{2} \left( \gamma_{0} \right)_{\alpha\beta} \tilde{Q}_{\beta}^{\pm}\,, \notag \\
\{ Q_{\alpha}^{+},Q_{\beta}^{+}\} &=&-\left(\gamma^{0}C \right)_{\alpha \beta} \left(\frac{1}{\ell} J + H \right)-\left(\gamma^{0}C\right)_{\alpha \beta}\left(U+\frac{1}{\ell}T\right)\,, \notag \\
\{ Q_{\alpha}^{+},Q_{\beta}^{-}\} &=&-\left(\gamma^{a}C \right)_{\alpha \beta} \left(\frac{1}{\ell} G_{a} + P_{a} \right)\,, \notag \\
\{ Q_{\alpha}^{-},Q_{\beta}^{-}\} &=&-\left(\gamma^{0}C \right)_{\alpha \beta} \left(\frac{1}{\ell} J + H \right)+\left(\gamma^{0}C\right)_{\alpha \beta}\left(U+\frac{1}{\ell}T\right)\,.
\label{sAdS2}
\end{eqnarray}

\subsection{Ultra-relativistic expansions}

The Carrollian regime of the kinematical superalgebras can be obtained by expanding the relativistic ones through a pertinent semigroup and by implementing an explicit reduction procedure. Interestingly, a family of UR superalgebras can be derived by considering larger semigroups. Before applying the UR expansion, we require to consider a particular subspace decomposition of the $\mathfrak{so}\left(2\right)$-extension of the $\mathfrak{osp}\left(2,2\right)\otimes \mathfrak{sp}\left(2\right)$ given by
\begin{align}
    V_{0}&=\{J,P_a,T\}\,, & V_{1}&=\{Q_{\alpha}^{+},Q_{\alpha}^{-}\}\,, & V_{2}&=\{H,G_{a},U\}\,. \label{sd3} 
\end{align}
Here, the subspaces $V_0$, $V_1$, and $V_2$ satisfy the following algebraic structure:
\begin{align}
   [V_0,V_0]&\subset V_0\,,  &  [V_1,V_1]&\subset V_0\oplus V_2\,,  \notag \\
   [V_0,V_1]&\subset V_1\,,  & [V_1,V_2]&\subset V_1\,,  \notag \\
   [V_0,V_2]&\subset V_2\,,  & [V_2,V_2]&\subset V_0\,. \label{AS}
\end{align}

\subsubsection{The $\mathcal{N}=2$ (AdS-)Carroll superalgebra}

Let $S_{E}^{(2)}=\{\lambda_0,\lambda_1,\lambda_2,\lambda_3\}$ be the finite Abelian semigroup whose elements obey the multiplication law \eqref{mlSE2} and whose zero element is given by $\lambda_3$. A subset decomposition of the semigroup $S_{E}^{(2)}=S_0\cup S_1\ S_2$, with
\begin{align}
    S_0&=\{\lambda_0,\lambda_2,\lambda_3\}\,, & S_1&=\{\lambda_1, \lambda_3\}\,, & S_2&=\{\lambda_2,\lambda_3\}\,, \label{}
\end{align}
is said to be resonant since it satisfies the same algebraic structure \eqref{AS} as the subspaces $V_0$, $V_1$, and $V_2$, namely
\begin{align}
    S_0\cdot S_0&\subset S_0\,, &  S_1\cdot S_1&\subset S_0\cap S_2\,,  \notag \\
    S_0\cdot S_1&\subset S_1\,, &  S_1\cdot S_2&\subset S_1\,,   \notag \\
    S_0\cdot S_2&\subset S_2\,, &  S_2\cdot S_2&\subset S_0\,. \label{rc2}
\end{align}
According to \cite{Izaurieta:2006zz}, the superalgebra $\mathfrak{G}_{R}=\left(S_0\times V_0\right)\oplus\left(S_1\times V_1\right)\oplus\left(S_2\times V_2\right)$
corresponds to a resonant super subalgebra of the $S_{E}^{(2)}$-expanded one. A smaller superalgebra can be extracted from the resonant one through a reduction procedure \cite{Izaurieta:2006zz}. Let us consider a partition of the subsets $S_p=\hat{S}_p\cup \check{S}_p$ with
\begin{align}
\check{S}_{0}&=\{\lambda_0\}\,, & \hat{S}_{0}&=\{\lambda_2,\lambda_3\}\,, \notag \\
\check{S}_{1}&=\{\lambda_1\}\,, & \hat{S}_{1}&=\{\lambda_3\}\,, \notag \\
\check{S}_{2}&=\{\lambda_2\}\,, & \hat{S}_{2}&=\{\lambda_3\}\,. \label{Sp}
\end{align}
Such partition satisfies $\hat{S}_p\cap \check{S}_p =\varnothing$ together with the following algebraic structure:
\begin{align}
    \check{S}_{0}\cdot \hat{S}_{0}&\subset\hat{S}_{0}\,, &  \check{S}_{1}\cdot \hat{S}_{1}&\subset \hat{S}_{0}\cap \hat{S}_{2}\,, \notag \\
    \check{S}_{0}\cdot \hat{S}_{1}&\subset\hat{S}_{1}\,, &  \check{S}_{1}\cdot \hat{S}_{2}&\subset \hat{S}_{1}\,, \notag \\
    \check{S}_{0}\cdot \hat{S}_{2}&\subset\hat{S}_{2}\,, &  \check{S}_{2}\cdot \hat{S}_{2}&\subset \hat{S}_{0}\,. \label{Part}
\end{align}
Then, we have
\begin{align}
    \check{\mathfrak{G}}_{R}&=\left(\check{S}_{0}\times V_0\right)\oplus\left(\check{S}_{1}\times V_1\right)\oplus \left(\check{S}_{2}\times V_2\right)\,, \notag \\
    \hat{\mathfrak{G}}_{R}&=\left(\hat{S}_{0}\times V_0\right)\oplus\left(\hat{S}_{1}\times V_1\right)\oplus \left(\hat{S}_{2}\times V_2\right)\,, 
\end{align}
where $\check{\mathfrak{G}}_{R}$ corresponds to a \emph{reduced} superalgebra of $\mathfrak{G}_{R}$ which satisfies
\begin{eqnarray}
    [\check{\mathfrak{G}}_{R},\hat{\mathfrak{G}}_{R}]&\subset&\hat{\mathfrak{G}}_{R}\,.
\end{eqnarray}
The expanded generators of the reduced superalgebra $\check{\mathfrak{G}}_{R}$ are written in terms of the $\mathfrak{so}\left(2\right)$-extension of the $\mathfrak{osp}\left(2,2\right)\otimes \mathfrak{sp}\left(2\right)$ ones and the semigroup elements as in Table \ref{Table5}.
\renewcommand{\arraystretch}{1.2}
\begin{table}[h]
\centering
    \begin{tabular}{l ||C{2.2cm}|C{2.2cm}|C{2.2cm}|}
    $\lambda_3$  & \cellcolor[gray]{0.8} & \cellcolor[gray]{0.8}& \cellcolor[gray]{0.8}   \\ \hline 
    $\lambda_2$  & \cellcolor[gray]{0.8} & \cellcolor[gray]{0.8}& $\texttt{H}$, \quad $\texttt{G}_{a}$, \quad $\texttt{U}$  \\ \hline 
    $\lambda_1$  & \cellcolor[gray]{0.8} & $\texttt{Q}_{\alpha}^{+}$, \quad $\texttt{Q}_{\alpha}^{-}$ & \cellcolor[gray]{0.8}  \\ \hline
    $\lambda_0$  &  $\texttt{J}$, \quad $\texttt{P}_{a}$, \quad $\texttt{T}$ & \cellcolor[gray]{0.8}  & \cellcolor[gray]{0.8}  \\ \hline
    & $J$,\ \ \, $P_{a}$, \ \, $T$ & $Q_{\alpha}^{+}$,\, \, $Q_{\alpha}^{-}$ & $H$, \ \,  $G_a$,\ \ \,  $U$  \\ 
    \end{tabular}
    \captionsetup{font=footnotesize}
    \caption{Expanded generators of the reduced superalgebra $\check{\mathfrak{G}}_{R}$ in terms of the relativistic ones and the semigroup elements.}
    \label{Table5}
    \end{table}

\begin{center}
 \begin{figure}[h!]
  \begin{center}
        \includegraphics[width=8.6cm, height=5.4cm]{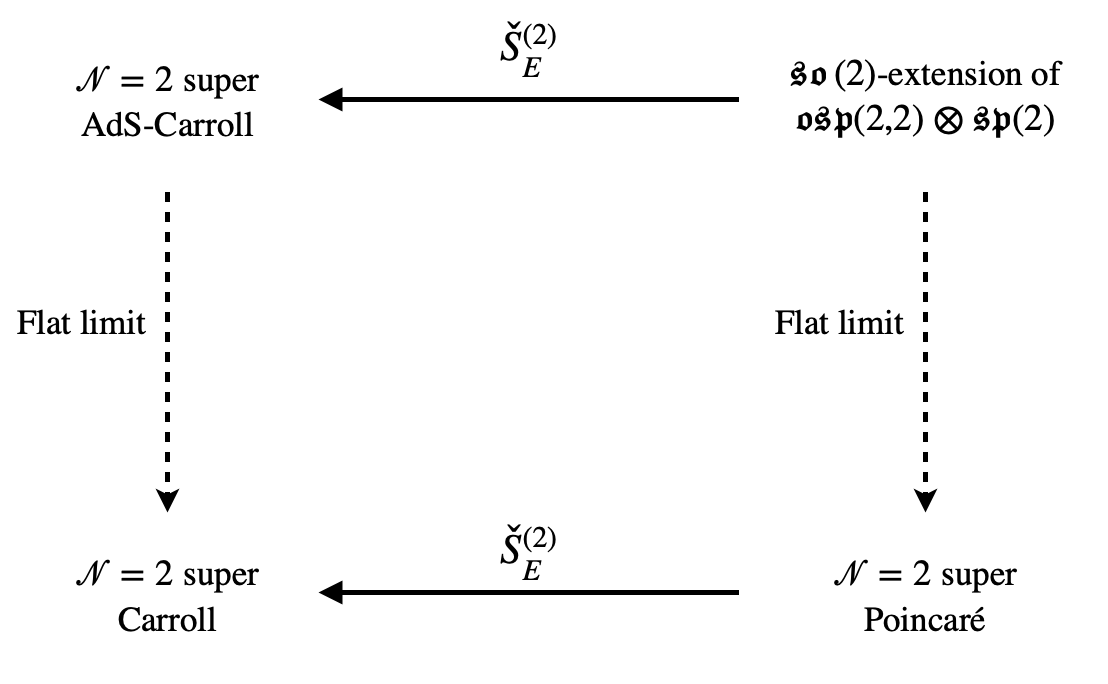}
        \captionsetup{font=footnotesize}
        \caption{Diagram summarizing the (reduced) UR expansion and flat limit starting from the $\mathfrak{so}\left(2\right)$-extension of the $\mathfrak{osp}\left(2,2\right)\otimes \mathfrak{sp}\left(2\right)$ superalgebra.}
        \label{fig2}
         \end{center}
        \end{figure}
    \end{center}

The (anti-)commutation relations of the reduced superalgebra $\check{\mathfrak{G}}_{R}$ are obtained by combining the multiplication law of the $S_{E}^{(2)}$ semigroup and the  (anti-)commutators of the $\mathfrak{so}\left(2\right)$-extension of the $\mathfrak{osp}\left(2,2\right)\otimes \mathfrak{sp}\left(2\right)$ superalgebra \eqref{sAdS2},
\begin{eqnarray}
\left[ \texttt{J},\texttt{G}_{a}\right] &=&\epsilon_{ab}\texttt{G}_{b}\,, \qquad  \qquad \qquad \quad
\left[ \texttt{J},\texttt{P}_{a}\right] =\epsilon_{ab}\texttt{P}_{b}\,,  \qquad \qquad \qquad
\left[ \texttt{P}_{a},\texttt{P}_{b}\right] =-\frac{1}{\ell^{2}}\epsilon_{ab}\texttt{J}\,,  \notag\\
\left[ \texttt{H},\texttt{P}_{a}\right] &=&\frac{1}{\ell^{2}}\epsilon_{ab}\texttt{G}_{b}\,, \qquad  \qquad \quad \  \,
\left[ \texttt{G}_{a},\texttt{P}_{a}\right] =-\epsilon_{ab}\texttt{H}\,,  \qquad \qquad \quad \ \, 
\left[ \texttt{J},\texttt{Q}_{\alpha}^{\pm}\right] =-\frac{1}{2}\left(\gamma_{0}\right)_{\alpha}^{\ \beta}\texttt{Q}_{\beta}^{\pm}\,,  \notag \\
\left[ \texttt{P}_{a},\texttt{Q}_{\alpha}^{\pm}\right] &=&-\frac{1}{2\ell}\left(\gamma_{a}\right)_{\alpha}^{\ \beta}\texttt{Q}_{\beta}^{\mp}\,, \qquad \ \ 
\left[ \texttt{T},\texttt{Q}_{\alpha}^{\pm}\right] =\pm\frac{1}{2}\left(\gamma_{0}\right)_{\alpha}^{\ \beta}\texttt{Q}_{\beta}^{\pm}\,, \notag \\
\{ \texttt{Q}_{\alpha}^{+},\texttt{Q}_{\beta}^{+}\} &=&-\left(\gamma^{0}C \right)_{\alpha \beta} \left( \texttt{H} + \texttt{U}\right)\,, \notag \\
\{ \texttt{Q}_{\alpha}^{+},\texttt{Q}_{\beta}^{-}\} &=&-\frac{1}{\ell}\left(\gamma^{a}C \right)_{\alpha \beta}  \texttt{G}_{a} \,, \notag \\
\{ \texttt{Q}_{\alpha}^{-},\texttt{Q}_{\beta}^{-}\} &=&-\left(\gamma^{0}C \right)_{\alpha \beta} \left( \texttt{H} - \texttt{U}\right)\,. \label{SADSC}
\end{eqnarray}
The reduced superalgebra $\check{\mathfrak{G}}_{R}$ corresponds to the $\mathcal{N}=2$ AdS-Carroll superalgebra\footnote{There are two inequivalent $\mathcal{N}=2$ AdS-Carroll superalgebras: the so-called $\mathcal{N}=(1,1)$ and the $\mathcal{N}=(2,0)$. Here, $\mathcal{N}=2$ refers to $\mathcal{N}=(2,0)$, since we started from the $\mathcal{N}=(2,0)$ AdS superalgebra. An $\mathcal{N}=(1,1)$ AdS-Carroll superalgebra can also be obtained from the $\mathcal{N}=(1,1)$ AdS one following the same procedure considered here.} introduced in \cite{Ali:2019jjp} and differs from the one presented in \cite{Bergshoeff:2015wma} due to the presence of the $\texttt{U}$ generator. In the ultra-relativistic regime, the $\texttt{U}$ generator becomes a central charge as in the Poincaré superalgebra. In particular, the $\mathcal{N}=2$ AdS-Carroll superalgebra \eqref{SADSC} is isomorphic to the $\mathcal{N}=2$ Poincaré superalgebra under exchange of the $\ell\texttt{P}_{a}$ and $\texttt{G}_{a}$ generators. In the vanishing cosmological constant limit $\ell\rightarrow\infty$, the superalgebra reproduces the $\mathcal{N}=2$ Carroll superalgebra. Interestingly, the $\mathcal{N}=2$ Carroll superalgebra can alternatively be recovered by $S$-expanding the $\mathcal{N}=2$ Poincaré one following the same procedure used to obtain the $\mathcal{N}=2$ AdS-Carroll superalgebra (see Figure \ref{fig2}).

\subsubsection{Generalized $\mathcal{N}=2$ AdS-Carroll superalgebras}

\begin{center}
 \begin{figure}[h!]
  \begin{center}
        \includegraphics[width=8.6cm, height=5.4cm]{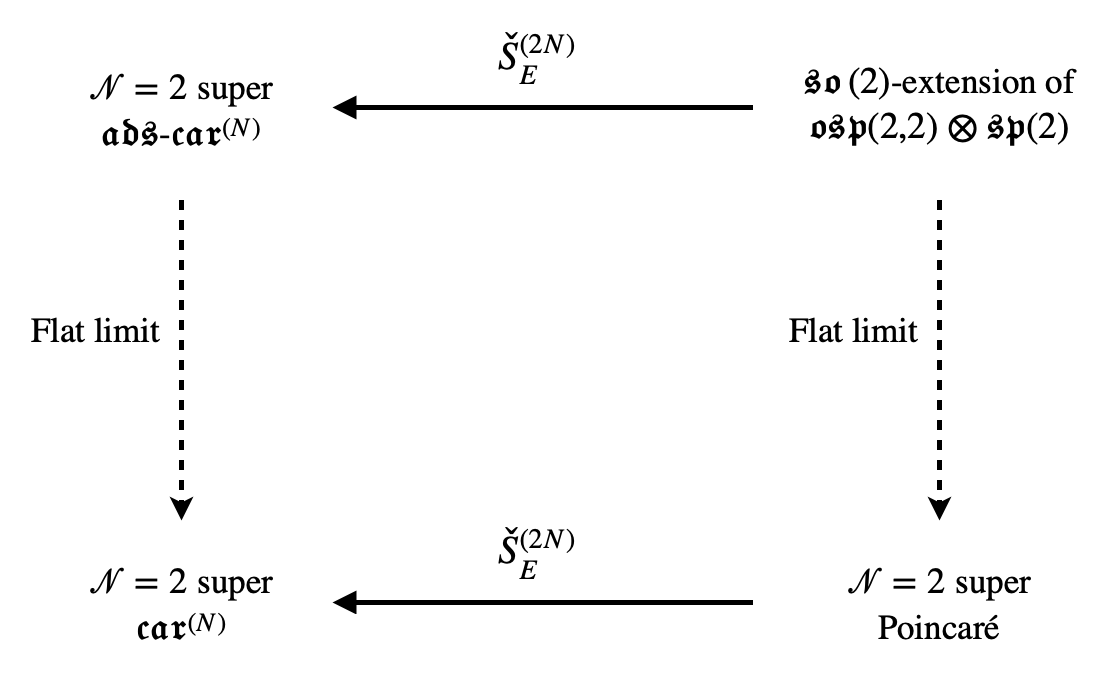}
        \captionsetup{font=footnotesize}
        \caption{Diagram summarizing the generalized (reduced) UR expansion and flat limit starting from the $\mathfrak{so}\left(2\right)$-extension of the $\mathfrak{osp}\left(2,2\right)\otimes \mathfrak{sp}\left(2\right)$ superalgebra.}
        \label{fig3}
         \end{center}
        \end{figure}
    \end{center}

A large semigroup $S_{E}^{\left(2N\right)}$ can be considered to obtain generalized $\mathcal{N}=2$ AdS-Carroll superalgebras from the $\mathfrak{osp}\left(2,2\right)\otimes \mathfrak{sp}\left(2\right)$ superalgebra. Let us consider first $S_{E}^{\left(2N\right)}=\{\lambda_0,\lambda_1,\lambda_2,\cdots,\lambda_{2N+1}\}$ as the finite semigroup whose elements satisfy
\begin{equation}
\lambda _{\alpha }\lambda _{\beta }=\left\{ 
\begin{array}{lcl}
\lambda _{\alpha +\beta }\,\,\,\, & \mathrm{if}\,\,\,\,\alpha +\beta \leq
2N+1\,, &  \\ 
\lambda _{2N+1}\,\,\, & \mathrm{if}\,\,\,\,\alpha +\beta >2N+1\,, & 
\end{array}%
\right.   \label{mlSEN}
\end{equation}%
with $\lambda_{2N+1}=0_S$ being the zero element of the semigroup. A subset decomposition of the semigroup $S_{E}^{\left(2N\right)}=S_{0}\cup S_{1}\cup S_{2}$ is given by
\begin{eqnarray}
S_{0} &=&\left\{ \lambda _{2m},\ \text{with }m=0,\ldots ,N \right\} \cup \{\lambda _{2N+1}\}\,, \notag \\
S_{1} &=&\left\{ \lambda _{2m+1},\ \text{with }m=0,\ldots ,N-1 \right\} \cup \{\lambda _{2N+1}\}\,, \notag \\
S_{2} &=&\left\{ \lambda _{2m+2},\ \text{with }m=0,\ldots , N-1 \right\} \cup \{\lambda _{2N+1}\}\,.
\label{sdN}
\end{eqnarray}%
The decomposition \eqref{sdN} is resonant since it satisfies the same algebraic structure as the subspace decomposition \eqref{AS} of the original $\mathfrak{so}\left(2\right)$-extension of the $\mathfrak{osp}\left(2,2\right)\otimes \mathfrak{sp}\left(2\right)$ superalgebra. The resonant expanded superalgebra is given by
\begin{eqnarray}
   \mathfrak{G}_{R}=\left(S_0\times V_0\right)\oplus\left(S_1\times V_1\right)\oplus\left(S_2\times V_2\right)\,, 
\end{eqnarray}
 where $V_0$, $V_{1}$, and $V_{2}$ are the subspaces of the starting relativistic superalgebra given in \eqref{sd3}. A smaller superalgebra can be extracted from $\mathfrak{G}_{R}$ by considering a partition of the subsets $S_p=\hat{S}_p\cup \check{S}_p$ with
\begin{align}
    \check{S}_{0}&=\left\{ \lambda _{2m},\ \text{with }m=0,\ldots ,N-1 \right\}\,, & \hat{S}_{0}&=\left\{\lambda_{2N},\lambda_{2N+1}\right\}\,, \notag \\
    \check{S}_{1}&=\left\{ \lambda _{2m+1},\ \text{with }m=0,\ldots ,N-1 \right\}\,, & \hat{S}_{1}&=\left\{\lambda_{2N+1}\right\}\,, \notag \\
    \check{S}_{2}&=\left\{ \lambda _{2m+2},\ \text{with }m=0,\ldots ,N-1 \right\}\,, & \hat{S}_{2}&=\left\{\lambda_{2N+1}\right\}\,. \label{Sp2}
\end{align}
One can check that this subset partition satisfies the conditions \eqref{Part} and $\hat{S}_p\cap \check{S}_p =\varnothing$. Both conditions ensure that 
\begin{eqnarray}
    [\check{\mathfrak{G}}_{R},\hat{\mathfrak{G}}_{R}]&\subset&\hat{\mathfrak{G}}_{R}\,,
\end{eqnarray}
where
\begin{align}
    \check{\mathfrak{G}}_{R}&=\left(\check{S}_{0}\times V_0\right)\oplus\left(\check{S}_{1}\times V_1\right)\oplus \left(\check{S}_{2}\times V_2\right)\,, \notag \\
    \hat{\mathfrak{G}}_{R}&=\left(\hat{S}_{0}\times V_0\right)\oplus\left(\hat{S}_{1}\times V_1\right)\oplus \left(\hat{S}_{2}\times V_2\right)\,.
\end{align}
Then, $\check{G}_{R}$ is a reduced superalgebra of the resonant superalgebra $\mathfrak{G}_{R}$ \cite{Izaurieta:2006zz}. The generators of the reduced superalgebra $\check{G}_{R}$ are related to the relativistic ones through the semigroup elements as follows:
\begin{align}
    \texttt{J}^{\left(m\right)}&=\lambda_{2m}J\,, & \texttt{P}_{a}^{\left(m\right)}&=\lambda_{2m}P_{a}\,, & \texttt{T}^{\left(m\right)}&=\lambda_{2m}T\,, \notag \\
    \texttt{H}^{\left(m\right)}&=\lambda_{2m+2}H\,, & \texttt{G}_{a}^{\left(m\right)}&=\lambda_{2m+2}G_{a}\,, & \texttt{U}^{\left(m\right)}&=\lambda_{2m+2}U\,, \notag \\
    \texttt{Q}_{\alpha}^{+\,\left(m\right)}&= \lambda_{2m+1}Q_{\alpha}^{+}\,, & \texttt{Q}_{\alpha}^{-\,\left(m\right)}&= \lambda_{2m+1}Q_{\alpha}^{-}\,,
\end{align}
where $m=0,\dots,N-1$. The $\check{G}_{R}$ generators satisfy the following commutation relations:
\begin{align}
\left[ \texttt{J}^{\left(m\right)},\texttt{G}_{a}^{\left(n\right)}\right] &=\epsilon_{ab}\texttt{G}_{b}^{\left(m+n\right)}\,, &
\left[ \texttt{G}_{a}^{\left(m\right)},\texttt{G}_{b}^{\left(n\right)}\right] &=-\epsilon_{ab}\texttt{J}^{\left(m+n+2\right)}\,, \notag \\
\left[ \texttt{J}^{\left(m\right)},\texttt{P}_{a}^{\left(n\right)}\right] &=\epsilon_{ab}\texttt{P}_{b}^{\left(m+n\right)}\,,  &
\left[ \texttt{G}_{a}^{\left(m\right)},\texttt{P}_{a}^{\left(n\right)}\right] &=-\epsilon_{ab}\texttt{H}^{\left(m+n\right)}\,, \notag \\
\left[ \texttt{H}^{\left(m\right)},\texttt{P}_{a}^{\left(n\right)}\right] &=\frac{1}{\ell^{2}}\epsilon_{ab}\texttt{G}_{b}^{\left(m+n\right)}\,, &
\left[ \texttt{P}_{a}^{\left(m\right)},\texttt{P}_{b}^{\left(n\right)}\right] &=-\frac{1}{\ell^{2}}\epsilon_{ab}\texttt{J}^{\left(m+n\right)}\,, \notag \\
\left[ \texttt{H}^{\left(m\right)},\texttt{G}_{a}^{\left(n\right)}\right] &=\epsilon_{ab}\texttt{P}_{b}^{\left(m+n+2\right)}\,,  &
\left[ \texttt{J}^{\left(m\right)},\texttt{Q}_{\alpha}^{\pm\,\left(n\right)}\right] &=-\frac{1}{2}\left(\gamma_{0}\right)_{\alpha}^{\ \beta}\texttt{Q}_{\beta}^{\pm \, \left(m+n\right)}\,, \notag \\ 
\left[ \texttt{H}^{\left(m\right)},\texttt{Q}_{\alpha}^{\pm\,\left(n\right)}\right] &=-\frac{1}{2\ell}\left(\gamma_{0}\right)_{\alpha}^{\ \beta}\texttt{Q}_{\beta}^{\pm \, \left(m+n+1\right)}\,, 
&
\left[ \texttt{P}_{a}^{\left(m\right)},\texttt{Q}_{\alpha}^{\pm\,\left(n\right)}\right] &=-\frac{1}{2\ell}\left(\gamma_{a}\right)_{\alpha}^{\ \beta}\texttt{Q}_{\beta}^{\mp \, \left(m+n\right)}\,, \notag \\
\left[ \texttt{G}_{a}^{\left(m\right)},\texttt{Q}_{\alpha}^{\pm\,\left(n\right)}\right] &=-\frac{1}{2}\left(\gamma_{a}\right)_{\alpha}^{\ \beta}\texttt{Q}_{\beta}^{\mp \, \left(m+n+1\right)}\,, &
\left[ \texttt{T}^{\left(m\right)},\texttt{Q}_{\alpha}^{\pm\,\left(n\right)}\right] &=\pm\frac{1}{2}\left(\gamma_{0}\right)_{\alpha}^{\ \beta}\texttt{Q}_{\beta}^{\pm \, \left(m+n\right)}\,,  \label{SADSCNa}
\end{align}
together with the following anti-commutators:
\begin{align}
\{ \texttt{Q}_{\alpha}^{+\,\left(m\right)},\texttt{Q}_{\beta}^{+\,\left(n\right)}\} &=-\left(\gamma^{0}C \right)_{\alpha \beta} \left(\frac{1}{\ell}\texttt{J}^{\left(m+n+1\right)}+ \texttt{H}^{\left(m+n\right)} \right) -\left(\gamma^{0}C \right)_{\alpha \beta} \left(\texttt{U}^{\left(m+n\right)} +\frac{1}{\ell}\texttt{T}^{\left(m+n+1\right)} \right)\,,  \notag \\
\{ \texttt{Q}_{\alpha}^{+\,\left(m\right)},\texttt{Q}_{\beta}^{-\,\left(n\right)}\} &=-\left(\gamma^{a}C \right)_{\alpha \beta} \left( \frac{1}{\ell} \texttt{G}_{a}^{\left(m+n\right)} + \texttt{P}_{a}^{\left(m+n+1\right)}\right) \,, &\notag \\
\{ \texttt{Q}_{\alpha}^{-\,\left(m\right)},\texttt{Q}_{\beta}^{-\,\left(n\right)}\} &=-\left(\gamma^{0}C \right)_{\alpha \beta} \left(\frac{1}{\ell}\texttt{J}^{\left(m+n+1\right)}+ \texttt{H}^{\left(m+n\right)} \right) +\left(\gamma^{0}C \right)_{\alpha \beta} \left(\texttt{U}^{\left(m+n\right)} +\frac{1}{\ell}\texttt{T}^{\left(m+n+1\right)} \right)\,.  \label{SADSCNb}
\end{align}
Here, we considered the multiplication laws of the $S_{E}^{\left(2N\right)}$ semigroup \eqref{mlSEN} and the (anti-)commutation relations of the $\mathfrak{so}\left(2\right)$-extension of $\mathfrak{osp}\left(2,2\right)\otimes \mathfrak{sp}\left(2\right)$ \eqref{sAdS2}. In particular, for $m+n+1>N-1$ we have $T_{A}^{(m+n+1)}=0$. The obtained $\mathcal{N}=2$ superalgebra corresponds to a generalization of the $\mathcal{N}=2$ AdS-Carroll superalgebra which we have denoted as the $\mathcal{N}=2$ $\mathfrak{ads}$-$\mathfrak{car}^{\left(N\right)}$ superalgebra.\footnote{The $\mathfrak{ads}$-$\mathfrak{car}^{\left(N\right)}$ superalgebra can also be seen as a $\mathcal{N}=2$ supersymmetric extension of the generalized para-Poincaré algebra introduced in \cite{Concha:2023bly}.} In the flat limit $\ell\rightarrow\infty$, we find a generalized $\mathcal{N}=2$ Carroll superalgebra denoted as $\mathcal{N}=2$ $\mathfrak{car}^{\left(N\right)}$, which can be seen as the higher-dimensional Carrollian analogue of supersymmetric extensions of the (post-)Newtonian symmetries discussed in \cite{Ozdemir:2019orp,Gomis:2019sqv,Concha:2019dqs}.  Alternatively, the $\mathcal{N}=2$ $\mathfrak{car}^{\left(N\right)}$ superalgebra can also be derived from the $\mathcal{N}=2$ Poincaré superalgebra considering the same semigroup and series of steps employed to obtain the $\mathfrak{ads}$-$\mathfrak{car}^{\left(N\right)}$ superalgebra (see Figure \ref{fig3}). 

One can check that, for $N=1$ (i.e., considering $S^{(2)}_E$ in the expansion procedure), the $\mathcal{N}=2$ super $\mathfrak{ads}$-$\mathfrak{car}^{\left(1\right)}$ reproduces the $\mathcal{N}=2$ AdS-Carroll superalgebra \eqref{SADSC}. A novel $\mathcal{N}=2$ AdS-Carroll superalgebra appears for $N=2$ which involves the inclusion of two additional Majorana fermionic charges $\texttt{F}_{\alpha}^{\pm}$. The generators of the $\mathcal{N}=2$ $\mathfrak{ads}$-$\mathfrak{car}^{\left(2\right)}$ superalgebra satisfy the following commutators:
\begin{align}
\left[ \texttt{J},\texttt{G}_{a}\right] &=\epsilon_{ab}\texttt{G}_{b}\,, &
\left[ \texttt{G}_{a},\texttt{P}_{a}\right] &=-\epsilon_{ab}\texttt{H}\,,  &
\left[ \texttt{J},\texttt{P}_{a}\right] &=\epsilon_{ab}\texttt{P}_{b}\,, \notag \\
\left[ \texttt{H},\texttt{P}_{a}\right] &=\frac{1}{\ell^{2}}\epsilon_{ab}\texttt{G}_{b}\,, &
\left[ \texttt{P}_{a},\texttt{P}_{b}\right] &=-\frac{1}{\ell^{2}}\epsilon_{ab}\texttt{J}\,,  &
\left[ \texttt{J},\texttt{M}_{a}\right] &=\epsilon_{ab}\texttt{M}_{b}\,,  \notag \\
\left[ \texttt{J},\texttt{Z}_{a}\right] &=\epsilon_{ab}\texttt{Z}_{b}\,, &
\left[ \texttt{P}_{a},\texttt{M}_{b}\right] &=-\frac{1}{\ell^{2}}\epsilon_{ab}\texttt{K}\,,  &
\left[ \texttt{K},\texttt{G}_{a}\right] &=\epsilon_{ab}\texttt{Z}_{b}\,,  \notag \\
\left[ \texttt{Z},\texttt{P}_{a}\right] &=\frac{1}{\ell^{2}}\epsilon_{ab}\texttt{Z}_{b}\,, &
\left[ \texttt{P}_{a},\texttt{Z}_{b}\right] &=-\epsilon_{ab}\texttt{Z}\,,  &
\left[ \texttt{K},\texttt{P}_{a}\right] &=\epsilon_{ab}\texttt{M}_{b}\,,  \notag \\
\left[ \texttt{H},\texttt{M}_{a}\right] &=\frac{1}{\ell^{2}}\epsilon_{ab}\texttt{Z}_{b}\,, &
\left[ \texttt{G}_{a},\texttt{M}_{b}\right] &=-\epsilon_{ab}\texttt{Z}\,,  & 
\left[ \texttt{J},\texttt{Q}_{\alpha}^{\pm}\right] &=-\frac{1}{2}\left(\gamma_{0}\right)_{\alpha}^{\ \beta}\texttt{Q}_{\beta}^{\pm}\,, \notag \\ 
\left[ \texttt{P}_{a},\texttt{Q}_{\alpha}^{\pm}\right] &=-\frac{1}{2\ell}\left(\gamma_{a}\right)_{\alpha}^{\ \beta}\texttt{Q}_{\beta}^{\mp}\,, &
\left[ \texttt{H},\texttt{Q}_{\alpha}^{\pm}\right] &=-\frac{1}{2\ell}\left(\gamma_{0}\right)_{\alpha}^{\ \beta}\texttt{F}_{\beta}^{\pm}\,, & 
\left[ \texttt{G}_{a},\texttt{Q}_{\alpha}^{\pm}\right] &=-\frac{1}{2}\left(\gamma_{a}\right)_{\alpha}^{\ \beta}\texttt{F}_{\beta}^{\mp}\,, \notag \\
\left[ \texttt{J},\texttt{F}_{\alpha}^{\pm}\right] &=-\frac{1}{2}\left(\gamma_{0}\right)_{\alpha}^{\ \beta}\texttt{F}_{\beta}^{\pm}\,, &
\left[ \texttt{P}_{a},\texttt{F}_{\alpha}^{\pm}\right] &=-\frac{1}{2\ell}\left(\gamma_{a}\right)_{\alpha}^{\ \beta}\texttt{F}_{\beta}^{\mp}\,, &
\left[ \texttt{K},\texttt{Q}_{\alpha}^{\pm}\right] &=-\frac{1}{2}\left(\gamma_{0}\right)_{\alpha}^{\ \beta}\texttt{F}_{\beta}^{\pm}\,, \notag \\
\left[ \texttt{M}_{a},\texttt{Q}_{\alpha}^{\pm}\right] &=-\frac{1}{2\ell}\left(\gamma_{a}\right)_{\alpha}^{\ \beta}\texttt{F}_{\beta}^{\mp}\,, &
\left[ \texttt{T}_1,\texttt{Q}_{\alpha}^{\pm}\right] &=\pm\frac{1}{2}\left(\gamma_{0}\right)_{\alpha}^{\ \beta}\texttt{Q}_{\beta}^{\pm}\,, &
\left[ \texttt{T}_1,\texttt{F}_{\alpha}^{\pm}\right] &=\pm\frac{1}{2}\left(\gamma_{0}\right)_{\alpha}^{\ \beta}\texttt{F}_{\beta}^{\pm}\,, \notag \\
\left[ \texttt{T}_2,\texttt{Q}_{\alpha}^{\pm}\right] &=\pm\frac{1}{2}\left(\gamma_{0}\right)_{\alpha}^{\ \beta}\texttt{F}_{\beta}^{\pm}\,, \label{SADSC2a}
\end{align}
and anti-commutators:
\begin{eqnarray}
\{ \texttt{Q}_{\alpha}^{\pm},\texttt{Q}_{\beta}^{\pm}\} &=&-\left(\gamma^{0}C \right)_{\alpha \beta} \left(\frac{1}{\ell}\texttt{K}+ \texttt{H} \right) \mp\left(\gamma^{0}C \right)_{\alpha \beta} \left(\texttt{U}_1 +\frac{1}{\ell}\texttt{T}_2 \right)\,,  \notag \\
\{ \texttt{Q}_{\alpha}^{+},\texttt{Q}_{\beta}^{-}\} &=&-\left(\gamma^{a}C \right)_{\alpha \beta} \left( \frac{1}{\ell} \texttt{G}_{a} + \texttt{M}_{a}\right) \,, \notag \\
\{ \texttt{Q}_{\alpha}^{\pm},\texttt{F}_{\beta}^{\pm}\} &=&-\left(\gamma^{0}C \right)_{\alpha \beta} \left(\texttt{Z}\pm\texttt{U}_{2} \right)\,,  \notag \\
\{ \texttt{Q}_{\alpha}^{\pm},\texttt{F}_{\beta}^{\mp}\} &=&-\frac{1}{\ell}\left(\gamma^{a}C \right)_{\alpha \beta} \texttt{Z}_{a}\,,  \notag \\
\label{SADSC2b}
\end{eqnarray}
where we have defined 
\begin{align}
    \texttt{J}^{\left(0\right)}&=\texttt{J}\,, & \texttt{J}^{\left(1\right)}&=\texttt{K}\,, & \texttt{H}^{\left(0\right)}&=\texttt{H}\,, & \texttt{H}^{\left(1\right)}&=\texttt{Z}\,, \notag \\
    \texttt{G}_{a}^{\left(0\right)}&=\texttt{G}_{a}\,, & \texttt{G}_{a}^{\left(1\right)}&=\texttt{Z}_{a}\,, & \texttt{P}_{a}^{\left(0\right)}&=\texttt{P}_{a}\,, & \texttt{P}_{a}^{\left(1\right)}&=\texttt{M}_{a}\,, \notag \\
    \texttt{T}^{\left(0\right)}&=\texttt{T}_1\,,  & \texttt{T}^{\left(1\right)}&=\texttt{T}_2\,, & \texttt{U}^{\left(0\right)}&=\texttt{U}_1\,, & \texttt{U}^{\left(1\right)}&=\texttt{U}_2\,, \notag \\
    \texttt{Q}_{\alpha}^{\pm \, \left(0\right)}&=\texttt{Q}_{\alpha}^{\pm}\,, & \texttt{Q}_{\alpha}^{\pm \, \left(1\right)}&=\texttt{F}_{\alpha}^{\pm}\,.
\end{align}
In the flat limit $\ell\rightarrow\infty$, the (anti-)commutation relations \eqref{SADSC2a} and \eqref{SADSC2b} reproduces an $\mathcal{N}=2$ $\mathfrak{car}^{\left(2\right)}$ superalgebra which can alternatively be obtained as a resonant $\check{S}_{E}^{\left(4\right)}$-expansion of the $\mathcal{N}=2$ Poincaré superalgebra. As an ending remark, one could consider $S_{E}^{(2N+1)}$ as the relevant semigroup applied to the $\mathfrak{so}\left(2\right)$-extension of the $\mathfrak{osp}\left(2,2\right)\otimes \mathfrak{sp}\left(2\right)$ superalgebra. Nonetheless, as we shall discuss in next section, the non-degeneracy of the invariant bilinear trace requires $\langle \texttt{T}^{\left(m\right)}\texttt{U}^{\left(m\right)}\rangle\neq0$, which is fulfilled only for superalgebras obtained with $S_{E}^{(2N)}$.

\subsection{Non-relativistic expansions}

\begin{center}
 \begin{figure}[h!]
  \begin{center}
        \includegraphics[width=9.3cm, height=8.3cm]{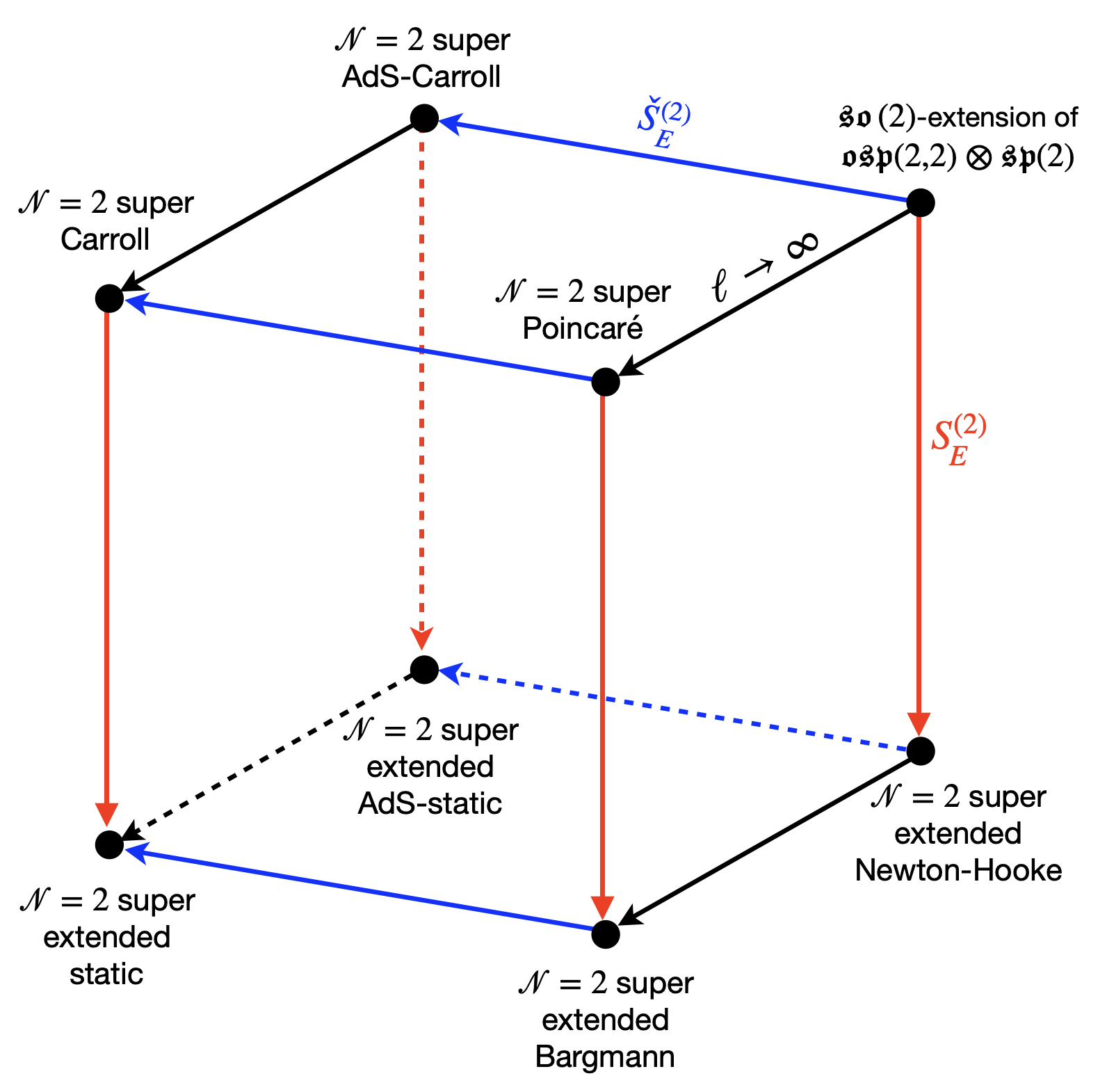}
        \captionsetup{font=footnotesize}
        \caption{This cube summarizes the different expansion relations starting from the $\mathfrak{so}\left(2\right)$-extension of the $\mathfrak{osp}\left(2,2\right)\otimes \mathfrak{sp}\left(2\right)$ superalgebra.}
        \label{fig4}
         \end{center}
        \end{figure}
    \end{center}

The NR sector of the kinematical superalgebra without degeneracy can be derived by expanding the  $\mathcal{N}=2$ relativistic superalgebras with $S_{E}^{\left(2\right)}$ as the relevant semigroup. As we shall see, the $S_{E}^{\left(2\right)}$-expansion can also be applied to the Carrollian superalgebra obtained previously, which reproduces an $\mathcal{N}=2$ extended AdS-static superalgebra and its flat limit (see Figure \ref{fig4}). The static sector can alternatively be recovered from the $\mathcal{N}=2$ NR superalgebras by applying a reduced $S$-expansion following the same procedure used to derive the $\mathcal{N}=2$ $\left(\mathfrak{ads}\right)$ $\mathfrak{car}^{\left(2\right)}$ superalgebra. Then, the expansion made with the $S_{E}^{\left(2\right)}$ semigroup along its reduced subset $\check{S}_{E}^{\left(2\right)}$ allows us to define the supersymmetric extension of the extended kinematical algebras \cite{Concha:2023bly}. Therefore, the NR expansions and UR ones not only extend the cube of Bacry and Lévy-Leblond \cite{Bacry:1968zf} to supersymmetry but also solve the degeneracy issue presented in the original cube.

\subsubsection{The $\mathcal{N}=2$ extended Newton-Hooke and extended Bargmann superalgebras}

Before applying the NR expansion, we first consider a subspace decomposition of the $\mathfrak{so}\left(2\right)$-extension of the $\mathfrak{osp}\left(2,2\right)\otimes \mathfrak{sp}\left(2\right)$ superalgebra different from the one considered in the UR expansion:
\begin{align}
    V_{0}&=\{J,H,T,U,Q_{\alpha}^{+}\}\,,  & V_{1}&=\{G_{a},P_{a},Q_{\alpha}^{-}\}\,, \label{sd4} 
\end{align}
 which satisfy a $\mathbb{Z}_2$-graded Lie algebra \eqref{sd}. Let $S_{E}^{\left(2\right)}$ be the finite semigroup whose elements satisfy the multiplication rule \eqref{mlSE2} and with $\lambda_3$ being the zero element of the semigroup. A resonant decomposition of the semigroup $S_{E}^{\left(2\right)}$ is given by
\begin{align}
    S_0&=\{\lambda_0,\lambda_2,\lambda_3\}\,, & S_1&=\{\lambda_1,\lambda_3\}\,,\label{sd2b}
\end{align}
 which satisfy the same algebraic structure as the subspaces of the $\mathfrak{so}\left(2\right)$-extension of the $\mathfrak{osp}\left(2,2\right)\otimes \mathfrak{sp}\left(2\right)$ superalgebra. Then, an NR superalgebra is obtained after performing a resonant $S_{E}^{\left(2\right)}$-expansion of the relativistic $\mathfrak{osp}\left(2,2\right)\otimes \mathfrak{sp}\left(2\right)$,
 \begin{eqnarray}
     \mathfrak{G}=\left(S_0\times V_0\right)\oplus\left(S_1\times V_1\right)\,,
 \end{eqnarray}
and considering the $0_S$-reduction $\lambda_3 T_{A}=0$. The expanded generators are related to the relativistic ones through the corresponding semigroup elements as in Table \ref{Table6}.
\begin{table}[h!]
\renewcommand{\arraystretch}{1.3}
\centering
    \begin{tabular}{l ||C{4cm}|C{4cm}|}
    $\lambda_3$  & \cellcolor[gray]{0.8} & \cellcolor[gray]{0.8}  \\ \hline 
    $\lambda_2$  & $\texttt{S}$, \ \, $\texttt{M}$, \ \, $\texttt{T}_{2}$,\ \, $\texttt{U}_{2}$, \ \,$\texttt{R}_{\alpha}$ & \cellcolor[gray]{0.8}  \\ \hline 
    $\lambda_1$  & \cellcolor[gray]{0.8} & $\texttt{G}_a$, \quad \, $\texttt{P}_a$, \quad \ \, $\texttt{Q}_{\alpha}^{-}$ \\ \hline
    $\lambda_0$  &  $\texttt{J}$, \ \, $\texttt{H}$, \, \,  $\texttt{T}_{1}$,\ \, $\texttt{U}_{1}$,\ \, $\texttt{Q}_{\alpha}^{+}$ & \cellcolor[gray]{0.8} \\ \hline
    & $J$, \ \ $H$, \ \ \,$T$, \ \ \,$U$,\ \, $Q_{\alpha}^{+}$ & $G_a$, \quad $P_a$, \quad \, $Q_{\alpha}^{-}$ \\ 
    \end{tabular}
    \captionsetup{font=footnotesize}
    \caption{Expanded generators in terms of the relativistic $\mathfrak{osp}\left(2,2\right)\otimes \mathfrak{sp}\left(2\right)$ ones and the semigroup elements.}
    \label{Table6}
    \end{table}

The (anti-)commutators of the resonant $S_{E}^{\left(2\right)}$-expansion of the $\mathfrak{so}\left(2\right)$-extension of the $\mathfrak{osp}\left(2,2\right)\otimes \mathfrak{sp}\left(2\right)$ superalgebra are obtained by combining the relativistic (anti-)commutation relations \eqref{sAdS2} and the multiplication rule of the $S_{E}^{\left(2\right)}$ semigroup \eqref{mlSE2}. In particular, the expanded generators satisfy the following commutation relations:
\begin{align}
\left[ \texttt{G}_{a},\texttt{G}_{b}\right] &=-\epsilon_{ab}\texttt{S}\,, &
\left[ \texttt{J},\texttt{G}_{a}\right] &=\epsilon_{ab}\texttt{G}_{b}\,, &
\left[ \texttt{J},\texttt{P}_{a}\right] &=\epsilon_{ab}\texttt{P}_{b}\,,  \notag\\
\left[ \texttt{G}_{a},\texttt{P}_{b}\right] &=-\epsilon_{ab}\texttt{M}\,,  &
\left[ \texttt{H},\texttt{G}_{a}\right] &=\epsilon_{ab}\texttt{P}_{b}\,, &
\left[ \texttt{H},\texttt{P}_{a}\right] &=\frac{1}{\ell^{2}}\epsilon_{ab}\texttt{G}_{b}\,, \notag \\
\left[ \texttt{P}_{a},\texttt{P}_{a}\right] &=-\frac{1}{\ell^{2}}\epsilon_{ab}\texttt{S}\,,  &
\left[ \texttt{J},\texttt{Q}_{\alpha}^{\pm}\right]&=-\frac{1}{2}\left(\gamma_{0}\right)_{\alpha}^{\ \beta}\texttt{Q}_{\beta}^{\pm}\,, &
\left[ \texttt{J},\texttt{R}_{\alpha}\right]&=-\frac{1}{2}\left(\gamma_{0}\right)_{\alpha}^{\ \beta}\texttt{R}_{\beta}\,, \notag \\
\left[ \texttt{H},\texttt{Q}_{\alpha}^{\pm}\right]&=-\frac{1}{2\ell}\left(\gamma_{0}\right)_{\alpha}^{\ \beta}\texttt{Q}_{\beta}^{\pm}\,, &
\left[ \texttt{H},\texttt{R}_{\alpha}\right]&=-\frac{1}{2\ell}\left(\gamma_{0}\right)_{\alpha}^{\ \beta}\texttt{R}_{\beta}\,, &
\left[ \texttt{S},\texttt{Q}_{\alpha}^{+}\right]&=-\frac{1}{2}\left(\gamma_{0}\right)_{\alpha}^{\ \beta}\texttt{R}_{\beta}\,, \notag \\
\left[ \texttt{M},\texttt{Q}_{\alpha}^{+}\right]&=-\frac{1}{2\ell}\left(\gamma_{0}\right)_{\alpha}^{\ \beta}\texttt{R}_{\beta}\,, &
\left[ \texttt{G}_{a},\texttt{Q}_{\alpha}^{+}\right] &=-\frac{1}{2}\left(\gamma_{a}\right)_{\alpha}^{\ \beta}\texttt{Q}_{\beta}^{-}\,, &
\left[ \texttt{G}_{a},\texttt{Q}_{\alpha}^{-}\right] &=-\frac{1}{2}\left(\gamma_{a}\right)_{\alpha}^{\ \beta}\texttt{R}_{\beta}\,, \notag \\
\left[ \texttt{P}_{a},\texttt{Q}_{\alpha}^{+}\right] &=-\frac{1}{2\ell}\left(\gamma_{a}\right)_{\alpha}^{\ \beta}\texttt{Q}_{\beta}^{-}\,, &
\left[ \texttt{P}_{a},\texttt{Q}_{\alpha}^{-}\right] &=-\frac{1}{2\ell}\left(\gamma_{a}\right)_{\alpha}^{\ \beta}\texttt{R}_{\beta}\,, &
\left[ \texttt{T}_1,\texttt{Q}_{\alpha}^{\pm}\right] &=\pm\frac{1}{2}\left(\gamma_{0}\right)_{\alpha}^{\ \beta}\texttt{Q}_{\beta}^{\pm}\,, \notag \\
\left[ \texttt{T}_1,\texttt{R}_{\alpha}\right] &=\pm\frac{1}{2}\left(\gamma_{0}\right)_{\alpha}^{\ \beta}\texttt{R}_{\beta}\,, &
\left[ \texttt{T}_2,\texttt{Q}_{\alpha}^{+}\right] &=\pm\frac{1}{2}\left(\gamma_{0}\right)_{\alpha}^{\ \beta}\texttt{R}_{\beta}\,, \label{SENH}
\end{align}
together with the following anti-commutation relations:
\begin{eqnarray}
\{ \texttt{Q}_{\alpha}^{+},\texttt{Q}_{\beta}^{+}\} &=&-\left(\gamma^{0}C \right)_{\alpha \beta} \left( \frac{1}{\ell}\texttt{J} +\texttt{H} \right)-\left(\gamma^{0}C \right)_{\alpha \beta} \left(\frac{1}{\ell}\texttt{T}_{1}+\texttt{U}_1\right)\,, \notag \\
\{ \texttt{Q}_{\alpha}^{+},\texttt{Q}_{\beta}^{-}\} &=&-\left(\gamma^{a}C \right)_{\alpha \beta} \left( \frac{1}{\ell} \texttt{G}_{a} + \texttt{P}_{a}\right) \,, \notag \\
\{ \texttt{Q}_{\alpha}^{-},\texttt{Q}_{\beta}^{-}\} &=&-\left(\gamma^{0}C \right)_{\alpha \beta} \left( \frac{1}{\ell}\texttt{S} +\texttt{M} \right)+\left(\gamma^{0}C \right)_{\alpha \beta} \left(\frac{1}{\ell}\texttt{T}_{2}+\texttt{U}_{2}\right)\,, \notag \\
\{ \texttt{Q}_{\alpha}^{+},\texttt{R}_{\beta}\} &=&-\left(\gamma^{0}C \right)_{\alpha \beta} \left( \frac{1}{\ell}\texttt{S} +\texttt{M} \right)-\left(\gamma^{0}C \right)_{\alpha \beta} \left(\frac{1}{\ell}\texttt{T}_{2}+\texttt{U}_{2}\right)\,.
\label{SENH2}
\end{eqnarray}
The obtained superalgebra corresponds to the $\mathcal{N}=2$ extended Newton-Hooke superalgebra \cite{Ozdemir:2019tby, Concha:2020eam}. The extended Newton-Hooke algebra\cite{Aldrovandi:1998im,Gibbons:2003rv,Brugues:2006yd,Alvarez:2007fw,Papageorgiou:2010ud,Duval:2011mi,Duval:2016tzi} appears as a subalgebra and is characterized by the central charges $\texttt{S}$ and $\texttt{M}$ which are crucial for avoiding degeneracy. In the presence of supersymmetry, the additional bosonic content $\{\texttt{T}_1,\texttt{T}_2,\texttt{U}_1,\texttt{U}_2\}$ ensures the non-degeneracy of the extended Bargmann superalgebra \cite{Bergshoeff:2016lwr} obtained in the vanishing cosmological constant limit $\ell\rightarrow\infty$. In particular, both $\texttt{T}_1$ and $\texttt{T}_2$ appear as expansion of the $R$-symmetry generator $T$ and act non-trivially on the fermionic charges. Let us note that the extended Bargmann superalgebra can also be derived by expanding the $\mathcal{N}=2$ Poincaré superalgebra with the $S_{E}^{\left(2\right)}$ semigroup, following the same procedure considered to obtain the extended Newton-Hooke superalgebra (see Figure \ref{fig4}).

\subsubsection{The $\mathcal{N}=2$ extended (AdS-)static superalgebra}

The $\mathcal{N}=2$ extended AdS-static superalgebra and its vanishing cosmological constant limit can be derived from the NR expansion of the $\mathcal{N}=2$ AdS-Carroll superalgebra \eqref{SADSC}. To this end, let us consider the following subspace decomposition of the $\mathcal{N}=2$ AdS-Carroll superalgebra:
\begin{align}
    V_{0}&=\{\texttt{J},\texttt{H},\texttt{T},\texttt{U},\texttt{Q}_{\alpha}^{+}\}\,,  & V_{1}&=\{\texttt{G}_{a},\texttt{P}_{a},\texttt{Q}_{\alpha}^{-}\}\,, \label{sd5} 
\end{align}
which satisfies \eqref{sd}. Let us consider the same semigroup used to obtain the extended Newton-Hooke superalgebra, namely $S_{E}^{\left(2\right)}=\{\lambda_0,\lambda_1,\lambda_2,\lambda_3\}$. The elements of $S_{E}^{\left(2\right)}$ obey the multiplication laws \eqref{mlSE2} and $\lambda_3$ corresponds to the zero element of the semigroup. A resonant decomposition of the semigroup $S_{E}^{\left(2\right)}$ is given by \eqref{sd2b}, which satisfies the same algebraic structure as the subspaces of the $\mathcal{N}=2$ AdS-Carroll superalgebra. Then, a supersymmetric extension of the extended AdS-static algebra\footnote{Originally denoted as the extended para-Bargmann in \cite{Concha:2023bly}, due to its isomorphism with the extended Bargmann algebra.} is obtained after performing a resonant $S_{E}^{\left(2\right)}$-expansion of the $\mathcal{N}=2$ AdS-Carroll superalgebra and imposing the $0_S$-reduction $\lambda_3 T_{A}=0$. The expanded generators are related to the Carrollian ones through the semigroup elements as in Table \ref{Table7}.
 \begin{table}[h!]
\renewcommand{\arraystretch}{1.3}
\centering
    \begin{tabular}{l ||C{4cm}|C{4cm}|}
    $\lambda_3$  & \cellcolor[gray]{0.8} & \cellcolor[gray]{0.8}  \\ \hline 
    $\lambda_2$  & $\texttt{S}$, \ \, $\texttt{M}$, \ \, $\texttt{T}_{2}$,\ \, $\texttt{U}_{2}$, \ \,$\texttt{R}_{\alpha}$ & \cellcolor[gray]{0.8}  \\ \hline 
    $\lambda_1$  & \cellcolor[gray]{0.8} & $\texttt{G}_a$, \quad \, $\texttt{P}_a$, \quad \ \, $\texttt{Q}_{\alpha}^{-}$ \\ \hline
    $\lambda_0$  &  $\texttt{J}$, \ \, $\texttt{H}$, \, \,  $\texttt{T}_{1}$,\ \, $\texttt{U}_{1}$,\ \, $\texttt{Q}_{\alpha}^{+}$ & \cellcolor[gray]{0.8} \\ \hline
    & $\texttt{J}$, \ \ \,$\texttt{H}$, \ \ \ \,$\texttt{T}$, \ \ \ $\texttt{U}$,\ \ \,  $\texttt{Q}_{\alpha}^{+}$ & $\texttt{G}_a$, \quad \ \, $\texttt{P}_a$, \quad \ \ $\texttt{Q}_{\alpha}^{-}$ \\ 
    \end{tabular}
    \captionsetup{font=footnotesize}
    \caption{Expanded generators in terms of the $\mathcal{N}=2$ super AdS-Carroll ones and the semigroup elements.}
    \label{Table7}
    \end{table}  

\noindent    
Then, the expanded superalgebra is given by
\begin{align}
\left[ \texttt{J},\texttt{G}_{a}\right] &=\epsilon_{ab}\texttt{G}_{b}\,, &
\left[ \texttt{J},\texttt{P}_{a}\right] &=\epsilon_{ab}\texttt{P}_{b}\,,  &
\left[ \texttt{P}_{a},\texttt{P}_{b}\right] &=-\frac{1}{\ell^{2}}\epsilon_{ab}\texttt{S}\,,  \notag\\
\left[ \texttt{H},\texttt{P}_{a}\right] &=\frac{1}{\ell^{2}}\epsilon_{ab}\texttt{G}_{b}\,, &
\left[ \texttt{G}_{a},\texttt{P}_{a}\right] &=-\epsilon_{ab}\texttt{M}\,,  &
\left[ \texttt{J},\texttt{Q}_{\alpha}^{\pm}\right] &=-\frac{1}{2}\left(\gamma_{0}\right)_{\alpha}^{\ \beta}\texttt{Q}_{\beta}^{\pm}\,,  \notag \\
\left[ \texttt{J},\texttt{R}_{\alpha}\right] &=-\frac{1}{2}\left(\gamma_{0}\right)_{\alpha}^{\ \beta}\texttt{R}_{\beta}\, &
\left[ \texttt{S},\texttt{Q}_{\alpha}^{+}\right] &=-\frac{1}{2}\left(\gamma_{0}\right)_{\alpha}^{\ \beta}\texttt{R}_{\beta}\, &
\left[ \texttt{P}_{a},\texttt{Q}_{\alpha}^{+}\right] &=-\frac{1}{2\ell}\left(\gamma_{a}\right)_{\alpha}^{\ \beta}\texttt{Q}_{\beta}^{-}\,, \notag \\
\left[ \texttt{P}_{a},\texttt{Q}_{\alpha}^{-}\right] &=-\frac{1}{2\ell}\left(\gamma_{a}\right)_{\alpha}^{\ \beta}\texttt{R}_{\beta}\,, &
\left[ \texttt{T}_1,\texttt{Q}_{\alpha}^{\pm}\right] &=\pm\frac{1}{2}\left(\gamma_{0}\right)_{\alpha}^{\ \beta}\texttt{Q}_{\beta}^{\pm}\,, &
\left[ \texttt{T}_1,\texttt{R}_{\alpha}\right] &=\frac{1}{2}\left(\gamma_{0}\right)_{\alpha}^{\ \beta}\texttt{R}_{\beta}\,, \notag \\
\left[ \texttt{T}_2,\texttt{Q}_{\alpha}^{+}\right] &=\frac{1}{2}\left(\gamma_{0}\right)_{\alpha}^{\ \beta}\texttt{R}_{\beta}\,, \label{SADSS}
\end{align}
and
\begin{align}
\{ \texttt{Q}_{\alpha}^{+},\texttt{Q}_{\beta}^{+}\} &=-\left(\gamma^{0}C \right)_{\alpha \beta} \left( \texttt{H} + \texttt{U}_1\right)\,, &
\{ \texttt{Q}_{\alpha}^{+},\texttt{Q}_{\beta}^{-}\} &=-\frac{1}{\ell}\left(\gamma^{a}C \right)_{\alpha \beta}  \texttt{G}_{a} \,, \notag \\
\{ \texttt{Q}_{\alpha}^{-},\texttt{Q}_{\beta}^{-}\} &=-\left(\gamma^{0}C \right)_{\alpha \beta} \left( \texttt{M} - \texttt{U}_2\right)\,, &
\{ \texttt{Q}_{\alpha}^{+},\texttt{R}_{\beta}\} &=-\left(\gamma^{0}C \right)_{\alpha \beta} \left( \texttt{M} + \texttt{U}_2\right)\,,\label{SADSS2}
\end{align}
where we have considered the multiplication laws of $S_{E}^{\left(2\right)}$ along the starting (anti-)commutation relations of the $\mathcal{N}=2$ super AdS-Carroll superalgebra given by \eqref{SADSC}. The obtained superalgebra corresponds to a $\mathcal{N}=2$ supersymmetric extension of the extended para-Bargmann algebra defined in \cite{Concha:2023bly}. As its bosonic sector, the $\mathcal{N}=2$ extended AdS-static superalgebra \eqref{SADSS} and \eqref{SADSS2} is isomorphic to the $\mathcal{N}=2$ extended Bargmann superalgebra \cite{Bergshoeff:2016lwr} by interchanging $\texttt{G}_{a}$ and $\ell\texttt{P}_{a}$. In the vanishing cosmological constant limit $\ell\rightarrow\infty$, we get the $\mathcal{N}=2$ supersymmetric extension of the extended static algebra introduced in \cite{Concha:2023bly}.

An alternative procedure to recover the $\mathcal{N}=2$ extended AdS-static superalgebra is starting from the $\mathcal{N}=2$ extended Newton-Hooke and apply a resonant $\check{S}_{E}^{\left(2\right)}$-expansion (see Figure \ref{fig4}). In such case, the subspace decomposition of the extended Newton-Hooke superalgebra that one has to consider is the following:
\begin{align}
    V_{0}&=\{\texttt{J},\texttt{S},\texttt{P}_a,\texttt{T}_1,\texttt{T}_2\}\,, & V_{1}&=\{\texttt{Q}_{\alpha}^{+},\texttt{Q}_{\alpha}^{-},\texttt{R}_{\alpha}\}\,, & V_{2}&=\{\texttt{H},\texttt{M},\texttt{G}_{a},\texttt{U}_{1},\texttt{U}_2\}\,. \label{sd2c} 
\end{align}
After performing a resonant $S_{E}^{(2)}$-expansion of the extended Newton-Hooke superalgebra and extracting a $\check{S}_{E}^{\left(2\right)}$ reduced subalgebra, we can express the $\mathcal{N}=2$ super extended AdS-static generators in terms of the extended Newton-Hooke ones as in Table \ref{Table8}.
\renewcommand{\arraystretch}{1.2}
\begin{table}[h]
\centering
    \begin{tabular}{l ||C{4cm}|C{3.5cm}|C{4cm}|}
    $\lambda_3$  & \cellcolor[gray]{0.8} & \cellcolor[gray]{0.8}& \cellcolor[gray]{0.8}   \\ \hline 
    $\lambda_2$  & \cellcolor[gray]{0.8} & \cellcolor[gray]{0.8}& $\texttt{H}$, \quad $\texttt{M}$, \quad $\texttt{G}_{a}$, \quad $\texttt{U}_{1}$, \quad $\texttt{U}_{2}$  \\ \hline 
    $\lambda_1$  & \cellcolor[gray]{0.8} & $\texttt{Q}_{\alpha}^{+}$, \quad $\texttt{Q}_{\alpha}^{-}$, \quad $\texttt{R}_{\alpha}$ & \cellcolor[gray]{0.8}  \\ \hline
    $\lambda_0$  &  $\texttt{J}$, \quad $\texttt{S}$, \quad $\texttt{P}_{a}$, \quad $\texttt{T}_{1}$, \quad $\texttt{T}_{2}$ & \cellcolor[gray]{0.8}  & \cellcolor[gray]{0.8}  \\ \hline
    & $\texttt{J}$, \quad $\texttt{S}$, \quad $\texttt{P}_{a}$, \quad $\texttt{T}_{1}$, \quad $\texttt{T}_{2}$ & $\texttt{Q}_{\alpha}^{+}$, \quad $\texttt{Q}_{\alpha}^{-}$, \quad $\texttt{R}_{\alpha}$ & $\texttt{H}$, \quad $\texttt{M}$, \quad $\texttt{G}_{a}$, \quad $\texttt{U}_{1}$, \quad $\texttt{U}_{2}$  \\ 
    \end{tabular}
    \captionsetup{font=footnotesize}
    \caption{$\mathcal{N}=2$ super extended AdS-static generators in terms of the extended Newton-Hooke ones and the semigroup elements.}
    \label{Table8}
    \end{table}

\subsubsection{Generalized Newton-Hooke and Galilean superalgebras}

Generalized kinematical superalgebras can be obtained by applying the NR expansion with a bigger semigroup $S_{E}^{\left(2N\right)}=\{\lambda_0,\lambda_1,\lambda_2,\cdots,\lambda_{2N+1}\}$. Indeed, $\mathcal{N}=2$ supersymmetric extension of the generalized Newton-Hooke algebra \cite{Gomis:2019nih,Concha:2023bly} can be derived from the $\mathfrak{so}\left(2\right)$-extension of the  $\mathfrak{osp}\left(2,2\right)\otimes \mathfrak{sp}\left(2\right)$ superalgebra. To this end, let us consider the subspace decomposition of the relativistic superalgebra given in \eqref{sd4}, which satisfies a $\mathbb{Z}_2$-graded Lie algebra \eqref{sd}. Let $S_{E}^{\left(2N\right)}=\{\lambda_0,\lambda_1,\lambda_2,\cdots,\lambda_{2N+1}\}$ be the finite semigroup whose elements obey \eqref{mlSEN}, and $\lambda_{2N+1}=0_S$ corresponds to the zero element of the semigroup. A resonant subset decomposition of the semigroup $S_{E}^{\left(2N\right)}=S_0\cup S_1$ reads 
\begin{eqnarray}
S_{0} &=&\left\{ \lambda _{2m},\ \text{with }m=0,\ldots ,N \right\} \cup \{\lambda _{2N+1}\}\,, \notag \\
S_{1} &=&\left\{ \lambda _{2m+1},\ \text{with }m=0,\ldots ,N-1 \right\} \cup \{\lambda _{2N+1}\}\,, \label{sdN2}
\end{eqnarray}
which satisfy the same algebraic structure \eqref{sd} as the $\mathfrak{osp}\left(2,2\right)\otimes \mathfrak{sp}\left(2\right)$ subspaces. A resonant expanded superalgebra is given by
\begin{eqnarray}
   \mathfrak{G}_{R}=\left(S_0\times V_0\right)\oplus\left(S_1\times V_1\right)\,, 
\end{eqnarray}
where $V_0$ and $V_{1}$ are the relativistic subspaces given in \eqref{sd4}. Then, generalized Newton-Hooke superalgebras are obtained after performing a $0_S$-reduction, namely $0_S T_A=0$. The generalized Newton-Hooke generators are related to the relativistic ones through the semigroup elements as follows
\begin{align}
    \texttt{J}^{\left(m\right)}&=\lambda_{2m}J\,, & \texttt{P}_{a}^{\left(m\right)}&=\lambda_{2m+1}P_{a}\,, & \texttt{T}^{\left(m\right)}&=\lambda_{2m}T\,, \notag \\
    \texttt{H}^{\left(m\right)}&=\lambda_{2m}H\,, & \texttt{G}_{a}^{\left(m\right)}&=\lambda_{2m+1}G_{a}\,, & \texttt{U}^{(m)}&=\lambda_{2m}U\,, \notag \\
    \texttt{Q}_{\alpha}^{+\,\left(m\right)}&= \lambda_{2m}Q_{\alpha}^{+}\,, & \texttt{Q}_{\alpha}^{-\,\left(m\right)}&= \lambda_{2m+1}Q_{\alpha}^{-}\,. \label{genSGNH}
\end{align}
Considering the (anti-)commutation relations of the starting $\mathfrak{so}\left(2\right)$-extension of the $\mathfrak{osp}\left(2,2\right)\otimes \mathfrak{sp}\left(2\right)$ superalgebra and the multiplication laws of the $S_{E}^{\left(2N\right)}$ semigroup, the generalized Newton-Hooke superalgebra is expressed as
\begin{align}
\left[ \texttt{J}^{\left(m\right)},\texttt{G}_{a}^{\left(n\right)}\right] &=\epsilon_{ab}\texttt{G}_{b}^{\left(m+n\right)}\,, &
\left[ \texttt{G}_{a}^{\left(m\right)},\texttt{G}_{b}^{\left(n\right)}\right] &=-\epsilon_{ab}\texttt{J}^{\left(m+n+1\right)}\,, \notag \\
\left[ \texttt{J}^{\left(m\right)},\texttt{P}_{a}^{\left(n\right)}\right] &=\epsilon_{ab}\texttt{P}_{b}^{\left(m+n\right)}\,,  &
\left[ \texttt{G}_{a}^{\left(m\right)},\texttt{P}_{a}^{\left(n\right)}\right] &=-\epsilon_{ab}\texttt{H}^{\left(m+n+1\right)}\,, \notag \\
\left[ \texttt{H}^{\left(m\right)},\texttt{P}_{a}^{\left(n\right)}\right] &=\frac{1}{\ell^{2}}\epsilon_{ab}\texttt{G}_{b}^{\left(m+n\right)}\,, &
\left[ \texttt{P}_{a}^{\left(m\right)},\texttt{P}_{b}^{\left(n\right)}\right] &=-\frac{1}{\ell^{2}}\epsilon_{ab}\texttt{J}^{\left(m+n+1\right)}\,, \notag \\
\left[ \texttt{H}^{\left(m\right)},\texttt{G}_{a}^{\left(n\right)}\right] &=\epsilon_{ab}\texttt{P}_{b}^{\left(m+n\right)}\,,  &
\left[ \texttt{J}^{\left(m\right)},\texttt{Q}_{\alpha}^{\pm\,\left(n\right)}\right] &=-\frac{1}{2}\left(\gamma_{0}\right)_{\alpha}^{\ \beta}\texttt{Q}_{\beta}^{\pm \, \left(m+n\right)}\,, \notag \\ 
\left[ \texttt{H}^{\left(m\right)},\texttt{Q}_{\alpha}^{\pm\,\left(n\right)}\right] &=-\frac{1}{2\ell}\left(\gamma_{0}\right)_{\alpha}^{\ \beta}\texttt{Q}_{\beta}^{\pm \, \left(m+n\right)}\,,  &
\left[ \texttt{P}_{a}^{\left(m\right)},\texttt{Q}_{\alpha}^{+\,\left(n\right)}\right] &=-\frac{1}{2\ell}\left(\gamma_{a}\right)_{\alpha}^{\ \beta}\texttt{Q}_{\beta}^{- \, \left(m+n\right)}\,, \notag \\
\left[ \texttt{P}_{a}^{\left(m\right)},\texttt{Q}_{\alpha}^{-\,\left(n\right)}\right] &=-\frac{1}{2\ell}\left(\gamma_{a}\right)_{\alpha}^{\ \beta}\texttt{Q}_{\beta}^{+ \, \left(m+n+1\right)}\,, &
\left[ \texttt{G}_{a}^{\left(m\right)},\texttt{Q}_{\alpha}^{+ \,\left(n\right)}\right] &=-\frac{1}{2}\left(\gamma_{a}\right)_{\alpha}^{\ \beta}\texttt{Q}_{\beta}^{- \, \left(m+n\right)}\,, \notag \\
\left[ \texttt{G}_{a}^{\left(m\right)},\texttt{Q}_{\alpha}^{- \,\left(n\right)}\right] &=-\frac{1}{2}\left(\gamma_{a}\right)_{\alpha}^{\ \beta}\texttt{Q}_{\beta}^{+ \, \left(m+n+1\right)}\,, &
\left[ \texttt{T}^{(m)},\texttt{Q}_{\alpha}^{\pm\,\left(n\right)}\right] &=\pm\frac{1}{2}\left(\gamma_{0}\right)_{\alpha}^{\ \beta}\texttt{Q}_{\beta}^{\pm \, \left(m+n\right)}\,,  \label{SGNH}
\end{align}
together with the following anti-commutators:
\begin{align}
    \{ \texttt{Q}_{\alpha}^{+\,\left(m\right)},\texttt{Q}_{\beta}^{+\,\left(n\right)}\} &=-\left(\gamma^{0}C \right)_{\alpha \beta} \left(\frac{1}{\ell}\texttt{J}^{\left(m+n\right)}+ \texttt{H}^{\left(m+n\right)} \right) -\left(\gamma^{0}C \right)_{\alpha \beta} \left(\texttt{U}^{\left(m+n\right)} +\frac{1}{\ell}\texttt{T}^{\left(m+n\right)} \right)\,,  \notag \\
\{ \texttt{Q}_{\alpha}^{+\,\left(m\right)},\texttt{Q}_{\beta}^{-\,\left(n\right)}\} &=-\left(\gamma^{a}C \right)_{\alpha \beta} \left( \frac{1}{\ell} \texttt{G}_{a}^{\left(m+n\right)} + \texttt{P}_{a}^{\left(m+n\right)}\right) \,, \notag \\
\{ \texttt{Q}_{\alpha}^{-\,\left(m\right)},\texttt{Q}_{\beta}^{-\,\left(n\right)}\} &=-\left(\gamma^{0}C \right)_{\alpha \beta} \left(\frac{1}{\ell}\texttt{J}^{\left(m+n+1\right)}+ \texttt{H}^{\left(m+n+1\right)} \right) +\left(\gamma^{0}C \right)_{\alpha \beta} \left(\texttt{U}^{\left(m+n+1\right)} +\frac{1}{\ell}\texttt{T}^{\left(m+n+1\right)} \right)\,.  \label{SGNH2}
\end{align}
Let us note that the (anti-)commutators are Abelian for $T_{A}^{\left(m+n\right)}=\lambda_{2N+1}T_{A}$ due to the properties of the semigroup $S_{E}^{\left(2N\right)}$. The $\mathcal{N}=2$ generalized Newton-Hooke superalgebra \eqref{SGNH}-\eqref{SGNH2} corresponds to the $\mathcal{N}=2$ supersymmetric extension of the generalized Newton-Hooke algebra \cite{Gomis:2019nih} which has been denoted as $\mathfrak{nh}^{\left(N\right)}$ in \cite{Concha:2023bly}. In the vanishing cosmological constant limit $\ell\rightarrow\infty$, we obtain the $\mathcal{N}=2$ supersymmetric extension of the generalized Galilean algebras, denoted as $\mathfrak{gal}^{\left(N\right)}$ in \cite{Concha:2023bly}. The super $\mathfrak{gal}^{\left(N\right)}$, obtained after taking the flat limit, can be viewed both as a generalization of the extended Newtonian superalgebra discussed in \cite{Ozdemir:2019orp,Concha:2021jos} and as a supersymmetric extension of the extended post-Newtonian symmetries \cite{Gomis:2019sqv,Gomis:2019nih}. Let us note that the super $\mathfrak{gal}^{\left(N\right)}$ can also be derived as a resonant $S_{E}^{\left(2N\right)}$-expansion of the $\mathcal{N}=2$ Poincaré superalgebra (see Figure \ref{fig5}).

The $\mathcal{N}=2$ extended Newton-Hooke superalgebra \eqref{SENH}-\eqref{SENH2} together with the extended Bargmann one are recovered for $N=1$, namely with $S^{(2)}_E$ as the semigroup involved in the expansion. For $N=2$, we obtain the exotic Newtonian superalgebra introduced in \cite{Concha:2021jos}, whose generators obey the $\mathcal{N}=2$ extended Newton-Hooke (anti-)commutation relations \eqref{SENH}-\eqref{SENH2}, together with the following commutation relations:
\begin{align}
\left[ \texttt{G}_{a},\texttt{B}_{b}\right] &=-\epsilon_{ab}\texttt{Z}\,, &
\left[ \texttt{J},\texttt{B}_{a}\right] &=\epsilon_{ab}\texttt{B}_{b}\,, &
\left[ \texttt{J},\texttt{T}_{a}\right] &=\epsilon_{ab}\texttt{T}_{b}\,,  \notag\\
\left[ \texttt{G}_{a},\texttt{T}_{b}\right] &=-\epsilon_{ab}\texttt{Y}\,,  &
\left[ \texttt{H},\texttt{B}_{a}\right] &=\epsilon_{ab}\texttt{T}_{b}\,, &
\left[ \texttt{H},\texttt{T}_{a}\right] &=\frac{1}{\ell^{2}}\epsilon_{ab}\texttt{B}_{b}\,, \notag \\
\left[ \texttt{P}_{a},\texttt{B}_{a}\right] &=-\epsilon_{ab}\texttt{Y}\,,  &
\left[ \texttt{S},\texttt{G}_{a}\right] &=\epsilon_{ab}\texttt{B}_{b}\,, &
\left[ \texttt{S},\texttt{P}_{a}\right] &=\epsilon_{ab}\texttt{T}_{b}\,, \notag \\
\left[ \texttt{P}_{a},\texttt{T}_{a}\right] &=-\frac{1}{\ell^{2}}\epsilon_{ab}\texttt{Z}\,,  &
\left[ \texttt{M},\texttt{G}_{a}\right] &=\epsilon_{ab}\texttt{T}_{b}\,, &
\left[ \texttt{M},\texttt{P}_{a}\right] &=\frac{1}{\ell^{2}}\epsilon_{ab}\texttt{B}_{b}\,, \notag \\
\left[ \texttt{J},\texttt{W}_{\alpha}^{\pm}\right]&=-\frac{1}{2}\left(\gamma_{0}\right)_{\alpha}^{\ \beta}\texttt{W}_{\beta}^{\pm}\,, &
\left[ \texttt{S},\texttt{Q}_{\alpha}^{-}\right]&=-\frac{1}{2}\left(\gamma_{0}\right)_{\alpha}^{\ \beta}\texttt{W}_{\beta}^{-}\,, &
\left[ \texttt{S},\texttt{R}_{\alpha}\right]&=-\frac{1}{2}\left(\gamma_{0}\right)_{\alpha}^{\ \beta}\texttt{W}_{\beta}^{+}\,, \notag \\
\left[ \texttt{H},\texttt{W}_{\alpha}^{\pm}\right]&=-\frac{1}{2\ell}\left(\gamma_{0}\right)_{\alpha}^{\ \beta}\texttt{W}_{\beta}^{\pm}\,, &
\left[ \texttt{M},\texttt{Q}_{\alpha}^{-}\right]&=-\frac{1}{2\ell}\left(\gamma_{0}\right)_{\alpha}^{\ \beta}\texttt{W}_{\beta}^{-}\,, &
\left[ \texttt{M},\texttt{R}_{\alpha}\right]&=-\frac{1}{2\ell}\left(\gamma_{0}\right)_{\alpha}^{\ \beta}\texttt{W}_{\beta}^{+}\,, \notag \\
\left[ \texttt{Z},\texttt{Q}_{\alpha}^{+}\right]&=-\frac{1}{2}\left(\gamma_{0}\right)_{\alpha}^{\ \beta}\texttt{W}_{\beta}^{+}\,, &
\left[ \texttt{Y},\texttt{Q}_{\alpha}^{+}\right]&=-\frac{1}{2\ell}\left(\gamma_{0}\right)_{\alpha}^{\ \beta}\texttt{W}_{\beta}^{+}\,, &
\left[ \texttt{G}_{a},\texttt{R}_{\alpha}\right] &=-\frac{1}{2}\left(\gamma_{a}\right)_{\alpha}^{\ \beta}\texttt{W}_{\beta}^{-}\,, \notag \\
\left[ \texttt{G}_{a},\texttt{W}_{\alpha}^{-}\right] &=-\frac{1}{2}\left(\gamma_{a}\right)_{\alpha}^{\ \beta}\texttt{W}_{\beta}^{+}\,, &
\left[ \texttt{P}_{a},\texttt{R}_{\alpha}\right] &=-\frac{1}{2\ell}\left(\gamma_{a}\right)_{\alpha}^{\ \beta}\texttt{W}_{\beta}^{-}\,, &
\left[ \texttt{P}_{a},\texttt{W}_{\alpha}^{-}\right] &=-\frac{1}{2\ell}\left(\gamma_{a}\right)_{\alpha}^{\ \beta}\texttt{W}_{\beta}^{+}\,, \notag \\
\left[ \texttt{B}_{a},\texttt{Q}_{\alpha}^{\pm}\right] &=-\frac{1}{2}\left(\gamma_{a}\right)_{\alpha}^{\ \beta}\texttt{W}_{\beta}^{\mp}\,, &
\left[ \texttt{T}_{a},\texttt{Q}_{\alpha}^{\pm}\right] &=-\frac{1}{2\ell}\left(\gamma_{a}\right)_{\alpha}^{\ \beta}\texttt{W}_{\beta}^{\mp}\,, &
\left[ \texttt{T}_1,\texttt{W}_{\alpha}^{\pm}\right] &=\pm\frac{1}{2}\left(\gamma_{0}\right)_{\alpha}^{\ \beta}\texttt{W}_{\beta}^{\pm}\,, \notag \\
\left[ \texttt{T}_2,\texttt{Q}_{\alpha}^{-}\right] &=-\frac{1}{2}\left(\gamma_{0}\right)_{\alpha}^{\ \beta}\texttt{W}_{\beta}^{-}\,, &
\left[ \texttt{T}_2,\texttt{R}_{\alpha}\right] &=\frac{1}{2}\left(\gamma_{0}\right)_{\alpha}^{\ \beta}\texttt{W}_{\beta}^{+}\,, &
\left[ \texttt{T}_3,\texttt{Q}_{\alpha}^{+}\right] &=\frac{1}{2}\left(\gamma_{0}\right)_{\alpha}^{\ \beta}\texttt{W}_{\beta}^{+}\,, &\label{SEN}
\end{align}
and anti-commutators:
\begin{eqnarray}
\{ \texttt{Q}_{\alpha}^{+},\texttt{W}_{\beta}^{+}\} &=&-\left(\gamma^{0}C \right)_{\alpha \beta} \left( \frac{1}{\ell}\texttt{Z} +\texttt{Y} \right)-\left(\gamma^{0}C \right)_{\alpha \beta} \left(\frac{1}{\ell}\texttt{T}_{3}+\texttt{U}_3\right)\,, \notag \\
\{ \texttt{Q}_{\alpha}^{+},\texttt{W}_{\beta}^{-}\} &=&-\left(\gamma^{a}C \right)_{\alpha \beta} \left( \frac{1}{\ell} \texttt{B}_{a} + \texttt{T}_{a}\right) \,, \notag \\
\{ \texttt{Q}_{\alpha}^{-},\texttt{R}_{\beta}\} &=&-\left(\gamma^{a}C \right)_{\alpha \beta} \left( \frac{1}{\ell} \texttt{B}_{a} + \texttt{T}_{a}\right) \,, \notag \\
\{ \texttt{Q}_{\alpha}^{-},\texttt{W}_{\beta}^{-}\} &=&-\left(\gamma^{0}C \right)_{\alpha \beta} \left( \frac{1}{\ell}\texttt{Z} +\texttt{Y} \right)+\left(\gamma^{0}C \right)_{\alpha \beta} \left(\frac{1}{\ell}\texttt{T}_{3}+\texttt{U}_3\right)\,, \notag \\
\{ \texttt{R}_{\alpha},\texttt{R}_{\beta}\} &=&-\left(\gamma^{0}C \right)_{\alpha \beta} \left( \frac{1}{\ell}\texttt{Z} +\texttt{Y} \right)-\left(\gamma^{0}C \right)_{\alpha \beta} \left(\frac{1}{\ell}\texttt{T}_{3}+\texttt{U}_3\right)\,. \label{SEN2}
\end{eqnarray}
Here the generators have been identified to the generalized Newton-Hooke ones \eqref{genSGNH} as follows:
\begin{align}
    \texttt{J}&=\texttt{J}^{\left(0\right)}\,, & \texttt{G}_{a}&=\texttt{G}_{a}^{\left(0\right)}\,, & \texttt{T}_{1}&=\texttt{T}^{\left(0\right)}\,, & \texttt{U}_{1}&=\texttt{U}^{\left(0\right)}\,, \notag \\
    \texttt{S}&=\texttt{J}^{\left(1\right)}\,, & \texttt{B}_{a}&=\texttt{G}_{a}^{\left(1\right)}\,, & \texttt{T}_{2}&=\texttt{T}^{\left(1\right)}\,, & \texttt{U}_{2}&=\texttt{U}^{\left(1\right)}\,, \notag \\
    \texttt{Z}&=\texttt{J}^{\left(2\right)}\,, & \texttt{P}_{a}&=\texttt{P}_{a}^{\left(0\right)}\,, & \texttt{T}_{3}&=\texttt{T}^{\left(2\right)}\,, & \texttt{U}_{3}&=\texttt{U}^{\left(2\right)}\,, \notag \\
    \texttt{H}&=\texttt{H}^{\left(0\right)}\,, & \texttt{T}_{a}&=\texttt{P}_{a}^{\left(1\right)}\,, & \texttt{Q}_{a}^{+}&=\texttt{Q}_{a}^{+ \,\left(0\right)}\,, & \texttt{R}_{a}&=\texttt{Q}_{a}^{+ \,\left(1\right)}\,,   \notag \\
    \texttt{M}&=\texttt{H}^{\left(1\right)}\,,  &  \texttt{W}_{a}^{+}&=\texttt{Q}_{a}^{+\,\left(2\right)}\,, & \texttt{Q}_{a}^{-}&=\texttt{Q}_{a}^{- \,\left(0\right)}\,, & \texttt{W}_{a}^{-}&=\texttt{Q}_{a}^{- \,\left(1\right)}\,, \notag \\
    \texttt{Y}&=\texttt{H}^{\left(2\right)}\,. 
\end{align}
In the vanishing cosmological constant limit $\ell\rightarrow\infty$, we find the extended Newtonian superalgebra introduced in \cite{Ozdemir:2019orp}. At the bosonic level, the central charges $\texttt{Y}$ and $\texttt{Z}$ are essential to ensure the non-degeneracy of the invariant bilinear trace of the three-dimensional extended Newtonian gravity. In the absence of such central charges, the algebra corresponds to the so-called Newtonian algebra, which has been first encountered as the underlying symmetry of an action principle for Newtonian gravity \cite{Hansen:2019vqf} in four spacetime dimensions. On the other hand, the Newton-Hooke version of the Newtonian algebra has been useful to construct a four-dimensional non-relativistic gravity action based on the MacDowell-Mansouri formalism \cite{Concha:2022jdc}.

\subsubsection{Generalized (AdS-)static superalgebras}

\begin{center}
 \begin{figure}[h!]
  \begin{center}
        \includegraphics[width=8.5cm, height=7.7cm]{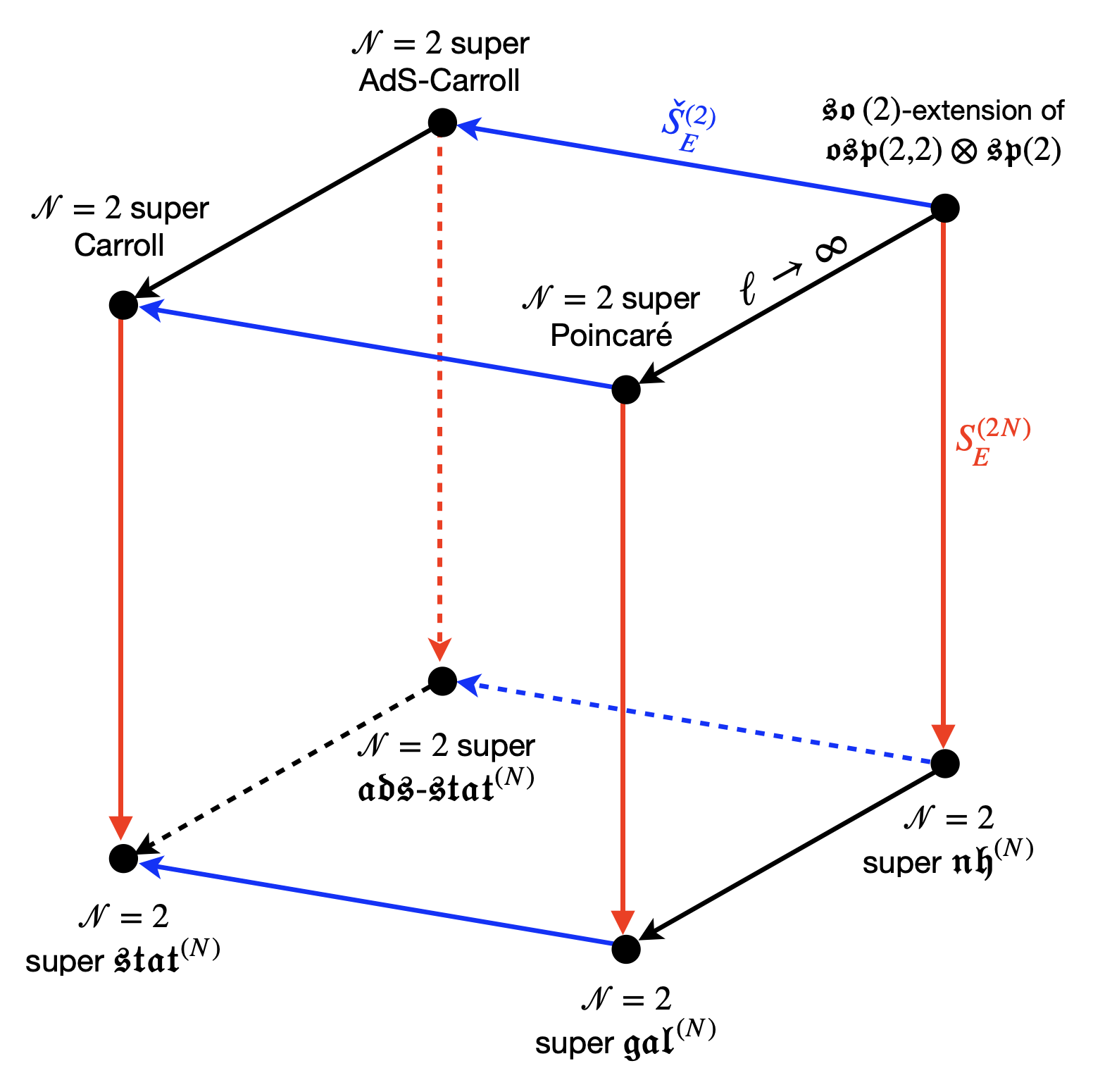}
        \captionsetup{font=footnotesize}
        \caption{This cube summarizes the generalized expansion processes starting from the $\mathfrak{so}\left(2\right)$-extension of the $\mathfrak{osp}\left(2,2\right)\otimes \mathfrak{sp}\left(2\right)$ superalgebra.}
        \label{fig5}
         \end{center}
        \end{figure}
    \end{center}

The $S_{E}^{\left(2N\right)}$ semigroup can also be applied to the Carrollian sector to reproduce generalizations of the (AdS-)static superalgebra. In particular, $\mathcal{N}=2$ generalized AdS-static superalgebras are obtained by expanding the $\mathcal{N}=2$ AdS-Carroll superalgebra \eqref{SADSC} with $S_{E}^{\left(2N\right)}$. Before applying the expansion, we first consider the subspace decomposition of the $\mathcal{N}=2$ AdS-Carroll superalgebra given in \eqref{sd5}, which satisfies a $\mathbb{Z}_2$-graded Lie algebra \eqref{sd}. A subset decomposition of the semigroup $S_{E}^{\left(2N\right)}=S_0\cup S_1$ resonant with \eqref{sd} is given by \eqref{sdN2}. Then, a $\mathcal{N}=2$ generalized AdS-static superalgebra is derived after performing a resonant $S_{E}^{\left(2N\right)}$-expansion of the $\mathcal{N}=2$ AdS-Carroll superalgebra \eqref{SADSC} and extracting a $0_S$-reduced subalgebra. The expanded generators can be expressed in terms of the AdS-Carroll ones through the semigroup elements as
\begin{align}
    \texttt{J}^{\left(m\right)}&=\lambda_{2m}\texttt{J}\,, & \texttt{P}_{a}^{\left(m\right)}&=\lambda_{2m+1}\texttt{P}_{a}\,, & \texttt{T}^{\left(m\right)}&=\lambda_{2m}\texttt{T}\,, &
    \texttt{Q}_{\alpha}^{+\,\left(m\right)}&= \lambda_{2m}\texttt{Q}_{\alpha}^{+}\,, \notag \\
    \texttt{H}^{\left(m\right)}&=\lambda_{2m}\texttt{H}\,, & \texttt{G}_{a}^{\left(m\right)}&=\lambda_{2m+1}\texttt{G}_{a}\,, & \texttt{U}^{(m)}&=\lambda_{2m}\texttt{U}\,, & \texttt{Q}_{\alpha}^{-\,\left(m\right)}&= \lambda_{2m+1}\texttt{Q}_{\alpha}^{-}\,. \label{genSGAdSS}
\end{align}
The explicit (anti-)commutation relations of the $\mathcal{N}=2$ generalized AdS-static superalgebra are obtained by considering the multiplication laws of $S_{E}^{\left(2N\right)}$ and the original (anti-)commutators of the $\mathcal{N}=2$ AdS-Carroll superalgebra. {Therefore} we get:
\begin{align}
\left[ \texttt{J}^{\left(m\right)},\texttt{G}_{a}^{\left(n\right)}\right] &=\epsilon_{ab}\texttt{G}_{b}^{\left(m+n\right)}\,, &
\left[ \texttt{J}^{\left(m\right)},\texttt{P}_{a}^{\left(n\right)}\right] &=\epsilon_{ab}\texttt{P}_{b}^{\left(m+n\right)}\,,  \notag\\
\left[ \texttt{P}_{a}^{\left(m\right)},\texttt{P}_{b}^{\left(n\right)}\right] &=-\frac{1}{\ell^{2}}\epsilon_{ab}\texttt{J}^{\left(m+n+1\right)}\,, &
\left[ \texttt{H}^{\left(m\right)},\texttt{P}_{a}^{\left(n\right)}\right] &=\frac{1}{\ell^{2}}\epsilon_{ab}\texttt{G}_{b}^{\left(m+n\right)}\,, \notag\\
\left[ \texttt{G}_{a}^{\left(m\right)},\texttt{P}_{a}^{\left(n\right)}\right] &=-\epsilon_{ab}\texttt{H}^{\left(m+n+1\right)}\,,   & 
\left[ \texttt{J}^{\left(m\right)},\texttt{Q}_{\alpha}^{\pm \, \left(n\right)}\right] &=-\frac{1}{2}\left(\gamma_{0}\right)_{\alpha}^{\ \beta}\texttt{Q}_{\beta}^{\pm \, \left(m+n\right)}\,,  \notag \\
\left[ \texttt{P}_{a}^{\left(m\right)},\texttt{Q}_{\alpha}^{+ \,\left(n\right)}\right] &=-\frac{1}{2\ell}\left(\gamma_{a}\right)_{\alpha}^{\ \beta}\texttt{Q}_{\beta}^{- \,\left(m+n\right)}\,, & \left[ \texttt{P}_{a}^{\left(m\right)},\texttt{Q}_{\alpha}^{- \,\left(n\right)}\right] &=-\frac{1}{2\ell}\left(\gamma_{a}\right)_{\alpha}^{\ \beta}\texttt{Q}_{\beta}^{+ \,\left(m+n+1\right)}\,, \notag \\
\left[ \texttt{T}^{\left(m\right)},\texttt{Q}_{\alpha}^{\pm \,\left(n\right)}\right] &=\pm\frac{1}{2}\left(\gamma_{0}\right)_{\alpha}^{\ \beta}\texttt{Q}_{\beta}^{\pm \,\left(m+n\right)}\,, & \notag \\
\{ \texttt{Q}_{\alpha}^{+ \,\left(m\right)},\texttt{Q}_{\beta}^{+ \,\left(n\right)}\} &=-\left(\gamma^{0}C \right)_{\alpha \beta} \left( \texttt{H}^{\left(m+n\right)} + \texttt{U}^{\left(m+n\right)}\right)\,, & \notag \\
\{ \texttt{Q}_{\alpha}^{+ \,\left(m\right)},\texttt{Q}_{\beta}^{- \,\left(n\right)}\} &=-\frac{1}{\ell}\left(\gamma^{a}C \right)_{\alpha \beta}  \texttt{G}_{a}^{\left(m+n\right)} \,, & \notag \\
\{ \texttt{Q}_{\alpha}^{- \,\left(m\right)},\texttt{Q}_{\beta}^{- \,\left(n\right)}\} &=-\left(\gamma^{0}C \right)_{\alpha \beta} \left( \texttt{H}^{\left(m+n+1\right)} - \texttt{U}^{\left(m+n+1\right)}\right)\,. & \label{SADSSN}
\end{align}
As in the generalized $\mathfrak{nh}^{\left(N\right)}$ superalgebra, the (anti-)commutators vanish for $T_{A}^{\left(m+n\right)}=\lambda_{2N+1}T_A$, due to the $0_S$-reduction condition imposed during the expansion. The obtained superalgebra \eqref{SADSSN}, which we have denoted as the $\mathcal{N}=2$ $\mathfrak{ads}$-$\mathfrak{stat}^{\left(N\right)}$ superalgebra, reproduces an $\mathcal{N}=2$ generalized static superalgebra in the vanishing cosmological constant limit $\ell\rightarrow\infty$. The $\mathcal{N}=2$ $\mathfrak{stat}^{\left(N\right)}$ superalgebra can also be recovered after imposing a resonant $S_{E}^{\left(2N\right)}$-expansion on the $\mathcal{N}=2$ Carroll superalgebra and applying a $0_S$-reduction. Alternatively, the $\mathcal{N}=2$ $\mathfrak{ads}$-$\mathfrak{stat}^{\left(N\right)}$ superalgebra together with its flat limit can be derived from the $\mathcal{N}=2$ super $\mathfrak{nh}^{\left(N\right)}$ one \eqref{SGNH}-\eqref{SGNH2} after performing a resonant $S_{E}^{\left(2\right)}$-expansion and extracting a $\check{S}_{E}^{\left(2N\right)}$ reduced subalgebra (see Figure \ref{fig5}).
As in the generalized $\mathfrak{nh}^{\left(N\right)}$ superalgebra, the (anti-)commutators vanishes for $T_{A}^{\left(m+n\right)}=\lambda_{2N+1}T_A$ due to the $0_S$-reduction condition imposed during the expansion. The obtained superalgebra \eqref{SADSSN}, which we have denoted as the $\mathcal{N}=2$ $\mathfrak{ads}$-$\mathfrak{stat}^{\left(N\right)}$ superalgebra, reproduces a $\mathcal{N}=2$ generalized static superalgebra in the vanishing cosmological constant limit $\ell\rightarrow\infty$. The $\mathcal{N}=2$ $\mathfrak{stat}^{\left(N\right)}$ superalgebra can also be recovered after imposing a resonant $S_{E}^{\left(2N\right)}$-expansion on the $\mathcal{N}=2$ Carroll superalgebra and applying a $0_S$-reduction. Alternatively, the $\mathcal{N}=2$ $\mathfrak{ads}$-$\mathfrak{stat}^{\left(N\right)}$ superalgebra together with its flat limit can be derived from the $\mathcal{N}=2$ super $\mathfrak{nh}^{\left(N\right)}$ one \eqref{SGNH}-\eqref{SGNH2} after performing a resonant $S_{E}^{\left(2\right)}$-expansion and extracting a $\check{S}_{E}^{\left(2N\right)}$ reduced subalgebra (see Figure \ref{fig5}).

\section{Three-dimensional non-Lorentzian supergravity theories based on extended kinematical superalgebras}\label{sec4}

In this section, we present three-dimensional NL supergravity actions invariant under NL superalgebras which have been obtained after a suitable sequence of expansions starting from the $\mathfrak{so}\left(2\right)$-extension of the $\mathfrak{osp}\left(2,2\right)\otimes \mathfrak{sp}\left(2\right)$ superalgebra and using the $S_{E}^{(2)}$ and $\check{S}_{E}^{(2)}$ semigroups. Our construction is based on the CS formalism, whose action reads
   \begin{eqnarray}
       I_{CS}&=&\frac{k}{4\pi}\int_{\mathcal{M}}\langle AdS + \frac{2}{3}A^{3}\rangle\,, \label{CS}
   \end{eqnarray}
where $A$ is the gauge connection one-form, $\langle\cdots\rangle$ denotes the invariant tensor and $k$ corresponds to the CS level of the theory, related to the gravitational constant $G$ through $k=1/(4G)$. The degeneracy of the invariant bilinear trace prevents to obtain a kinematical term for each gauge field and some fields could not be determined by the field equations. Remarkably, the $\mathcal{N}=2$ Carrollian superalgebras and the extended kinematical superalgebras discussed in the NR regime admit a non-degenerate invariant metric which, in three spacetime dimensions, implies the vanishing of the curvatures as field equations in the CS formulation.  

Here, we will present the CS supergravity action based on each $\mathcal{N}=2$ kinematical superalgebra of the cube \ref{fig4} and their generalizations discussed in the previous section. Fore this purpose, we will divide our results into three parts. First, we will review the $\mathcal{N}=2$ AdS-Carroll supergravity and explore novel generalizations. Then, we will study the non-relativistic regime of the $\mathfrak{osp}\left(2,2\right)\otimes \mathfrak{sp}\left(2\right)$ supergravity along with its vanishing cosmological constant limit. Finally, we will analyze NL supergravity actions based on the $\mathcal{N}=2$ (AdS-)static superalgebra and its generalization.

\subsection{$\mathcal{N}=2$ AdS-Carroll Chern-Simons supergravity and beyond}

The $\mathcal{N}=2$ AdS-Carroll supergravity defined à la CS in three spacetime dimensions has been first discussed in \cite{Ali:2019jjp}. Here, we briefly review its construction and present its generalization: the $\mathcal{N}=2$ $\mathfrak{ads}$-$\mathfrak{car}^{\left(N\right)}$ supergravity.

Let $A$ be the gauge connection one-form for the $\mathcal{N}=2$ AdS-Carroll superalgebra \eqref{SADSC},
\begin{eqnarray}
    A&=&\omega \texttt{J}+\omega^{a}\texttt{G}_{a}+\tau \texttt{H}+e^{a}\texttt{P}_{a}+t\texttt{T}+u\texttt{U}+\bar{\psi}^{+}\texttt{Q}^{+}+\bar{\psi}^{-}\texttt{Q}^{-}\,, \label{1fA}
\end{eqnarray}
where $\omega$ and $\omega_{a}$ are the time and spatial components of the spin-connection. The time and spatial components of the dreibein are given by $\tau$ and $e^{a}$, respectively. On the other hand, $t$ and $u$ are the gauge field dual to the generators $\texttt{T}$ and $\texttt{U}$, respectively. The curvature two-form $F=dA+\frac{1}{2}\left[A,A\right]$ reads
\begin{align}
F&=F\left(\omega\right)\texttt{J}+F^{a}\left(\omega^{b}\right)\texttt{G}_{a}+F\left(\tau\right)\texttt{H}+F^{a}\left(e^{b}\right)\texttt{P}_{a}+F\left(t\right)\texttt{T}+F\left(u\right)\texttt{U}+\nabla\bar{\psi}^{+}\texttt{Q}^{+}+\nabla\bar{\psi}^{-}\texttt{Q}^{-}\,, \label{2fF}
\end{align}
whose components are given by
\begin{align}
    F\left(\omega\right)&=d\omega+\frac{1}{2\ell^{2}}\epsilon^{ac}e_{a}e_{c}=R\left(\omega\right)+\frac{1}{2\ell^{2}}\epsilon^{ac}e_{a}e_{c}\,, \notag \\
    F^{a}\left(\omega^{b}\right)&=d\omega^{a}+\epsilon^{ac}\omega\omega_{c}+\frac{1}{\ell^{2}}\epsilon^{ac}\tau e_{c}+\frac{1}{\ell}\bar{\psi}^{+}\gamma^{a}\psi^{-}=R^{a}\left(\omega^{b}\right)+\frac{1}{\ell^{2}}\epsilon^{ac}\tau e_{c}+\frac{1}{\ell}\bar{\psi}^{+}\gamma^{a}\psi^{-}\,, \notag \\
    F\left(\tau\right)&=d\tau+\epsilon^{ac}\omega_a e_{c}+\frac{1}{2}\bar{\psi}^{+}\gamma^{0}\psi^{+}+\frac{1}{2}\bar{\psi}^{-}\gamma^{0}\psi^{-}=R\left(\tau\right)+\frac{1}{2}\bar{\psi}^{+}\gamma^{0}\psi^{+}+\frac{1}{2}\bar{\psi}^{-}\gamma^{0}\psi^{-}\,, \notag \\
    F^{a}\left(e^{b}\right)&=de^{a}+\epsilon^{ac}\omega e_{c}=R^{a}\left(e^{b}\right)\,, \notag \\
    F\left(t\right)&=dt\,, \notag \\
    F\left(u\right)&=du+\frac{1}{2}\bar{\psi}^{+}\gamma^{0}\psi^{+}-\frac{1}{2}\bar{\psi}^{-}\gamma^{0}\psi^{-}\,,\notag \\
    \nabla\psi^{+}&=d\psi^{+}+\frac{1}{2}\omega\gamma_0\psi^{+}+\frac{1}{2\ell}e^{a}\gamma_{a}\psi^{-}-\frac{1}{2}t\gamma_{0}\psi^{+}\,,\notag\\
    \nabla\psi^{-}&=d\psi^{-}+\frac{1}{2}\omega\gamma_0\psi^{-}+\frac{1}{2\ell}e^{a}\gamma_{a}\psi^{+}+\frac{1}{2}t\gamma_{0}\psi^{-}\,. \label{2fFs}
\end{align}
Let us note that the curvature two-forms \eqref{2fFs} reduce to the $\mathcal{N}=2$ Carroll ones in the vanishing cosmological constant limit $\ell\rightarrow\infty$. The $\mathcal{N}=2$ AdS-Carroll superalgebra admits a non-degenerate invariant tensor whose non-vanishing components are given by\cite{Ali:2019jjp}
\begin{align}
    \langle \texttt{J} \texttt{J} \rangle &=-\alpha_0\,, & \langle \texttt{P}_{a} \texttt{P}_{b} \rangle&=\frac{\alpha_0}{\ell^{2}}\delta_{ab}\,,\notag \\
    \langle \texttt{J} \texttt{H} \rangle &=-\alpha_1\,, & \langle \texttt{G}_{a} \texttt{P}_{b} \rangle&=\alpha_1\delta_{ab}\,,\notag \\
    \langle \texttt{T} \texttt{T} \rangle &=\alpha_0\,, & \langle \texttt{T} \texttt{U} \rangle&=\alpha_1\,,\notag \\
    \langle \texttt{Q}_{\alpha}^{+}\texttt{Q}_{\beta}^{+} \rangle &=2\alpha_1 C_{\alpha\beta}\,, &
    \langle \texttt{Q}_{\alpha}^{-}\texttt{Q}_{\beta}^{-} \rangle &=2\alpha_1 C_{\alpha\beta}\,, \label{IT2}
\end{align}
where $\alpha_0$ and $\alpha_1$ are arbitrary constants which can be related to the relativistic ones \eqref{IT} through the $S_{E}^{\left(2\right)}$ elements as
\begin{align}
    \alpha_0&=\lambda_0 \tilde{\alpha_0}\,, & \alpha_1&=\lambda_2 \tilde{\alpha_1}\,.
\end{align}
Here, the non-degeneracy of the invariant tensor requires $\alpha_1\neq 0$. 

The CS supergravity action for the $\mathcal{N}=2$ AdS-Carroll superalgebra \eqref{SADSC} is obtained considering the non-vanishing components of the invariant tensor \eqref{IT2} along the gauge connection one-form \eqref{1fA} into the general CS expression \eqref{CS}. Thus, we get \cite{Ali:2019jjp}
\begin{eqnarray}
    I_{\mathfrak{ads}\text{-}\mathfrak{car}}^{\mathcal{N}=2}&=&\frac{k}{4\pi}\int \alpha_0\left[\frac{1}{\ell^{2}}e_a R^{a}\left(e^{b}\right)- \omega R\left(\omega\right)+tdt\right]\notag\\ 
    &&+\alpha_1 \left[2e_aR^{a}\left(\omega^{b}\right)-2\tau R\left(\omega\right)+\frac{1}{\ell^2}\epsilon^{ac}\tau e_{a}e_{c}+2tdu-2\bar{\psi}^{+}\nabla\psi^{+}-2\bar{\psi}^{-}\nabla\psi^{-}\right]\,.\label{CSADSC}
\end{eqnarray}
The $\mathcal{N}=2$ AdS-Carroll CS supergravity action contains two independent terms. The term proportional to $\alpha_0$ corresponds to the UR version of the exotic Lagrangian \cite{Witten:1988hc}. On the other hand, the sector along $\alpha_1$ corresponds to the Carrollian regime of the $SO(2)$ extension of the $\mathcal{N}=2$ AdS supergravity Lagrangian \cite{Howe:1995zm}. Let us note that the $\mathcal{N}=2$ Carroll CS supergravity theory is derived in the vanishing cosmological constant limit $\ell\rightarrow\infty$. Since the supersymmetric terms appear only in the term involving $\alpha_1$, we can, without loss of generality, set $\alpha_0=0$ as this does not affect the non-degeneracy. In particular, the non-degeneracy ensures that the field equations are given by the vanishing of the curvature two-forms \eqref{2fFs}, which transform covariantly with respect to the supersymmetry transformation rules:
\begin{align}
    \delta \omega&=0\,, \notag\\
    \delta \omega^{a}&=\frac{1}{\ell}\bar{\varepsilon}^{+}\gamma^{a}\psi^{-}+\frac{1}{\ell}\bar{\varepsilon}^{-}\gamma^{a}\psi^{+}\,,\notag\\
    \delta \tau&=\bar{\varepsilon}^{+}\gamma^{0}\psi^{+}+\bar{\varepsilon}^{-}\gamma^{0}\psi^{-}\,,\notag\\
    \delta e^{a}&=0\,, \notag \\
    \delta t&=0\,, \notag \\
    \delta u&=\bar{\varepsilon}^{+}\gamma^{0}\psi^{+}-\bar{\varepsilon}^{-}\gamma^{0}\psi^{-}\,,\notag\\
    \delta \psi^{+}&=d\varepsilon^{+}+\frac{1}{2}\omega\gamma_{0}\varepsilon^{+}+\frac{1}{2\ell}e^{a}\gamma_{a}\varepsilon^{-}-\frac{1}{2}t\gamma_{0}\varepsilon^{+}\,,\notag \\
    \delta \psi^{-}&=d\varepsilon^{-}+\frac{1}{2}\omega\gamma_{0}\varepsilon^{-}+\frac{1}{2\ell}e^{a}\gamma_{a}\varepsilon^{+}+\frac{1}{2}t\gamma_{0}\varepsilon^{-}\,.
\end{align}
Here, $\varepsilon^{\pm}$ are the gauge parameters related to the fermionic charges $Q^{\pm}$. The supersymmetry transformation rules for the $\mathcal{N}=2$ Carroll CS supergravity are recovered in the flat limit $\ell\rightarrow\infty$.

\subsubsection*{Generalized $\mathcal{N}=2$ AdS-Carroll supergravity theory and its flat limit}

A generalized UR CS supergravity theory can be constructed from the $\mathcal{N}=2$ generalized AdS-Carroll superalgebra \eqref{SADSCNa}-\eqref{SADSCNb}. To this end, let us consider the gauge connection one-form for the $\mathcal{N}=2$ $\mathfrak{ads}$-$\mathfrak{car}^{\left(N\right)}$,
\begin{align}
    A&=\sum_{m=0}^{N-1}\left(\omega^{\left(m\right)}\texttt{J}^{\left(m\right)}+\omega^{a \,\left(m\right)}\texttt{G}_{a}^{\left(m\right)}+\tau^{\left(m\right)} \texttt{H}^{\left(m\right)}+ e^{a \,\left(m\right)}\texttt{P}_{a}^{\left(m\right)}+t^{\left(m\right)}\texttt{T}^{\left(m\right)}+u^{\left(m\right)}\texttt{U}^{\left(m\right)}\right. \notag\\
    &\left. +\bar{\psi}^{+ \,\left(m\right)}\texttt{Q}^{+ \,\left(m\right)}+\bar{\psi}^{- \,\left(m\right)}\texttt{Q}^{- \,\left(m\right)}\right)\,.\label{1fAN}
\end{align}
The corresponding curvature two-form reads
\begin{align}
    F&=\sum_{m=0}^{N-1}\left[F\left(\omega^{\left(m\right)}\right)\texttt{J}^{\left(m\right)}+F^{a}\left(\omega^{b \,\left(m\right)}\right)\texttt{G}_{a}^{\left(m\right)}+F\left(\tau^{\left(m\right)}\right)\texttt{H}^{\left(m\right)}+F^{a}\left(e^{b \,\left(m\right)}\right)\texttt{P}_{a}^{\left(m\right)}\right.\notag\\
&\left.+F\left(t^{\left(m\right)}\right)\texttt{T}^{\left(m\right)}+F\left(u^{\left(m\right)}\right)\texttt{U}^{\left(m\right)}+\nabla\bar{\psi}^{+ \,\left(m\right)}\texttt{Q}^{+ \,\left(m\right)}+\nabla\bar{\psi}^{- \,\left(m\right)}\texttt{Q}^{- \,\left(m\right)}\right]\,,
\end{align}
where the curvature components are directly derived from the $\mathcal{N}=2$ $\mathfrak{ads}$-$\mathfrak{car}^{\left(N\right)}$ superalgebra and are given by
\begin{align}
    F\left(\omega^{\left(m\right)}\right)&=d\omega^{\left(m\right)}+\frac{1}{2\ell^{2}}\sum_{n,p=0}^{N-1}\delta^{m}_{n+p}\,\epsilon^{ac}\,e_{a}^{\left(n\right)}e_{c}^{\left(p\right)}+\frac{1}{2\ell}\sum_{n,p=0}^{N-1}\delta^{m}_{n+p+1}\bar{\psi}^{\pm \,\left(n\right)}\gamma^{0}\psi^{\pm \,\left(p\right)}\,,\notag\\
    &+\frac{1}{2}\sum_{n,p=0}^{N-1}\delta^{m}_{n+p+2}\,\epsilon^{ac}\,\omega_{a}^{\left(n\right)}\omega_{c}^{\left(p\right)}\,,\notag\\
    F^{a}\left(\omega^{b \,\left(m\right)}\right)&=d\omega^{a \,\left(m\right)}+\sum_{n,p=0}^{N-1}\delta^{m}_{n+p}\left(\epsilon^{ac}\omega^{\left(n\right)}\omega_{c}^{\left(p\right)}+\frac{1}{\ell^{2}}\epsilon^{ac}\tau^{\left(n\right)} e_{c}^{\left(p\right)}+\frac{1}{\ell}\bar{\psi}^{+ \,\left(n\right)}\gamma^{a}\psi^{- \,\left(p\right)}\right)\,, \notag \\
    F\left(\tau^{\left(m\right)}\right)&=d\tau^{\left(m\right)}+\sum_{n,p=0}^{N-1}\delta^{m}_{n+p}\left(\epsilon^{ac}\omega_a^{\left(n\right)} e_{c}^{\left(p\right)}+\frac{1}{2}\bar{\psi}^{+ \,\left(n\right)}\gamma^{0}\psi^{+ \,\left(p\right)}+\frac{1}{2}\bar{\psi}^{- \,\left(n\right)}\gamma^{0}\psi^{- \,\left(p\right)}\right) \,,\notag \\
    F^{a}\left(e^{b \,\left(m\right)}\right)&=de^{a \,\left(m\right)}+\sum_{n,p=0}^{N-1}\delta^{m}_{n+p}\,\epsilon^{ac}\omega^{\left(n\right)} e_{c}^{\left(p\right)}+\sum_{n,p=0}^{N-1}\delta^{m}_{n+p+2}\,\epsilon^{ac}\,\tau^{\left(n\right)}\omega_{c}^{\left(p\right)}\ \,,\notag \\
    F\left(t^{\left(m\right)}\right)&=dt^{\left(m\right)}+\frac{1}{2\ell}\sum_{n,p=0}^{N-1}\delta^{m}_{n+p+1}\,\bar{\psi}^{\pm \,\left(n\right)}\gamma^{0}\psi^{\pm \,\left(p\right)} \,, \notag \\
    F\left(u^{\left(m\right)}\right)&=du^{\left(m\right)}+\frac{1}{2}\sum_{n,p=0}^{N-1}\delta^{m}_{n+p}\,\bar{\psi}^{\pm \,\left(n\right)}\gamma^{0}\psi^{\pm \,\left(p\right)}  \,,\notag\\
    \nabla\psi^{\pm \,\left(m\right)}&=d\psi^{\pm \,\left(m\right)}+\frac{1}{2}\sum_{n,p=0}^{N-1}\delta^{m}_{n+p}\left(\omega^{\left(n\right)}\gamma_0\psi^{\pm \,\left(p\right)}-\frac{1}{\ell}e^{a \,\left(n\right)}\gamma_{a}\psi^{\mp \,\left(p\right)}\mp t^{\left(n\right)}\gamma_{0}\psi^{\pm \,\left(p\right)}\right) \notag\\
    &+\frac{1}{2}\sum_{n,p=0}^{N-1}\delta^{m}_{n+p+1}\left(\omega^{a \,\left(n\right)}\gamma_{a}\psi^{\mp \,\left(p\right)}+\frac{1}{\ell}\tau^{\left(n\right)}\gamma_0\psi^{\pm \,\left(p\right)}\right) \,. \label{2fFN}
\end{align}
In the vanishing cosmological constant limit $\ell\rightarrow\infty$, we obtain the curvature two-forms of the $\mathcal{N}=2$ generalized Carroll superalgebra. The $\mathcal{N}=2$ $\mathfrak{ads}$-$\mathfrak{car}^{\left(N\right)}$ superalgebra admits the following components of an invariant tensor:
\begin{align}
    \langle \texttt{J}^{\left(m\right)} \texttt{H}^{\left(n\right)} \rangle &=-\alpha_{m+n+1}\,, & \langle \texttt{G}_{a}^{\left(m\right)} \texttt{P}_{b}^{\left(n\right)} \rangle&=\alpha_{m+n+1}\delta_{ab}\,,\notag \\
    \langle \texttt{T}^{\left(m\right)} \texttt{U}^{\left(n\right)} \rangle&=\alpha_{m+n+1}\,,&  \langle \texttt{U}^{\left(m\right)} \texttt{U}^{\left(n\right)} \rangle&=-\frac{1}{\ell}\alpha_{m+n+2}\,,\notag\\
    \langle \texttt{Q}_{\alpha}^{+ \,\left(m\right)}\texttt{Q}_{\beta}^{+ \,\left(n\right)} \rangle &=2\alpha_{m+n+1} C_{\alpha\beta}\,, &
    \langle \texttt{Q}_{\alpha}^{- \,\left(m\right)}\texttt{Q}_{\beta}^{- \,\left(n\right)} \rangle &=2\alpha_{n+m+1} C_{\alpha\beta}\,, \label{IT3}
\end{align}
where $i\leq N$ for $\alpha_i$. Here, we have intentionally omitted the components related to the exotic sector of the theory. The non-degeneracy of the invariant tensor is guaranteed for $\alpha_{N}\neq 0$. Let us note that, for $N=1$, we recover the invariant tensor for the $\mathcal{N}=2$ AdS-Carroll superalgebra.

Then, by considering the invariant tensor \eqref{IT3} and the connection one-form \eqref{1fAN} in the CS expression \eqref{CS}, we get an UR CS supergravity action for the $\mathcal{N}=2$ generalized AdS-Carroll superalgebra \eqref{SADSCNa}-\eqref{SADSCNb},
\begin{eqnarray}
    I_{\mathfrak{ads}\text{-}\mathfrak{car}^{(N)}}^{\mathcal{N}=2}&=&\frac{k}{4\pi}\int \sum_{p=1}^{N} \alpha_{p} \,\delta_{m+n+1}^{p}\left[2e_a^{\left(m\right)}R^{a}\left(\omega^{b \,\left(n\right)}\right)-2\tau^{\left(m\right)} R\left(\omega^{\left(n\right)}\right)+\frac{\delta_{q+r}^{n}}{\ell^2}\epsilon^{ac}\tau^{\left(m\right)} e_{a}^{\left(q\right)}e_{c}^{\left(r\right)}\right.\notag\\
    &&\left.+2t^{\left(m\right)}du^{\left(n\right)}-\frac{\delta^{n}_{q+1}}{\ell}u^{\left(m\right)}du^{\left(q\right)}-2\bar{\psi}^{+ \,\left(m\right)}\nabla\psi^{+ \,\left(n\right)}-2\bar{\psi}^{- \,\left(m\right)}\nabla\psi^{- \,\left(n\right)}\right]\,,\label{CSADSCN}
\end{eqnarray}
where we have defined
\begin{align}
    R^{a}\left(\omega^{b \,\left(m\right)}\right)&=d\omega^{a \,\left(m\right)}+\sum_{n,p=0}^{N-1}\delta_{n+p}^{m}\epsilon^{ac}\omega^{\left(n\right)}\omega_c^{\left(p\right)}\,, \notag\\
    R\left(\omega^{\left(m\right)}\right)&=d\omega^{\left(m\right)}+\frac{1}{2}\sum_{n,p=0}^{N-1}\delta^{m}_{n+p+2}\,\epsilon^{ac}\,\omega_{a}^{\left(n\right)}\omega_{c}^{\left(p\right)}\,.
\end{align}
Here, we can set $\alpha_{i}=0$ for $i\neq p$ without affecting the non-degeneracy condition $\alpha_{N}\neq 0$. The non-degeneracy of the invariant tensor \eqref{IT3} ensures that the field equations are given by the vanishing of the curvature two-forms \eqref{2fFN}. One can notice that the $\mathcal{N}=2$ $\mathfrak{ads}$-$\mathfrak{car}^{\left(N\right)}$ CS supergravity action is invariant under the following supersymmetry transformation rules:
\begin{align}
    \delta \omega^{\left(m\right)}&=\frac{1}{\ell}\sum_{n,p=0}^{N-1}\delta^{m}_{n+p+1}\bar{\varepsilon}^{\pm \,\left(n\right)}\gamma^{0}\psi^{\pm \,\left(p\right)}\,, \notag\\
    \delta \omega^{a \,\left(m\right)}&=\frac{1}{\ell}\sum_{n,p=0}^{N-1}\delta^{m}_{n+p}\bar{\varepsilon}^{\pm \,\left(n\right)}\gamma^{a}\psi^{\mp \,\left(p\right)}\,,\notag\\
    \delta \tau^{\left(m\right)}&=\sum_{n,p=0}^{N-1}\delta^{m}_{n+p}\bar{\varepsilon}^{\pm \,\left(n\right)}\gamma^{0}\psi^{\pm \,\left(p\right)}\,,\notag\\
    \delta e^{a \,\left(m\right)}&=\sum_{n,p=0}^{N-1}\delta^{m}_{n+p+1}\bar{\varepsilon}^{\pm \,\left(n\right)}\gamma^{a}\psi^{\mp \,\left(p\right)}\,, \notag \\
    \delta t^{\left(m\right)}&=\pm\frac{1}{\ell}\sum_{n,p=0}^{N-1}\delta^{m}_{n+p+1}\bar{\varepsilon}^{\pm \,\left(n\right)}\gamma^{0}\psi^{\pm \,\left(p\right)}\,, \notag \\
    \delta u^{\left(m\right)}&=\pm\frac{1}{\ell}\sum_{n,p=0}^{N-1}\delta^{m}_{n+p}\bar{\varepsilon}^{\pm \,\left(n\right)}\gamma^{0}\psi^{\pm \,\left(p\right)}\,,\notag\\
    \delta \psi^{\pm \,\left(m\right)}&=d\varepsilon^{\pm \,\left(m\right)}+\sum_{n,p=0}^{N-1}\delta^{m}_{n+p}\left(\frac{1}{2}\omega^{\left(n\right)}\gamma_{0}\varepsilon^{\pm \,\left(p\right)}+\frac{1}{2\ell}e^{a \,\left(n\right)}\gamma_{a}\varepsilon^{\mp \,\left(p\right)}\mp \frac{1}{2}t^{\left(n\right)}\gamma_{0}\varepsilon^{\pm \,\left(p\right)}\right)\notag\\
    &+\sum_{n,p=0}^{N-1}\delta^{m}_{n+p+1}\left(\frac{1}{2}\omega^{a \,\left(n\right)}\gamma_{a}\varepsilon^{\mp \,\left(p\right)}+\frac{1}{2\ell}\tau^{\left(n\right)}\gamma_0\varepsilon^{\pm \,\left(p\right)}\right)\,,
\end{align}
with $\varepsilon^{\pm \,\left(m\right)}$ being the gauge parameters related to the fermionic generators $Q^{\pm \,\left(m\right)}$. Let us note that the CS supergravity action \eqref{CSADSCN} can be written as the sum of individual Lagrangians based on $\mathcal{N}=2$ $\mathfrak{ads}$-$\mathfrak{car}^{\left(p\right)}$ for $p=1,2,\cdots,N$,
\begin{eqnarray}
    I_{\mathfrak{ads}\text{-}\mathfrak{car}^{(N)}}^{\mathcal{N}=2}&=&\frac{k}{4\pi}\int \sum_{p=1}^{N} \alpha_{p} \mathcal{L}_{\mathfrak{ads}\text{-}\mathfrak{car}^{\left(p\right)}}^{\mathcal{N}=2}\,.
\end{eqnarray}
In particular, $p=1$ reproduces the $\mathcal{N}=2$ AdS-Carroll supergravity Lagrangian proportional to $\alpha_1$ in \eqref{CSADSC}. For $p=2$, the Lagrangian is invariant under the $\mathcal{N}=2$ $\mathfrak{ads}$-$\mathfrak{car}^{\left(2\right)}$ superalgebra \eqref{SADSC2a}-\eqref{SADSC2b},
\begin{eqnarray}
    \mathcal{L}_{\mathfrak{ads}\text{-}\mathfrak{car}^{(2)}}^{\mathcal{N}=2}&=&\left[2e_a R^{a}\left(k^{b}\right)+2m_a R^{a}\left(\omega^{b}\right)-2\tau R\left(k\right)-2h R\left(\omega\right)+\frac{2}{\ell^2}\epsilon^{ac}\tau e_{a}m_{c}+\frac{1}{\ell^2}\epsilon^{ac}h e_{a}e_{c}\right.\notag\\
    &&\left.+2t_{1}du_{2}+2t_{2}du_{1}-\frac{1}{\ell}u_{1}du_{1}-2\bar{\psi}^{\pm}\nabla\chi^{\pm}-2\bar{\chi}^{\pm}\nabla\psi^{\pm}\right]\,,\label{CSADSC2}
\end{eqnarray}
with
\begin{align}
    R^{a}\left(k^{b}\right)&=dk^{a}+\epsilon^{ac}\omega k_{c}+\epsilon^{ac}k\omega_{c}\,,\notag\\
    R\left(k\right)&=dk\,, \notag\\
    \nabla\chi^{\pm}&=d\chi^{\pm}+\frac{1}{2}\omega\gamma_{0}\chi^{\pm}+\frac{1}{2}\omega^{a}\gamma_{a}\psi^{\mp}+\frac{1}{2}k\gamma_{0}\psi^{\pm}+\frac{1}{2\ell}e^{a}\gamma_{a}\chi^{mp}+\frac{1}{2\ell}\tau\gamma_0\psi^{\pm}\notag\\
    &+\frac{1}{2\ell}m^{a}\gamma_{a}\psi^{\pm}\mp\frac{1}{2}t_{1}\gamma_{0}\chi^{\pm}\mp\frac{1}{2}t_{2}\gamma_{0}\psi^{\pm}\,.
\end{align}
On the other hand, $R^{a}\left(\omega^{a}\right)$, $R\left(\omega\right)$ and $\nabla\psi^{\pm}$ are given in \eqref{2fFs}. Here we have defined
\begin{align}
    \omega^{\left(0\right)}&=\omega\,, & \omega^{\left(1\right)}&=k\,, & \tau^{\left(0\right)}&=\tau\,, & \tau^{\left(1\right)}&=h\,, \notag \\
    \omega_{a}^{\left(0\right)}&=\omega_{a}\,, & \omega_{a}^{\left(1\right)}&=k_{a}\,, & e_{a}^{\left(0\right)}&=e_{a}\,, & e_{a}^{\left(1\right)}&=m_{a}\,, \notag \\
    t^{\left(0\right)}&=t_1\,,  & t^{\left(1\right)}&=t_2\,, & u^{\left(0\right)}&=u_1\,, & u^{\left(1\right)}&=u_2\,, \notag \\
    \psi_{\alpha}^{\pm \, \left(0\right)}&=\psi_{\alpha}^{\pm}\,, & \psi_{\alpha}^{\pm \, \left(1\right)}&=\chi_{\alpha}^{\pm}\,.
\end{align}
In the vanishing cosmological constant limit $\ell\rightarrow\infty$, the theory reduces to the $\mathcal{N}=2$ $\mathfrak{car}^{\left(2\right)}$ supergravity, which extends the $\mathcal{N}=2$ Carroll supergravity appearing along the $\alpha_1$ term. Post-Carrollian extensions of the $\mathcal{N}=2$ AdS-Carroll supergravity \eqref{CSADSC} appear at higher order in $p$.

\subsection{$\mathcal{N}=2$ extended Newton-Hooke Chern-Simons supergravity and its generalization}

The three-dimensional $\mathcal{N}=2$ extended Newton-Hooke supergravity has been first introduced in \cite{Ozdemir:2019tby} and subsequently recovered in \cite{Concha:2020eam,Concha:2021jos}. Here, we start with a brief review of the extended Newton-Hooke supergravity theory discussed in \cite{Concha:2020eam,Concha:2021jos}. Then, we show that the NR expansion, as in the Carrollian case, allows us to derive novel post-Newtonian generalizations of the Newton-Hooke supergravity.

The gauge connection one-form for the $\mathcal{N}=2$ extended Newton-Hooke reads
\begin{eqnarray}
    A&=&\omega \texttt{J}+\omega^{a}\texttt{G}_{a}+s \texttt{S}+\tau \texttt{H}+e^{a}\texttt{P}_{a}+m\, \texttt{M}+t_1\texttt{T}_{1}+t_{2}\texttt{T}_{2}+u_{1}\texttt{U}_{1}+u_{2}\texttt{U}_{2}\notag\\
    &&+\bar{\psi}^{+}\texttt{Q}^{+}+\bar{\psi}^{-}\texttt{Q}^{-}+\bar{\rho}\,\texttt{R}\,, \label{1fANH}
\end{eqnarray}
with $s$ and $m$ being the gauge field related to the central charges $S$ and $M$. The extended Newton-Hooke superalgebra admits a non-degenerate bilinear invariant tensor whose components are given by
\begin{align}
    \langle \texttt{J} \texttt{S} \rangle &=-\sigma_0\,, & \langle \texttt{G}_{a} \texttt{G}_{b} \rangle&=\sigma_0\delta_{ab}\,,\notag \\
    \langle \texttt{J} \texttt{M} \rangle &=  \langle \texttt{H} \texttt{S} \rangle =-\beta_1\,,  & \langle \texttt{G}_{a} \texttt{P}_{b} \rangle&=\beta_1\delta_{ab}\,,\notag \\
    \langle \texttt{H} \texttt{M} \rangle &=-\frac{\sigma_0}{\ell^{2}}\,, & \langle \texttt{P}_{a} \texttt{P}_{b} \rangle&=\frac{\sigma_0}{\ell^{2}}\delta_{ab}\,,\notag \\
    \langle \texttt{T}_{1} \texttt{T}_{2} \rangle &=\sigma_0\,, &  \langle \texttt{T}_{1} \texttt{U}_{2} \rangle&= \langle \texttt{T}_{2} \texttt{U}_{1} \rangle=\beta_1\,,\notag \\
     \langle \texttt{U}_{1} \texttt{U}_{2} \rangle &=-\frac{\beta_{1}}{\ell}\,, &
    \langle \texttt{Q}_{\alpha}^{-}\texttt{Q}_{\beta}^{-} \rangle &= \langle \texttt{Q}_{\alpha}^{+}\texttt{R}_{\beta} \rangle=2\left(\beta_1+\frac{\sigma_{0}}{\ell}\right) C_{\alpha\beta}\,. \label{IT4}
\end{align}
Here, $\sigma_0$ and $\beta_1$ are related to the relativistic $\mathfrak{osp}\left(2,2\right)\otimes\mathfrak{sp}\left(2\right)$ constants through the $S_{E}^{\left(2\right)}$ elements:
\begin{align}
    \sigma_0&=\lambda_2\tilde{\alpha}_{0}\,, & \beta_{1}&=\lambda_2\tilde{\alpha_{1}}\,.
\end{align}
Here, $\sigma_0$ is related to an exotic NR sector\cite{Concha:2020eam,Concha:2021jos} which can be set to zero since the non-degeneracy is satisfied for $\sigma_0\neq\pm \ell \beta_{1}$ and $\beta_1\neq0$. Then, the NR CS supergravity action based on the $\mathcal{N}=2$ extended Newton-Hooke superalgebra \eqref{SENH} is given by
\begin{eqnarray}
    I_{\text{extended}\,\mathfrak{nh}}^{\mathcal{N}=2}&=&\frac{k}{4\pi}\int \beta_1 \left[2e_aR^{a}\left(\omega^{b}\right)-2m R\left(\omega\right)-2\tau R\left(s\right)+\frac{1}{\ell^2}\epsilon^{ac}\tau e_{a}e_{c}+2t_{1}du_{2}+2t_{2}du_{1}\right.\notag\\
    &&\left.-\frac{2}{\ell}u_{1}du_{2}-2\bar{\psi}^{-}\nabla\psi^{-}-2\bar{\psi}^{+}\nabla\rho-2\bar{\rho}\nabla\psi^{+}\right]\,,\label{CSENH}
\end{eqnarray}
where
\begin{eqnarray}
    R\left(\omega\right)&=&d\omega\,,\notag\\
    R^{a}\left(\omega^{a}\right)&=&d\omega^{a}+\epsilon^{ac}\omega\omega_{c}\,,\notag\\
    R\left(s\right)&=&ds+\frac{1}{2}\epsilon^{ac}\omega_{a}\omega_{c}\,,\notag\\
    \nabla\psi^{+}&=&d\psi^{+}+\frac{1}{2}\omega\gamma_0\psi^{+}+\frac{1}{2\ell}\tau\gamma_0\psi^{+}-\frac{1}{2}t_{1}\gamma_0\psi^{+}\,,\notag\\
    \nabla\psi^{-}&=&d\psi^{-}+\frac{1}{2}\omega\gamma_0\psi^{-}+\frac{1}{2\ell}\tau\gamma_0\psi^{-}+\frac{1}{2}\omega^{a}\gamma_{a}\psi^{+}+\frac{1}{2\ell}e^{a}\gamma_a\psi^{+}+\frac{1}{2}t_{1}\gamma_0\psi^{-}\,,\notag\\
    \nabla\rho&=&d\rho+\frac{1}{2}\omega\gamma_0\rho+\frac{1}{2}s\gamma_0\psi^{+}+\frac{1}{2}\omega^{a}\gamma_{a}\psi^{-}+\frac{1}{2\ell}e^{a}\gamma_a\psi^{-}+\frac{1}{2\ell}\tau\gamma_0\rho\notag\\
    &&+\frac{1}{2\ell}m\gamma_0\psi^{+}-\frac{1}{2}t_{1}\gamma_0\rho-\frac{1}{2}t_{2}\gamma_0\psi^{+}\,. \label{2fNH}
\end{eqnarray}
In the vanishing cosmological constant limit $\ell\rightarrow\infty$, the CS action reproduces the extended Bargmann theory in the presence of the additional bosonic content $t_1$, $t_2$, $u_1$, and $u_2$ \cite{Concha:2020eam}. The presence of such gauge fields is not whimsical but is due to the non-degeneracy requirement. Indeed, they are related to the bosonic generators $\{\texttt{T}_{1},\texttt{T}_2,\texttt{U}_{1},\texttt{U}_{2}\}$, which appear as expansions of the $\mathfrak{so}\left(2\right)$ automorphism generator $T$ and central charge $U$ of the $\mathfrak{so}\left(2\right)$-extension of the $\mathfrak{osp}\left(2,2\right)\otimes\mathfrak{sp}\left(2\right)$ superalgebra \eqref{sAdS2}. As in the relativistic case \cite{Howe:1995zm}, the extra generators ensure the non-degeneracy of the bilinear invariant trace, implying the vanishing of the curvature two-forms as equations of motion even in the flat limit. In particular, the bosonic curvature two-forms read
\begin{eqnarray}
     F\left(\omega\right)&=&R\left(\omega\right)+\frac{1}{2\ell}\bar{\psi}^{+}\gamma^{0}\psi^{+}\,, \notag \\
    F\left(\omega^{a}\right)&=&R^{a}\left(\omega^{b}\right)+\frac{1}{\ell^{2}}\epsilon^{ac}\tau e_{c}+\frac{1}{\ell}\bar{\psi}^{+}\gamma^{a}\psi^{-}\,, \notag \\
    F\left(\tau\right)&=&d\tau+\frac{1}{2}\bar{\psi}^{+}\gamma^{0}\psi^{+}\,, \notag \\
    F\left(e^{a}\right)&=&de^{a}+\epsilon^{ac}\omega e_{c}+\epsilon^{ac}\tau \omega_c+\bar{\psi}^{+}\gamma^{a}\psi^{-}\,, \notag \\
    F\left(s\right)&=&R\left(s\right)+\frac{1}{2\ell^{2}}\epsilon^{ac}e_{a}e_{c}+\frac{1}{2\ell}\bar{\psi}^{-}\gamma^{0}\psi^{-}+\frac{1}{\ell}\bar{\psi}^{+}\gamma^{0}\rho\,,\notag\\
    F\left(m\right)&=&dm+\epsilon^{ac}\omega_{a}e_{c}+\frac{1}{2}\bar{\psi}^{-}\gamma^{0}\psi^{-}+\bar{\psi}^{+}\gamma^{0}\rho\,,\notag\\
    F\left(t_{1}\right)&=&dt_{1}+\frac{1}{2\ell}\bar{\psi}^{+}\gamma^{0}\psi^{+}\,, \notag \\
    F\left(t_{2}\right)&=&dt_{2}-\frac{1}{2\ell}\bar{\psi}^{-}\gamma^{0}\psi^{-}+\frac{1}{\ell}\bar{\psi}^{+}\gamma^{0}\rho\,, \notag \\
    F\left(u_{1}\right)&=&du_{1}+\frac{1}{2}\bar{\psi}^{+}\gamma^{0}\psi^{+}\,,\notag \\
    F\left(u_{2}\right)&=&du_{2}-\frac{1}{2}\bar{\psi}^{-}\gamma^{0}\psi^{-}+\bar{\psi}^{+}\gamma^{0}\rho\,. \label{2fENH}
\end{eqnarray}
On the other hand, the fermionic curvature two-forms are defined in \eqref{2fNH}. 

\subsubsection*{Generalized $\mathcal{N}=2$ Newton-Hooke and Galilean CS supergravity theories}

A CS supergravity action based on the $\mathcal{N}=2$ $\mathfrak{nh}^{\left(N\right)}$ superalgebra \eqref{SGNH}-\eqref{SGNH2} can be constructed from the gauge connection one-form,
\begin{align}
    A&=\sum_{m=0}^{N}\left(\omega^{\left(m\right)}\texttt{J}^{\left(m\right)}+\tau^{\left(m\right)} \texttt{H}^{\left(m\right)}+t^{\left(m\right)}\texttt{T}^{\left(m\right)}+u^{\left(m\right)}\texttt{U}^{\left(m\right)}+\bar{\psi}^{+ \,\left(m\right)}\texttt{Q}^{+ \,\left(m\right)}\right)\notag\\
    &+\sum_{m=0}^{N-1}\left(\omega^{a \,\left(m\right)}\texttt{G}_{a}^{\left(m\right)}+ e^{a \,\left(m\right)}\texttt{P}_{a}^{\left(m\right)}+\bar{\psi}^{- \,\left(m\right)}\texttt{Q}^{- \,\left(m\right)}\right)\,,\label{1fANHN}
\end{align}
and the non-vanishing components of the invariant tensor,
\begin{align}
    \langle \texttt{J}^{\left(m\right)} \texttt{H}^{\left(n\right)} \rangle &=-\beta_{m+n}\,, & \langle \texttt{G}_{a}^{\left(m\right)} \texttt{P}_{b}^{\left(n\right)} \rangle&=\beta_{m+n+1}\delta_{ab}\,,\notag \\
    \langle \texttt{T}^{\left(m\right)} \texttt{U}^{\left(n\right)} \rangle&=\beta_{m+n}\,,& \langle \texttt{U}^{\left(m\right)} \texttt{U}^{\left(n\right)} \rangle&=-\frac{1}{\ell}\beta_{m+n}\,,\notag\\
    \langle \texttt{Q}_{\alpha}^{- \,\left(m\right)}\texttt{Q}_{\beta}^{- \,\left(n\right)} \rangle &=2\beta_{m+n+1} C_{\alpha\beta}\,, &
    \langle \texttt{Q}_{\alpha}^{+ \,\left(m\right)}\texttt{Q}_{\beta}^{+ \,\left(n\right)} \rangle &=2\beta_{n+m} C_{\alpha\beta}\,, \label{IT5}
\end{align}
with $1\leq p\leq N$ for the $\beta_p$ constants. Here, the $\beta_p$ constants are related to the relativistic ones \eqref{IT} through the $S_{E}^{\left(2N\right)}$ semigroup elements as
\begin{align}
    \beta_p&=\lambda_{2p}\tilde{\alpha}_{1}\,.
\end{align}
Without loss of generality, we can set $\beta_0=0$, $\beta_0$ being related to the three-dimensional Galilean gravity term \cite{Bergshoeff:2017btm,Gomis:2019nih}.

A generalized NR CS supergravity action constructed from the gauge connection one-form \eqref{1fANHN} is then given by
\begin{eqnarray}
    I_{\mathfrak{nh}^{(N)}}^{\mathcal{N}=2}&=&\frac{k}{4\pi}\int \sum_{p=1}^{N} \beta_{p} \,\delta_{m+n}^{p}\,\left[-2\tau^{\left(m\right)} R\left(\omega^{\left(n\right)}\right)+\frac{\delta_{q+r+1}^{n}}{\ell^2}\epsilon^{ac}\tau^{\left(m\right)} e_{a}^{\left(q\right)}e_{c}^{\left(r\right)}+2t^{\left(m\right)}du^{\left(n\right)}-\frac{1}{\ell}u^{\left(m\right)}du^{\left(n\right)}\right.\notag\\
    &&\left.-2\bar{\psi}^{+ \,\left(m\right)}\nabla\psi^{+ \,\left(n\right)}\right]+\beta_p\,\delta_{m+n+1}^{p}\left[2e_a^{\left(m\right)}R^{a}\left(\omega^{b \,\left(n\right)}\right)-2\bar{\psi}^{- \,\left(m\right)}\nabla\psi^{- \,\left(n\right)}\right]\,,\label{CSNHN}
\end{eqnarray}
with
\begin{eqnarray}
    R\left(\omega^{\left(m\right)}\right)&=&d\omega^{\left(m\right)}+\frac{1}{2}\sum_{p,q=0}^{N-1}\delta^{m}_{p+q+1}\epsilon^{ac}\omega_{a}^{\left(p\right)}\omega_{c}^{\left(q\right)}\,,\notag\\
    R^{a}\left(\omega^{a \,\left(m\right)}\right)&=&d\omega^{a \,\left(m\right)}+\sum_{n,p=0}^{N}\delta_{n+p}^{m}\epsilon^{ac}\omega^{\left(n\right)}\omega_c^{\left(p\right)}\,,\notag\\
    \nabla\psi^{+\,\left(m\right)}&=&d\psi^{+ \,\left(m\right)}+\frac{1}{2}\sum_{n,p=0}^{N}\delta^{m}_{n+p}\left(\omega^{\left(n\right)}\gamma_0\psi^{+ \,\left(p\right)}+\frac{1}{\ell}\tau^{\left(n\right)}\gamma_0\psi^{+ \,\left(p\right)}- t^{\left(n\right)}\gamma_{0}\psi^{+ \,\left(p\right)}\right) \notag\\
    &&+\sum_{n,p=0}^{N-1}\delta^{m}_{n+p+1}\left(\omega^{a \,\left(n\right)}\gamma_{a}\psi^{- \,\left(p\right)}+\frac{1}{\ell}e^{a \,\left(n\right)}\gamma_{a}\psi^{- \,\left(p\right)}\right) \,,\notag\\
    \nabla\psi^{-\,\left(m\right)}&=&d\psi^{- \,\left(m\right)}+\frac{1}{2}\sum_{n,p=0}^{N-1}\delta^{m}_{n+p}\left(\omega^{\left(n\right)}\gamma_0\psi^{- \,\left(p\right)}+\frac{1}{\ell}\tau^{\left(n\right)}\gamma_0\psi^{- \,\left(p\right)}+ t^{\left(n\right)}\gamma_{0}\psi^{- \,\left(p\right)}\right. \notag\\
    &&+\left.\omega^{a \,\left(n\right)}\gamma_{a}\psi^{+ \,\left(p\right)}+\frac{1}{\ell}e^{a \,\left(n\right)}\gamma_{a}\psi^{+ \,\left(p\right)}\right) \,.\label{2fsNH} 
\end{eqnarray}
The obtained NR CS supergravity action is based on the $\mathcal{N}=2$ $\mathfrak{nh}^{\left(N\right)}$ superalgebra and can be seen as the supersymmetric extension of the generalized Newton-Hooke gravity action introduced in \cite{Gomis:2019nih}. In the vanishing cosmological constant limit $\ell\rightarrow\infty$ the CS action reproduces a generalized Galilean supergravity. As in the generalized super AdS-Carroll case, the CS supergravity action can be written as the sum of Lagrangians invariant under the $\mathcal{N}=2$ $\mathfrak{nh}^{\left(p\right)}$ superalgebra for $p=1,2,\cdots,N$. Indeed, the CS action \eqref{CSNHN} can be expressed as follows:
\begin{eqnarray}
    I_{\mathfrak{nh}^{\left(N\right)}}^{\mathcal{N}=2}&=&\frac{k}{4\pi}\int \sum_{p=1}^{N} \beta_{p} \mathcal{L}_{\mathfrak{nh^{\left(p\right)}}}^{\mathcal{N}=2}\,.
\end{eqnarray}
For each Lagrangian we have set $\beta_{i}=0$ for $i\neq p$ since the non-degeneracy condition for $\mathfrak{nh}^{\left(p\right)}$ is given by $\beta_p\neq 0$. Then, for an arbitrary $p$, the equations of motion for the $\mathfrak{nh}^{\left(p\right)}$ supergravity are given by the vanishing of the following curvature two-forms (field strengths of the bosonic one-form gauge potentials):
\begin{eqnarray}
    F\left(\omega^{\left(m\right)}\right)&=&R\left(\omega^{\left(m\right)}\right)+\frac{1}{2}\sum_{n,p=0}^{N-1}\delta^{m}_{n+p+1}\,\left(\frac{1}{\ell^{2}}\epsilon^{ac}\,e_{a}^{\left(n\right)}e_{c}^{\left(p\right)}+\frac{1}{\ell}\bar{\psi}^{- \,\left(n\right)}\gamma^{0}\psi^{- \,\left(p\right)}\right)\notag\\
    &&+\frac{1}{2\ell}\sum_{n,p=0}^{N}\delta^{m}_{n+p}\bar{\psi}^{+ \,\left(n\right)}\gamma^{0}\psi^{+ \,\left(p\right)}\,, \notag\\
    F^{a}\left(\omega^{b \,\left(m\right)}\right)&=&R\left(\omega^{a \,\left(m\right)}\right)+\sum_{n,p=0}^{N-1}\delta^{m}_{n+p}\left(\frac{1}{\ell^{2}}\epsilon^{ac}\tau^{\left(n\right)} e_{c}^{\left(p\right)}+\frac{1}{\ell}\bar{\psi}^{+ \,\left(n\right)}\gamma^{a}\psi^{- \,\left(p\right)}\right)\,, \notag \\
    F\left(\tau^{\left(m\right)}\right)&=&d\tau^{\left(m\right)}+\sum_{n,p=0}^{N-1}\delta^{m}_{n+p+1}\left(\epsilon^{ac}\omega_a^{\left(n\right)} e_{c}^{\left(p\right)}+\frac{1}{2}\bar{\psi}^{- \,\left(n\right)}\gamma^{0}\psi^{- \,\left(p\right)}\right) \notag \\
    &&+\frac{1}{2}\sum_{n,p=0}^{N}\delta^{m}_{n+p}\,\bar{\psi}^{+ \,\left(n\right)}\gamma^{0}\psi^{+ \,\left(p\right)}\,, \notag\\
    F^{a}\left(e^{b \,\left(m\right)}\right)&=&de^{a \,\left(m\right)}+\sum_{n,p=0}^{N-1}\delta^{m}_{n+p}\,\left(\epsilon^{ac}\omega^{\left(n\right)} e_{c}^{\left(p\right)}+\epsilon^{ac}\,\tau^{\left(n\right)}\omega_{c}^{\left(p\right)}+\bar{\psi}^{+ \,\left(n\right)}\gamma^{a}\psi^{- \,\left(p\right)}\right)\,, \notag \\
    F\left(t^{\left(m\right)}\right)&=&dt^{\left(m\right)}+\frac{1}{2\ell}\sum_{n,p=0}^{N}\delta^{m}_{n+p}\,\bar{\psi}^{+ \,\left(n\right)}\gamma^{0}\psi^{+ \,\left(p\right)}+\frac{1}{2\ell}\sum_{n,p=0}^{N-1}\delta^{m}_{n+p+1}\bar{\psi}^{- \,\left(n\right)}\gamma^{0}\psi^{- \,\left(p\right)}\,, \notag\\
    F\left(u^{\left(m\right)}\right)&=&du^{\left(m\right)}+\frac{1}{2}\sum_{n,p=0}^{N}\delta^{m}_{n+p}\,\bar{\psi}^{+ \,\left(n\right)}\gamma^{0}\psi^{+ \,\left(p\right)}+\frac{1}{2}\sum_{n,p=0}^{N-1}\delta^{m}_{n+p+1}\bar{\psi}^{- \,\left(n\right)}\gamma^{0}\psi^{- \,\left(p\right)}  \,;
\end{eqnarray}
the fermionic curvature two-forms are defined in \eqref{2fsNH}. One can notice that the Lagrangian for $p=1$ corresponds to the extended Newton-Hooke supergravity Lagrangian obtained in \eqref{CSENH}. For $p\geq 2$, the Lagrangian reproduces post-Newtonian extensions of the extended Newton-Hooke supergravity. For instance, $p=2$ reproduces the so-called exotic Newtonian supergravity \cite{Concha:2021jos}. Indeed, by considering the following redefinition of the gauge fields:
\begin{align}
    \omega^{\left(0\right)}&=\omega\,, &  \omega^{\left(1\right)}&=s\,, & \omega^{\left(2\right)}&=z\,,   \notag \\
    \tau^{\left(0\right)}&=\tau\,, & \tau^{\left(1\right)}&=m\,, & \tau^{\left(2\right)}&=t\,,   \notag \\
     t^{\left(0\right)}&=t_{1}\,, & t^{\left(1\right)}&=t_{2}\,,& t^{\left(2\right)}&=t_{3}\,,    \notag \\
    u^{\left(0\right)}&=u_{1}\,, & u^{\left(1\right)}&=u_{2}\,, & u^{\left(2\right)}&=u_{3}\,,     \notag \\
    \psi_{\alpha}^{+ \, \left(0\right)}&=\psi_{\alpha}^{+}\,, &  \psi_{\alpha}^{+ \, \left(1\right)}&=\rho_{\alpha}\,, & \psi_{\alpha}^{+ \, \left(2\right)}&=\phi_{\alpha}^{+}\,,  \notag \\
    \omega_{a}^{\left(0\right)}&=\omega_{a}\,, &  e_{a}^{\left(0\right)}&=e_{a}\,, & \psi_{\alpha}^{- \, \left(0\right)}&=\psi_{\alpha}^{-}\,, \notag \\
     \omega_{a}^{\left(1\right)}&=b_{a}\,, & e_{a}^{\left(1\right)}&=t_{a}\,, & \psi_{\alpha}^{- \, \left(1\right)}&=\phi_{\alpha}^{-}\,, 
\end{align}
we get
\begin{align}
\mathcal{L}_{\mathfrak{nh^{\left(2\right)}}}^{\mathcal{N}=2}&=2e^{a}R^{a}\left(b^{b}\right)+2t^{a}R^{a}\left(\omega^{b}\right)-2\tau R\left(z\right)-2mR\left(s\right)-2yR\left(\omega\right)+\frac{2}{\ell^{2}}\epsilon^{ac}\tau e_{a} t_{c}+\frac{1}{\ell^{2}}\epsilon^{ac} m e_{a} e_{c}\notag\\
    &+2t_{1}du_{3}+2t_{2}du_{2}+2t_{3}du_{1}-\frac{2}{\ell}u_{1}du_{3}-\frac{1}{\ell}u_{2}du_{2}-2\bar\psi^{\pm}\nabla\phi^{\pm}-2\bar\phi^{\pm}\nabla\psi^{\pm}-2\bar\rho\nabla\rho\,.
\end{align}
Here, the curvature two-forms read
\begin{eqnarray}
    R^{a}\left(b^{b}\right)&=&db^{a}+\epsilon^{ac}\omega b_{c}+\epsilon^{ac}s \omega_{c}\,,\notag\\
    R\left(z\right)&=&dz+\epsilon^{ac}\omega_{a}b_{c}\,,\notag\\
    \nabla\phi^{+}&=&d\phi^{+}+\frac{1}{2}\omega\gamma_0\phi^{+}+\frac{1}{2}s\gamma_0\rho+\frac{1}{2}\omega^{a}\gamma_{a}\phi^{-}+\frac{1}{2}b^{a}\gamma_{a}\psi^{-}+\frac{1}{2\ell}e^{a}\gamma_a\phi^{-}+\frac{1}{2\ell}\tau\gamma_0\phi^{+}\notag\\
    &&+\frac{1}{2\ell}m\gamma_0\rho+\frac{1}{2\ell}t^{a}\gamma_a\psi^{-}+\frac{1}{2}z\gamma_{0}\psi^{+}+\frac{1}{2\ell}y\gamma_{0}\psi^{+}-\frac{1}{2}t_{1}\gamma_0\phi^{+}-\frac{1}{2}t_{2}\gamma_0\rho-\frac{1}{2}t_{3}\gamma_{0}\psi^{+}\,,\notag\\
    \nabla\phi^{-}&=&d\phi^{-}\frac{1}{2}\omega\gamma_0\phi^{-}+\frac{1}{2}s\gamma_0\psi^{-}+\frac{1}{2}\omega^{a}\gamma_{a}\rho+\frac{1}{2}b^{a}\gamma_{a}\psi^{+}+\frac{1}{2\ell}e^{a}\gamma_a\rho+\frac{1}{2\ell}\tau\gamma_0\phi^{-}\notag\\
    &&+\frac{1}{2\ell}m\gamma_0\psi^{-}+\frac{1}{2\ell}t^{a}\gamma_a\psi^{+}-\frac{1}{2}t_{1}\gamma_0\phi^{-}-\frac{1}{2}t_{2}\gamma_0\psi^{-}\,.
\end{eqnarray}
Moreover, $R\left(\omega\right)$, $R^{a}\left(\omega^{b}\right)$, $R\left(s\right)$, $\nabla\psi^{\pm}$, and $\nabla\rho$ have been defined in \eqref{2fNH}. In the flat limit $\ell\rightarrow\infty$, we recover the extended Newtonian supergravity introduced in \cite{Ozdemir:2019orp}, which can also be understood as post-Newtonian corrections \cite{Gomis:2019fdh}. At higher order in $p$, the Lagrangians $\mathcal{L}_{\mathfrak{nh^{\left(2\right)}}}^{\mathcal{N}=2}$ extends to the supersymmetric case the family of Newton-Hooke gravity theories discussed in \cite{Gomis:2019nih}.

Interestingly, the $\mathcal{N}=2$ generalized Newton-Hooke CS supergravity action \eqref{CSNHN} can alternatively be obtained from the relativistic one,
\begin{eqnarray}
I_{\text{AdS}}^{\mathcal{N}=2}=\int \tilde{\alpha}_1\left[2e_AR^{A}+\frac{1}{3\ell^2}\epsilon_{ABC}e^{A}e^{B}e^{C}+\mathtt{t}d\mathtt{u}-\frac{1}{\ell^2}\mathtt{u}d\mathtt{u}-2\bar{\Psi}^{i}\nabla\Psi^{i}\right]\,,\label{AdSCS}
\end{eqnarray}
where
\begin{eqnarray}
R^{A}&=&d\omega^{A}+\frac{1}{2}\epsilon^{ABC}\omega_B\omega_C\,,\notag \\
T^{A}&=&de^{A}+\frac{1}{2}\epsilon^{ABC}\omega_{B}e_{C}\,,\notag \\
\nabla\Psi^{i}&=&d\Psi^{i}+\frac{1}{2}\omega^{A}\gamma_A\Psi^{i}+\frac{1}{2\ell}e^{A}\gamma_{A}\Psi^{i}+\mathtt{t}\epsilon^{ij}\Psi^{j}\,.\label{curv3}
\end{eqnarray}
Indeed, the NR supergravity action \eqref{CSENH} is obtained by expanding the relativistic theory \eqref{AdSCS} after identifying the NR gauge fields in terms of the relativistic ones through the $S_{E}^{\left(2\right)}$ elements as
\begin{align}
  \omega^{\left(m\right)}&=\lambda_{2m} \omega_{0}\,, & \tau^{\left(m\right)}&=\lambda_{2m} e_{0} & \psi^{+ \,\left(m\right)}&=\lambda_{2m} \Psi^{+}\,, & t^{\left(m\right)}&=\lambda_{2m} \texttt{t}\,, & u^{\left( m\right)}&=\lambda_{2m} \texttt{u}\,,  \notag\\
  \omega_{a}^{\left(m\right)}&=\lambda_{2m+1}\omega_{a} & e_{a}^{\left(m\right)}&=\lambda_{2m+1} e_{a} & \psi^{- \,\left(m\right)}&=\lambda_{2m+1}\Psi^{-}\,,
\end{align}
where we have defined
\begin{align}
    \Psi^{\pm}_{\alpha}&=\frac{1}{\sqrt{2}}\left(\Psi_{\alpha}^{1}\pm\left(\gamma^{0}\right)_{\alpha\beta}\Psi_{\beta}^{2}\right)\,.
\end{align}

\subsection{$\mathcal{N}=2$ (AdS-)static Chern-Simons supergravity and its generalization}

A third class of kinematical superalgebras are the static one which, as we have shown in the previous section, can be either be obtained from the UR superalgebras or from the NR ones. Here, we present the three-dimensional CS supergravity action based on the $\mathcal{N}=2$ extended AdS-static superalgebra \eqref{SADSS}-\eqref{SADSS2} and its flat limit.

The gauge connection one-form for the $\mathcal{N}=2$ AdS-static superalgebra is given by
\begin{eqnarray}
    A&=&\omega \texttt{J}+\omega^{a}\texttt{G}_{a}+s \texttt{S}+\tau \texttt{H}+e^{a}\texttt{P}_{a}+m\, \texttt{M}+t_1\texttt{T}_{1}+t_{2}\texttt{T}_{2}+u_{1}\texttt{U}_{1}+u_{2}\texttt{U}_{2}\notag\\
    &&+\bar{\psi}^{+}\texttt{Q}^{+}+\bar{\psi}^{-}\texttt{Q}^{-}+\bar{\rho}\,\texttt{R}\,. \label{1fAdSS}
\end{eqnarray}
The corresponding curvature two-form reads
\begin{align}
F&=\mathcal{F}\left(\omega\right)\texttt{J}+\mathcal{F}^{a}\left(\omega^{b}\right)\texttt{G}_{a}+\mathcal{F}\left(\tau\right)\texttt{H}+\mathcal{F}^{a}\left(e^{b}\right)\texttt{P}_{a}+\mathcal{F}\left(t_{1}\right)\texttt{T}_{1}+\mathcal{F}\left(t_{2}\right)\texttt{T}_{2}\notag\\
&+\mathcal{F}\left(u_{1}\right)\texttt{U}_{1}+\mathcal{F}\left(u_{2}\right)\texttt{U}_{2}+\nabla\bar{\psi}^{+}\texttt{Q}^{+}+\nabla\bar{\psi}^{-}\texttt{Q}^{-}+\nabla\bar{\rho}\,\texttt{R}\,, \label{2fADSS}
\end{align}
with
\begin{eqnarray}
    \mathcal{F}\left(\omega\right)&=&d\omega=\mathcal{R}\left(\omega\right)\,,\notag\\
    \mathcal{F}^{a}\left(\omega^{b}\right)&=&d\omega^{a}+\epsilon^{ac}\omega\omega_{c}+\frac{1}{\ell^{2}}\epsilon^{ac}\tau e_{c}+\frac{1}{\ell}\bar{\psi}^{+}\gamma^{a}\psi^{-}=\mathcal{R}^{a}\left(\omega^{b}\right)+\frac{1}{\ell^{2}}\epsilon^{ac}\tau e_{c}+\frac{1}{\ell}\bar{\psi}^{+}\gamma^{a}\psi^{-}\,,\notag\\
    \mathcal{F}\left(\tau\right)&=&d\tau+\frac{1}{2}\bar{\psi}^{+}\gamma^{0}\psi^{+}\,,\notag\\
    \mathcal{F}\left(e^{a}\right)&=&de^{a}+\epsilon^{ac}\omega e_{c}\,,\notag\\
    \mathcal{F}\left(s\right)&=&ds+\frac{1}{2\ell^{2}}\epsilon^{ac}e_{a}e_{c}=\mathcal{R}\left(s\right)+\frac{1}{2\ell^{2}}\epsilon^{ac}e_{a}e_{c}\,,\notag\\
    \mathcal{F}\left(m\right)&=&dm+\epsilon^{ac}\omega_{a}e_{c}+\frac{1}{2}\bar{\psi}^{-}\gamma^{0}\psi^{-}\,,\notag\\
    \mathcal{F}\left(t_{1}\right)&=&dt_{1}\,,\notag\\
    \mathcal{F}\left(t_{2}\right)&=&dt_{2}\,,\notag\\
    \mathcal{F}\left(u_{1}\right)&=&du_{1}+\frac{1}{2}\bar{\psi}^{+}\gamma^{0}\psi^{+}\,,\notag\\
    \mathcal{F}\left(u_{2}\right)&=&du_{2}+\frac{1}{2}\bar{\psi}^{-}\gamma^{0}\psi^{-}+\bar{\psi}^{+}\gamma^{0}\rho\,,\notag\\
    \nabla\psi^{+}&=&d\psi^{+}+\frac{1}{2}\omega\gamma_0\psi^{+}-\frac{1}{2}t_{1}\gamma_0\psi^{+}\,,\notag\\
    \nabla\psi^{-}&=&d\psi^{-}+\frac{1}{2}\omega\gamma_0\psi^{-}+\frac{1}{2\ell}e^{a}\gamma_a\psi^{+}+\frac{1}{2}t_{1}\gamma_0\psi^{-}\,,\notag\\
    \nabla\rho&=&d\rho+\frac{1}{2}\omega\gamma_0\rho+\frac{1}{2}s\gamma_0\psi^{+}+\frac{1}{2\ell}e^{a}\gamma_a\psi^{-}-\frac{1}{2}t_{1}\gamma_0\rho-\frac{1}{2}t_{2}\gamma_0\psi^{+}\,.
    \label{2fFADSS}
\end{eqnarray}
Let us note that the curvature two-forms above are quite different from the extended Newton-Hooke and extended Bargmann ones, although the gauge connection one-form \eqref{1fAdSS} is identical to the extended Newton-Hooke one \eqref{1fANH}. The non-vanishing components of the invariant tensor for the $\mathcal{N}=2$ AdS-static superalgebra are given by
\begin{align}
    \langle \texttt{J} \texttt{S} \rangle &=-\nu_0\,, & \langle \texttt{P}_{a} \texttt{P}_{b} \rangle&=\frac{\nu_0}{\ell^{2}}\delta_{ab}\,,\notag \\
    \langle \texttt{J} \texttt{M} \rangle &=  \langle \texttt{H} \texttt{S} \rangle =-\mu_1\,,  & \langle \texttt{G}_{a} \texttt{P}_{b} \rangle&=\mu_1\delta_{ab}\,,\notag \\
    \langle \texttt{T}_{1} \texttt{T}_{2} \rangle &=\nu_0\,, &  \langle \texttt{T}_{1} \texttt{U}_{2} \rangle&= \langle \texttt{T}_{2} \texttt{U}_{1} \rangle=\mu_1\,,\notag \\
    \langle \texttt{Q}_{\alpha}^{-}\texttt{Q}_{\beta}^{-} \rangle &=2\mu_1 C_{\alpha\beta}\,,  & \langle \texttt{Q}_{\alpha}^{+}\texttt{R}_{\beta} \rangle&=2\mu_1 C_{\alpha\beta}\,, \label{IT6}
\end{align}
where $\nu_0$ and $\mu_1$ can be written in terms of the $\mathcal{N}=2$ AdS-Carroll constants \eqref{IT2} and the element of the $S_{E}^{\left(2\right)}$ semigroup as
\begin{align}
    \nu_0&=\lambda_{2}\alpha_{0}\,, & \mu_1&=\lambda_2\alpha_{1}\,.\notag
\end{align}
Alternatively, $\nu_0$ and $\mu_1$ can also be recovered from the extended Newton-Hooke ones \eqref{IT4} with
\begin{align}
    \nu_0&=\lambda_{0}\sigma_{0}\,, & \mu_1&=\lambda_2\beta_{1}\,.\notag
\end{align}
The exotic sector proportional to $\nu_0$ can be set to zero without affecting the non-degeneracy criterion, which requires $\mu_1\neq 0$. The $\mathcal{N}=2$ AdS-static CS supergravity action constructed from the connection one-form \eqref{1fAdSS} and the invariant tensor \eqref{IT6} is then given by
\begin{eqnarray}
    I_{\mathfrak{ads}\text{-}\mathfrak{stat}}^{\mathcal{N}=2}&=&\frac{k}{4\pi}\int \mu_1 \left[2e_a\mathcal{R}^{a}\left(\omega^{b}\right)-2 m \mathcal{R}\left(\omega\right)-2 \tau \mathcal{R}\left(s\right)+\frac{1}{\ell^2}\epsilon^{ac}\tau e_{a}e_{c}+2t_{1}du_{2}+2t_{2}du_{1}\right.\notag\\
    &&\left.-2\bar{\psi}^{+}\nabla\rho-2\bar{\rho}\nabla\psi^{+}-2\bar{\psi}^{-}\nabla\psi^{-}\right]\,.\label{CSADSS}
\end{eqnarray}
The obtained CS action differs from the $\mathcal{N}=2$ extended Bargmann one (see \eqref{CSENH} for $\ell\rightarrow\infty$) in the definition of the curvature two-forms \eqref{2fFADSS}. The non-degeneracy of the bilinear invariant tensor \eqref{IT6} ensures the vanishing of the AdS-static curvature two-forms \eqref{2fFADSS} as field equations which are different from the extended Bargmann ones \eqref{2fENH}. Nonetheless, a reinterpretation of the $e^{a}$ and $\omega^{a}$ gauge fields allows us to recover the known non-relativistic supergravity action \cite{Bergshoeff:2016lwr, Ozdemir:2019tby, Concha:2020eam}. Indeed, the $\mathcal{N}=2$ extended Bargmann CS supergravity action appears by interchanging the roles of $e^{a}$ and $\ell\omega^{a}$. On the other hand, the vanishing cosmological constant limit $\ell\rightarrow\infty$  reproduces an $\mathcal{N}=2$ static CS supergravity action. The flat limit is free of degeneracy due to the presence of the bosonic generators $\{\texttt{T}_{1},\texttt{T}_2,\texttt{U}_{1},\texttt{U}_{2}\}$, ensuring the vanishing of the proper curvature two-forms as field equations both in the AdS-static and static supergravity. The CS supergravity action \eqref{CSADSS} is invariant under the following supersymmetry transformation rules:
\begin{align}
    \delta \omega&=0\,, \notag\\
    \delta \omega^{a}&=\frac{1}{\ell}\bar{\varepsilon}^{+}\gamma^{a}\psi^{-}+\frac{1}{\ell}\bar{\varepsilon}^{-}\gamma^{a}\psi^{+}\,,\notag\\
    \delta \tau&=\bar{\varepsilon}^{+}\gamma^{0}\psi^{+}\,,\notag\\
    \delta e^{a}&=0\,, \notag \\
    \delta s&=0\,, \notag\\
    \delta m&=\bar{\varepsilon}^{-}\gamma^{0}\psi^{-}+\bar{\varepsilon}^{+}\gamma^{0}\rho+\bar{\varrho}\gamma^{0}\psi^{+}\,,\notag\\
    \delta t_{1}&=0\,, \notag \\
    \delta t_{2}&=0\,, \notag \\
    \delta u_{1}&=\bar{\varepsilon}^{+}\gamma^{0}\psi^{+}\,, \notag \\
    \delta u_{2}&=\bar{\varepsilon}^{+}\gamma^{0}\rho+\bar{\varrho}^{+}\gamma^{0}\psi^{+}-\bar{\varepsilon}^{-}\gamma^{0}\psi^{-}\,, \notag \\
    \delta \psi^{+}&=d\varepsilon^{+}+\frac{1}{2}\omega\gamma_{0}\varepsilon^{+}+\frac{1}{2\ell}e^{a}\gamma_{a}\varepsilon^{-}-\frac{1}{2}t_{1}\gamma_{0}\varepsilon^{+}\,,\notag \\
    \delta \psi^{-}&=d\varepsilon^{-}+\frac{1}{2}\omega\gamma_{0}\varepsilon^{-}+\frac{1}{2\ell}e^{a}\gamma_{a}\varepsilon^{+}+\frac{1}{2}t_{1}\gamma_{0}\varepsilon^{-}\,,\notag\\
    \delta \rho&=d\varrho+\frac{1}{2}\omega\gamma_{0}\varrho+\frac{1}{2}s\gamma_{0}\varepsilon^{+}+\frac{1}{2\ell}e^{a}\gamma_{a}\varepsilon^{-}-\frac{1}{2}t_{1}\gamma_{0}\varrho-\frac{1}{2}t_{2}\gamma_{0}\varepsilon^{+}\,,
\end{align}
where the gauge parameters $\varepsilon^{\pm}$ and $\varrho$  are related to the fermionic charges $Q^{\pm}$ and $\rho$, respectively. The transformation rules for the $\mathcal{N}=2$ static supergravity theory are derived in the flat limit $\ell\rightarrow\infty$.

\subsubsection*{Generalized $\mathcal{N}=2$ (AdS-)static supergravity}

The $\mathcal{N}=2$ $\mathfrak{ads}$-$\mathfrak{stat}^{\left(N\right)}$ superalgebra \eqref{SADSSN} admits a non-degenerate invariant tensor whose non-vanishing components read
\begin{align}
    \langle \texttt{J}^{\left(m\right)} \texttt{H}^{\left(n\right)} \rangle &=-\mu_{m+n}\,, & \langle \texttt{G}_{a}^{\left(m\right)} \texttt{P}_{b}^{\left(n\right)} \rangle&=\mu_{m+n+1}\delta_{ab}\,,\notag \\
    \langle \texttt{T}^{\left(m\right)} \texttt{U}^{\left(n\right)} \rangle&=\mu_{m+n}\,,& 
    \langle \texttt{Q}_{\alpha}^{- \,\left(m\right)}\texttt{Q}_{\beta}^{- \,\left(n\right)} \rangle &=2\mu_{m+n+1} C_{\alpha\beta}\,, \notag\\
    \langle \texttt{Q}_{\alpha}^{+ \,\left(m\right)}\texttt{Q}_{\beta}^{+ \,\left(n\right)} \rangle &=2\mu_{n+m} C_{\alpha\beta}\,, \label{IT7}
\end{align}
with $1\leq p\leq N$ for the $\mu_p$ constants.\footnote{Here we have intentionally omitted the exotic sector.} Here, the $\mu_p$ constants can be either be obtained from the $\mathcal{N}=2$ AdS-Carroll ones \eqref{IT2},
\begin{align}
    \mu_p&=\lambda_{2p}\alpha_1\,, 
\end{align}
or from the $\mathcal{N}=2$ $\mathfrak{nh}^{\left(N\right)}$ constants \eqref{IT5},
\begin{align}
    \mu_p&=\lambda_{2}\beta_p\,. 
\end{align}
A generalized $\mathcal{N}=2$ (AdS-)static CS supergravity action can be constructed from the invariant tensor \eqref{IT7} and the gauge connection one-form for the $\mathcal{N}=2$ $\mathfrak{ads}$-$\mathfrak{stat}^{\left(N\right)}$ superalgebra,
\begin{align}
    A&=\sum_{m=0}^{N}\left(\omega^{\left(m\right)}\texttt{J}^{\left(m\right)}+\tau^{\left(m\right)} \texttt{H}^{\left(m\right)}+t^{\left(m\right)}\texttt{T}^{\left(m\right)}+u^{\left(m\right)}\texttt{U}^{\left(m\right)}+\bar{\psi}^{+ \,\left(m\right)}\texttt{Q}^{+ \,\left(m\right)}\right)\notag\\
    &+\sum_{m=0}^{N-1}\left(\omega^{a \,\left(m\right)}\texttt{G}_{a}^{\left(m\right)}+ e^{a \,\left(m\right)}\texttt{P}_{a}^{\left(m\right)}+\bar{\psi}^{- \,\left(m\right)}\texttt{Q}^{- \,\left(m\right)}\right)\,,\label{1fADSS}
\end{align}
and the corresponding curvature two-forms,
\begin{eqnarray}
    \mathcal{F}\left(\omega^{\left(m\right)}\right)&=&d\omega^{\left(m\right)}+\frac{1}{2\ell^{2}}\sum_{n,p=0}^{N-1}\delta^{m}_{n+p+1}\,\epsilon^{ac}\,e_{a}^{\left(n\right)}e_{c}^{\left(p\right)}\,,\notag\\
    \mathcal{F}^{a}\left(\omega^{b \,\left(m\right)}\right)&=&d\omega^{a \,\left(m\right)}+\sum_{n,p=0}^{N-1}\delta_{n+p}^{m}\left(\epsilon^{ac}\omega^{\left(n\right)}\omega_c^{\left(p\right)}+\frac{1}{\ell^{2}}\epsilon^{ac}\tau^{\left(n\right)} e_{c}^{\left(p\right)}+\frac{1}{\ell}\bar{\psi}^{+ \,\left(n\right)}\gamma^{a}\psi^{- \,\left(p\right)}\right)\,, \notag \\
    \mathcal{F}\left(\tau^{\left(m\right)}\right)&=&d\tau^{\left(m\right)}+\sum_{n,p=0}^{N-1}\delta^{m}_{n+p+1}\left(\epsilon^{ac}\omega_a^{\left(n\right)} e_{c}^{\left(p\right)}+\frac{1}{2}\bar{\psi}^{- \,\left(n\right)}\gamma^{0}\psi^{- \,\left(p\right)}\right) \notag \\
    &&+\frac{1}{2}\sum_{n,p=0}^{N}\delta^{m}_{n+p}\,\bar{\psi}^{+ \,\left(n\right)}\gamma^{0}\psi^{+ \,\left(p\right)}\,, \notag\\
    \mathcal{F}^{a}\left(e^{b \,\left(m\right)}\right)&=&de^{a \,\left(m\right)}+\sum_{n,p=0}^{N-1}\delta^{m}_{n+p}\,\epsilon^{ac}\omega^{\left(n\right)} e_{c}^{\left(p\right)}\,, \notag \\
    \mathcal{F}\left(t^{\left(m\right)}\right)&=&dt^{\left(m\right)}\,, \notag\\
    \mathcal{F}\left(u^{\left(m\right)}\right)&=&du^{\left(m\right)}+\frac{1}{2}\sum_{n,p=0}^{N}\delta^{m}_{n+p}\,\bar{\psi}^{+ \,\left(n\right)}\gamma^{0}\psi^{+ \,\left(p\right)}+\frac{1}{2}\sum_{n,p=0}^{N-1}\delta^{m}_{n+p+1}\bar{\psi}^{- \,\left(n\right)}\gamma^{0}\psi^{- \,\left(p\right)}  \,,\notag\\
     \nabla\psi^{+\,\left(m\right)}&=&d\psi^{+ \,\left(m\right)}+\frac{1}{2}\sum_{n,p=0}^{N}\delta^{m}_{n+p}\left(\omega^{\left(n\right)}\gamma_0\psi^{+ \,\left(p\right)}- t^{\left(n\right)}\gamma_{0}\psi^{+ \,\left(p\right)}\right) \notag\\
    &&+\frac{1}{\ell}\sum_{n,p=0}^{N-1}\delta^{m}_{n+p+1}e^{a \,\left(n\right)}\gamma_{a}\psi^{- \,\left(p\right)} \,,\notag\\
    \nabla\psi^{-\,\left(m\right)}&=&d\psi^{- \,\left(m\right)}+\frac{1}{2}\sum_{n,p=0}^{N-1}\delta^{m}_{n+p}\left(\omega^{\left(n\right)}\gamma_0\psi^{- \,\left(p\right)}+ t^{\left(n\right)}\gamma_{0}\psi^{- \,\left(p\right)}+\frac{1}{\ell}e^{a \,\left(n\right)}\gamma_{a}\psi^{+ \,\left(p\right)}\right) \,. \notag\\
\end{eqnarray}

Then, the general expression for the CS supergravity action based on the $\mathcal{N}=2$ $\mathfrak{ads}$-$\mathfrak{stat}^{\left(N\right)}$ superalgebra \eqref{SADSSN} is given by
\begin{align}
    I_{\mathfrak{ads}\text{-}\mathfrak{stat}^{(N)}}^{\mathcal{N}=2}&=\frac{k}{4\pi}\int \sum_{p=1}^{N}  \mu_p\,\delta_{m+n+1}^{p}\left[2e_a^{\left(m\right)}\mathcal{R}^{a}\left(\omega^{b \,\left(n\right)}\right)-2\bar{\psi}^{- \,\left(m\right)}\nabla\psi^{- \,\left(n\right)}\right]\notag\\
    &+\mu_{p} \,\delta_{m+n}^{p}\,\left[-2\tau^{\left(m\right)} \mathcal{R}\left(\omega^{\left(n\right)}\right)+\frac{\delta_{q+r+1}^{n}}{\ell^2}\epsilon^{ac}\tau^{\left(m\right)} e_{a}^{\left(q\right)}e_{c}^{\left(r\right)}+2t^{\left(m\right)}du^{\left(n\right)}-2\bar{\psi}^{+ \,\left(m\right)}\nabla\psi^{+ \,\left(n\right)}\right]\,,\notag\\ \label{CSADSSN}
\end{align}
where
\begin{align}
    \mathcal{R}\left(\omega^{\left(n\right)}\right)&=d\omega^{\left(n\right)}\,, \notag \\
    \mathcal{R}^{a}\left(\omega^{b \,\left(n\right)}\right)&=d\omega^{a \,\left(m\right)}+\sum_{n,p=0}^{N-1}\delta_{n+p}^{m}\, \epsilon^{ac}\omega^{\left(n\right)}\omega_c^{\left(p\right)}\,.
\end{align}
The $\mathcal{N}=2$ $\mathfrak{ads}$-$\mathfrak{stat}^{\left(N\right)}$ CS supergravity action \eqref{CSADSSN} can be expressed as the sum of individual Lagrangians $\mathcal{L}_{\mathfrak{ads}\text{-}\mathfrak{stat}^{\left(p\right)}}^{\mathcal{N}=2}$, each of which is invariant under the $\mathfrak{ads}$-$\mathfrak{stat}^{\left(p\right)}$ superalgebra,
\begin{eqnarray}
    I_{\mathfrak{ads}\text{-}\mathfrak{stat}^{\left(N\right)}}^{\mathcal{N}=2}&=&\frac{k}{4\pi}\int \sum_{p=1}^{N} \mu_{p} \,\mathcal{L}_{\mathfrak{ads}\text{-}\mathfrak{stat}^{\left(p\right)}}^{\mathcal{N}=2}\,.
\end{eqnarray}
For $p=1$, we recover the $\mathcal{N}=2$ AdS-static CS Lagrangian \eqref{CSADSS} along the $\mu_1$ constant. Generalized AdS-static supergravity actions are derived for $p\geq 2$.  According to Figures \ref{fig4} and \ref{fig5}, the corresponding $\mathcal{N}=2$ static CS supergravity theory together with its generalizations are obtained either as a vanishing cosmological constant limit $\ell\rightarrow\infty$ of the $\mathfrak{ads}$-$\mathfrak{stat}^{\left(N\right)}$ theory, or by expanding the $\mathcal{N}=2$ $\mathfrak{gal}^{\left(N\right)}$/$\mathfrak{car}^{\left(N\right)}$ supergravity models. In the expansion procedure, the $\mathcal{N}=2$ $\mathfrak{stat}^{\left(N\right)}$ CS supergravity action can be obtained from the generalized Galilean/Carrollian supergravity action. Indeed, the generalized static gauge fields can be expressed in terms of the $\mathfrak{gal}^{\left(N\right)}$ ones and the $S_{E}^{\left(2\right)}$ elements as
\begin{align}
  \omega^{\left(m\right)}&=\lambda_{0} \omega^{\left(m\right)}\,, & \tau^{\left(m\right)}&=\lambda_{2} \tau^{\left(m\right)} & \psi^{+ \,\left(m\right)}&=\lambda_{1} \psi^{+ \,\left(m\right)}\,, & t^{\left(m\right)}&=\lambda_{0} t^{\left(m\right)}\,, & u^{\left( m\right)}&=\lambda_{2} u^{\left( m\right)}\,,  \notag\\
  \omega_{a}^{\left(m\right)}&=\lambda_{2}\omega_{a}^{\left(m\right)} & e_{a}^{\left(m\right)}&=\lambda_{0} e_{a}^{\left(m\right)} & \psi^{- \,\left(m\right)}&=\lambda_{1}\psi^{-\,\left(m\right)}\,,
\end{align}
or in terms of the $\mathfrak{car}^{\left(N\right)}$ ones and the $S_{E}^{\left(2N\right)}$ elements,
\begin{align}
  \omega^{\left(m\right)}&=\lambda_{2m} \omega\,, & \tau^{\left(m\right)}&=\lambda_{2m} \tau & \psi^{+ \,\left(m\right)}&=\lambda_{2m} \psi^{+}\,, & t^{\left(m\right)}&=\lambda_{2m} t\,, & u^{\left( m\right)}&=\lambda_{2m} u\,,  \notag\\
  \omega_{a}^{\left(m\right)}&=\lambda_{2m+1}\omega_{a} & e_{a}^{\left(m\right)}&=\lambda_{2m+1} e_{a} & \psi^{- \,\left(m\right)}&=\lambda_{2m+1}\psi^{-}\,.
\end{align}

\section{Discussion}\label{sec5}

In this work, we have presented the supersymmetric extension of kinematical Lie algebras by applying systematic $S$-expansions starting from a $\mathfrak{so}\left(2\right)$-extension of the $\mathfrak{osp}\left(2,2\right)\times\mathfrak{sp}\left(2\right)$ superalgebra. We have classified both NR and UR regimes of the relativistic $\mathcal{N}=2$ AdS and Poincaré superalgebras constrained by the non-degeneracy condition. Such condition ensures that the field equations derived from the corresponding CS supergravity action are given by the vanishing of the curvatures for each supersymmetric (extended) kinematical Lie algebra. Our results can be summarized through the cube \ref{fig4}, in which both the relativistic and the NL superalgebras admit a non-degenerate invariant tensor. To this end, each NL regime requires particular considerations at the level of the semigroups and the subspaces of the original superalgebra. We then extend our results to post-Newtonian and post-Carrollian superalgebras, whose expansion relations have been summarized in cube \ref{fig5}. We have also considered the construction of CS supergravity theories for each (extended) kinematical superalgebra and their generalizations.

The general framework to derive NL regimes of known relativistic supergravity theories presented here could serve as a starting point for approaching various open issues. Of particular interest is the extension of our results to higher spacetime dimensions. In odd spacetime dimensions, the expansion method could be applied to distinct CS (super)gravity models defined in arbitrary higher spacetime dimensions \cite{Zanelli:2005sa}. Nonetheless, a geometric approach à la MacDowell-Mansouri \cite{MacDowell:1977jt} is required to explore the NL counterpart of (super)gravity theories defined in even spacetime dimensions. At the bosonic level, it would be interesting to explore the NL version of the Lovelock gravity \cite{Lovelock:1971yv,Lovelock:1972vz} which reproduces the CS and Born-Infeld (BI) gravity for particular values of the Lovelock parameters. One could explore whether, as in the relativistic case, the obtained NL CS and BI gravity theories can be recovered from a NL version of the Lovelock gravity model. Our results could be useful to explore the NL regime of the applications of higher-curvature terms in holography \cite{Brigante:2008gz,Camanho:2013pda}.
A first step toward holographic developments involves the potential extension to higher dimensions and, above all, the study of the resulting theory in the presence of a non-trivial spacetime boundary, following the approach of \cite{Andrianopoli:2014aqa,Concha:2018ywv,Banaudi:2018zmh,Eder:2021rgt,Andrianopoli:2021rdk}. This shall be followed by an investigation of the asymptotic symmetries and the theory emerging at the boundary. Further applications in the context of holography and analogue modeling include, for instance, the study of hydrodynamic systems -- e.g., the ``Carrollian fluids" \cite{Ciambelli:2018wre} -- within the framework of gauge-gravity duality applied within relativistic and non-relativistic physics.

Another aspect that deserves to be explored is the conformal extension of NL (super)algebras. The $S$-expansion method employed here could be useful to classify diverse NL (super)algebras and extend to conformal symmetries both the original cube of Bacry and Lévy-Leblond \cite{Bacry:1968zf} and its supersymmetric extension presented here. One could expect to obtain the Galilean conformal ($\mathfrak{gca}$) and Carrollian conformal ($\mathfrak{cca}$) (super)algebras \cite{Bagchi:2009my,Bagchi:2009pe,Bagchi:2009ke,Mandal:2010gx} by expanding the conformal one, after considering suitable partitions of the semigroups and decompositions of the relativistic subspaces. In analogy to the results obtained here, bigger semigroups should allow us to derive post-Newtonian and post-Carrollian extensions of the $\mathfrak{gca}$ and $\mathfrak{cca}$ (super)algebras, whose (anti-)commutators should satisfy a Schrödinger-like (super)algebra. Motivated by the $\mathfrak{bms}_{3}/\mathfrak{gca}_{2}$ (or $\mathfrak{bms}_{3}/\mathfrak{cca}_{2}$) duality \cite{Bagchi:2010zz,Duval:2014uva}, it would be interesting to explore the existence of a duality between two-dimensional extended NL conformal (super)algebras and infinite-dimensional asymptotic (super)algebra in higher spacetime dimensions.

It would be also interesting to extend our results to higher-spin gravity and bigger symmetries. The NL regime of gravity consistently coupled to spin-3 gauge fields has already been discussed in \cite{Bergshoeff:2016soe,Concha:2022muu,Caroca:2022byi}. One could go further and study the distinct NL versions of gravity coupled to spin-$5/2$ gauge fields, also refereed to as hypergravity \cite{Aragone:1983sz,Zinoviev:2014sza,Henneaux:2015tar,Henneaux:2015ywa,Fuentealba:2015jma,Fuentealba:2015wza}. One should expect to find hypersymmetric extensions of the Galilean and Carrollian algebras. Our method should allow us to select the ``good" hyperalgebras provided with non-degenerate bilinear invariant traces useful to construct CS hypergravity actions. In this direction, extended kinematical hyperalgebras should appear in the NR counterpart. On the other hand, one could extend the study of kinematical Lie (super)algebras to the Maxwellian case. The Maxwell algebra, first introduced in \cite{Schrader:1972zd,Bacry:1970du,Gomis:2017cmt} has received a growing interest due to its different applications in the (super)gravity context \cite{Salgado:2014jka,Hoseinzadeh:2014bla,deAzcarraga:2010sw,Durka:2011nf,Concha:2014xfa,Concha:2014tca,Penafiel:2017wfr,Ravera:2018vra,Concha:2018zeb,Concha:2018jxx,Chernyavsky:2020fqs,Kibaroglu:2020tbr,Caroca:2021bjo,Matulich:2023xpw}. A Maxwellian generalization of the cube presented in \cite{Concha:2023bly} and its supersymmetric extensions introduced here could require a subtle treatment. In particular, as it was pointed out in \cite{Concha:2021jnn}, the Maxwellian Carroll algebra obtained as a UR limit of the Maxwell algebra suffers from degeneracy, as in the NR limit. We expect to solve the degeneracy issue appearing in the Carrollian regime with a larger semigroup, analogously to the strategy employed here, in the NR expansion.

\section*{Acknowledgments}

The authors would like to thank Evelyn Rodríguez for valuable comments. This work was funded by the National Agency for Research and Development ANID - FONDECYT grants No. 1211077 and 11220328. This work was supported by FAA 2023 of the Universidad Católica de la Santísima Concepción (P.C.). L.R. would like to thank the DISAT of the Polytechnic of Turin and the INFN for financial support. P.C. and L.R. would like to thank to the Dirección de Investigación and Vice-rectoría de Investigación of the Universidad Católica de la Santísima Concepción, Chile, for their constant support.


\bibliographystyle{fullsort.bst}
 
\bibliography{Supergravity_based_on_kinematical_superalgebras}

\end{document}